\let\DRAGKernelLabel\label
\documentclass[aps,prx,reprint,floatfix,raggedbottom,superscriptaddress,longbibliography]{revtex4-2}
\let\label\DRAGKernelLabel
\usepackage{amsmath,amssymb,bm,graphicx,booktabs,xcolor,needspace,placeins}
\usepackage{capt-of}
\usepackage[stretch=30,shrink=30]{microtype}
\usepackage[colorlinks=true,allcolors=blue]{hyperref}
\hypersetup{
 pdftitle={Leakage suppression for high-fidelity exchange gates and quantum simulation in superconducting circuits},
 pdfauthor={Yiwen Li; Xinsheng Tan; Yang Yu},
 pdfsubject={Unified parametric control for quantum gates and simulation},
 pdfkeywords={PDRAG, superconducting qubits, parametric exchange, transmon leakage, quantum simulation}
}
\graphicspath{{figures/}}
\newcommand{\ket}[1]{\lvert #1\rangle}
\newcommand{\bra}[1]{\langle #1\rvert}
\newcommand{\Tr}{\operatorname{Tr}}
\newcommand{\ii}{\mathrm{i}}
\newcommand{\dd}{\mathrm{d}}

\providecommand{\wrap}{\operatorname{wrap}}

\newcommand{\DuffingMaxError}{0.000002}
\newcommand{\ChargeMaxError}{0.160063}

\newcommand{\DuffingMapDifference}{$5.47\times10^{-10}$}

\begin{document}
\title{Leakage suppression for high-fidelity exchange gates and quantum simulation in superconducting circuits}
\author{Yiwen Li}
\affiliation{National Laboratory of Solid State Microstructures, School of Physics, Nanjing University, Nanjing 210093, China}
\affiliation{Shishan Laboratory, Suzhou Campus of Nanjing University, Suzhou 215000, China}
\affiliation{Jiangsu Key Laboratory of Quantum Information Science and Technology, Nanjing University, Suzhou 215163, China}
\affiliation{Synergetic Innovation Center of Quantum Information and Quantum Physics, University of Science and Technology of China, Hefei, Anhui 230026, China}
\affiliation{Hefei National Laboratory, Hefei 230088, China}

\author{Xinsheng Tan}
\email[Contact author: ]{tanxs@nju.edu.cn}
\affiliation{National Laboratory of Solid State Microstructures, School of Physics, Nanjing University, Nanjing 210093, China}
\affiliation{Shishan Laboratory, Suzhou Campus of Nanjing University, Suzhou 215000, China}
\affiliation{Jiangsu Key Laboratory of Quantum Information Science and Technology, Nanjing University, Suzhou 215163, China}
\affiliation{Synergetic Innovation Center of Quantum Information and Quantum Physics, University of Science and Technology of China, Hefei, Anhui 230026, China}
\affiliation{Hefei National Laboratory, Hefei 230088, China}

\author{Yang Yu}
\email[Contact author: ]{yuyang@nju.edu.cn}
\affiliation{National Laboratory of Solid State Microstructures, School of Physics, Nanjing University, Nanjing 210093, China}
\affiliation{Shishan Laboratory, Suzhou Campus of Nanjing University, Suzhou 215000, China}
\affiliation{Jiangsu Key Laboratory of Quantum Information Science and Technology, Nanjing University, Suzhou 215163, China}
\affiliation{Synergetic Innovation Center of Quantum Information and Quantum Physics, University of Science and Technology of China, Hefei, Anhui 230026, China}
\affiliation{Hefei National Laboratory, Hefei 230088, China}
\date{\today}
\begin{abstract}
Parametric frequency modulation makes superconducting interactions programmable for quantum gates and quantum simulation.
Leakage into higher transmon levels and coupler modes degrades gate fidelity and takes simulated dynamics outside the intended state space.
Simple pulse shaping reduces off-resonant excitation, while strong coupling and short pulses demand simultaneous control of multiple leakage pathways.
Controlling these pathways across individual exchange operations and simultaneous couplings is essential for fast gates and faithful programmable dynamics.
We introduce parametric derivative removal by adiabatic gate (PDRAG), combining derivative shaping and finite-pulse calibration for qubit and coupler modulation.
At an interqubit coupling of $g/2\pi=100$ MHz, simulations demonstrate 13-ns exchange operations, approaching the timescale of fast single-qubit gates.
Combined with calibrated phase accumulation, these pulses realize a two-parameter fSim gate family with infidelity below $1.8\times10^{-7}$.
For quantum simulation, PDRAG suppresses leakage and improves fidelity in continuous and digital Floquet simulations of Su--Schrieffer--Heeger chains.
Floquet analysis identifies finite-pulse return interference as the calibration mechanism; realistic-circuit transfer preserves suppression.
PDRAG establishes a unified framework spanning quantum gates and quantum simulation.
\end{abstract}
\maketitle
\raggedbottom

\section{Introduction}
\label{sec:introduction}
Parametric frequency modulation makes superconducting interactions
programmable for quantum computation and simulation.
Frequency control enables high-fidelity entangling gates
\cite{Barends2014}, while periodic modulation selects the strength
and phase of exchange between detuned qubits
\cite{Bertet2006,Didier2018,Caldwell2018,Reagor2018}.
These capabilities support state transfer \cite{Li2018} and digital
simulations of spin and fermionic models \cite{Salathe2015,Barends2015}.
Programmable superconducting processors also realize synthetic
electromagnetic fields and Hall dynamics \cite{Rosen2024}, topological
phases \cite{Cai2019,Qian2025Topology}, and prethermal dynamics
\cite{Liu2026Prethermal}.
The exchange waveform therefore controls fSim gates and programs simulated
Hamiltonians, including phase-controlled hopping for Aharonov--Bohm interference
[Fig.~\ref{fig:mechanism}(a)] \cite{Rosen2024,Foxen2020}.

Stronger coupling accelerates these operations but also increases
unwanted excitation of higher transmon levels.
Because transmons are weakly anharmonic \cite{Koch2007}, the interaction
that exchanges $\ket{01}$ and $\ket{10}$ also couples $\ket{11}$
to $\ket{20}$ and $\ket{02}$.
Short pulses broaden the excitation spectrum and drive these
off-resonant transitions.
Leakage lowers gate fidelity \cite{Wood2018,Chen2016} and can persist
through subsequent operations, producing correlated errors that
degrade quantum error correction
\cite{McEwen2021Leakage,Miao2023Leakage}.
In spin and hard-core-boson simulations, occupation of higher transmon
levels changes the local state space assumed by the target model
\cite{Salathe2015,Karamlou2024Entanglement}.
Leakage suppression is therefore essential for converting faster
interactions into accurate gates and faithful simulated dynamics.

Programmable exchange in qubit--coupler--qubit (Q--C--Q) circuits
faces the same leakage problem. Tunable couplers allow independently
chosen qubit operating frequencies \cite{McKay2016,Yan2018}.
A separate coupler port activates exchange, shapes conditional
interactions, and suppresses idle crosstalk
\cite{Roth2017,Mundada2019,Stehlik2021}.
This architecture supports parametric entanglement
\cite{Ganzhorn2019,Han2020} and fast, high-fidelity exchange gates
\cite{Sung2021}.
Hybridization also opens transitions involving excited coupler levels.
Direct qubit--qubit (Q--Q) and Q--C--Q circuits must therefore retain
the desired exchange while suppressing multilevel transitions driven
by the same frequency modulation.

Smooth frequency trajectories reduce nonadiabatic excitation,
and interference can further suppress residual leakage
\cite{Sung2021,Martinis2014,Rol2019}.
For microwave control, derivative removal by adiabatic gate (DRAG)
uses a quadrature correction to cancel leading leakage
\cite{Motzoi2009,Gambetta2011}.
Higher-order constructions improve spectral selectivity
\cite{Motzoi2013,Hyyppa2024} and address additional transitions under
strong driving \cite{Jesus2026Blueprint}.
Related advances include parametric two-qubit control \cite{Jin2025},
analytical suppression for flux-tuned exchange \cite{Georgiadis2026},
and Magnus-based cancellation with a single baseband field
\cite{Heunisch2026}.
For parametric exchange, the complex envelope and its conjugate
drive Q--Q leakage at different signed detunings, while Q--C--Q
circuits add dressed coupler transitions.
Suppressing these channels simultaneously requires selective spectral
shaping and control of population return during a finite pulse.

We introduce parametric DRAG (PDRAG), which combines channel-selective
shaping with control of finite-pulse population return.
First- and second-derivative corrections jointly address the
conjugate Q--Q channels, even with unequal anharmonicities;
five- and seven-root constructions extend the same principle to
Q--C--Q leakage.
Across the Q--Q duration scan, joint derivative control lowers
endpoint leakage relative to both independently optimized
single-derivative controls at the same exchange angle.
Strong-coupling simulations reach full exchange in 13 ns,
comparable to the duration of fast single-qubit gates
\cite{Hyyppa2024,Jesus2026Blueprint}.
Combining these pulses with calibrated phase accumulation realizes
a two-parameter fermionic simulation (fSim) gate family with independently specified
exchange angles and conditional phases.
A moving Floquet basis \cite{Weinberg2017Floquet} resolves the origin
of the calibration gain. Transferring the pulses to a realistic circuit
Hamiltonian preserves their advantage at short gate times.

We also apply PDRAG to quantum simulation in seven- to thirteen-qubit
Su--Schrieffer--Heeger (SSH) chains, using continuous exchange
schedules and digital Floquet sequences.
Smoothly switched half-sine and quarter-cosine envelopes provide
additional waveform tests motivated by quantum simulation and
state-transfer protocols \cite{Wang2018Topological,Dangelis2020Transfer}.
Continuous modulation follows the SSH zero-mode transfer route
\cite{Mei2018Transfer}, whereas digital Floquet simulation builds
SSH dynamics from alternating odd- and even-bond layers
\cite{Asboth2014DrivenSSH}.
In both settings, derivative shaping reduces target-state errors
and residual coupler leakage, extending the method from individual
exchange operations to programmable lattice dynamics.
The Supplemental Material provides derivations, waveform tests,
control parameters, calibration procedures, and numerical checks
\cite{SupplementalMaterial}.

\section{Signed-channel construction of parametric DRAG}
\label{sec:rc:construction}

PDRAG associates each targeted leakage response with a signed detuning
$r_\ell$ from the relevant control sideband.
For a smooth base envelope $A(t)$, we construct
\begin{equation}
 \mathcal E_m(t)=s\prod_{\ell=1}^{m}
 \left(1-\frac{\ii}{r_\ell}\frac{\dd}{\dd t}\right)A(t),
 \label{eq:general:derivative}
\end{equation}
Here $m$ is the number of targeted spectral zeros, $r_\ell\ne0$ are
angular-frequency detunings, and $s$ scales the envelope amplitude.
The amplitude scale gives $\mathcal E_m$ angular-frequency units,
and $T$ denotes the pulse duration.
For $A^{(k)}(0)=A^{(k)}(T)=0$, $k=0,\ldots,m-1$, integration by
parts with the Fourier convention
$\widetilde{\mathcal E}(\delta)=\int_0^T\mathcal E(t)e^{\ii\delta t}\dd t$
gives
\begin{equation}
 \widetilde{\mathcal E}_m(\delta)
 =s\prod_{\ell=1}^{m}\left(1-\frac{\delta}{r_\ell}\right)
 \widetilde A(\delta).
 \label{eq:general:roots}
\end{equation}
Each factor cancels the spectral amplitude at a signed detuning.
The matrix elements of the modulated operator identify the associated transition.
Figures~\ref{fig:mechanism}(b),(c) show the Q--Q and Q--C--Q
architectures and their leakage classes.

\subsection{Exchange and conjugate leakage channels}

\begin{figure*}[!t]
 \centering
 \includegraphics[width=0.94\textwidth]{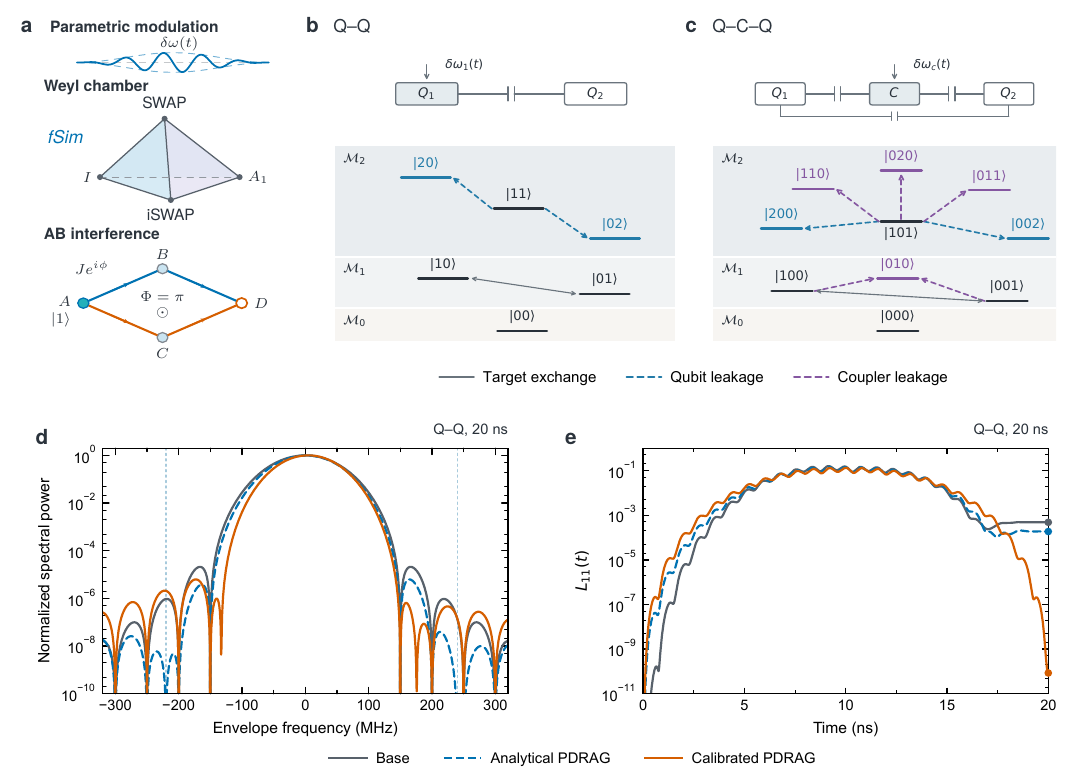}
 \caption{Parametric quantum gates, quantum simulation, and leakage suppression.
(a) Parametric frequency modulation $\delta\omega(t)$ (top, with a dashed envelope)
enables fSim gates (Weyl chamber) and quantum simulation, illustrated by
four-site Aharonov--Bohm interference (bottom)
\cite{Rosen2024,Foxen2020}.
The fSim family occupies the shaded $I$--iSWAP--SWAP,
$I$--$A_1$--SWAP and $A_1$--iSWAP--SWAP faces;
$A_1$ is locally equivalent to $I$.
At $\Phi=\pi$, the two paths interfere destructively, leaving D dark (open circle).
(b),(c) Capacitively coupled Q--Q and Q--C--Q circuits, with modulated
elements shaded; the lower link in (c) denotes weak direct qubit--qubit coupling.
Computational levels are dark; blue and purple denote qubit double-excitation
and coupler-associated leakage.
Solid and dashed arrows mark exchange and leakage-entry responses.
Idle dressed states are labeled by bare occupations, ordered as
$(Q_1,C,Q_2)$ in (c), and grouped into excitation manifolds $\mathcal M_N$;
level spacings are schematic.
(d) Fourier power of complex Q--Q envelopes, normalized at zero frequency;
dotted lines mark analytical roots at $-220$ and $+240$ MHz.
(e) Leakage $L_{11}(t)$ from $\ket{11}_{\rm d}$; filled circles mark final values.
Both use 20 ns Q--Q pulses with a $\sin^4$ base and $g/2\pi=100$ MHz.}
 \label{fig:mechanism}
\end{figure*}

Two capacitively coupled Duffing oscillators obey
\begin{align}
 H(t)/\hbar={}&\sum_{j=1}^{2}\left[\omega_j n_j+
 \frac{\alpha_j}{2}n_j(n_j-1)\right]
 +\delta\omega_1(t)n_1\notag\\
 &+g(a_1^\dagger a_2+a_1a_2^\dagger),
 \label{eq:rc:duffing}
\end{align}
Here $a_j$ is the annihilation operator of transmon $j$,
$n_j=a_j^\dagger a_j$, $\omega_j$ is the idle frequency, and
$\alpha_j<0$ is the anharmonicity. The exchange coupling is $g$,
and $\delta\omega_1(t)$ is the frequency excursion applied to transmon 1.
We define the idle detuning as $\Delta=\omega_1-\omega_2>0$.
Hamiltonian frequencies are angular frequencies; GHz and MHz values
include division by $2\pi$. We use
$(\omega_1,\omega_2)/2\pi=(5.10,4.50)$ GHz,
$(\alpha_1,\alpha_2)/2\pi=(-220,-240)$ MHz, and
$g/2\pi=100$ MHz, so that $g=2\pi\times0.100$ rad/ns.
The static exchange is treated within the rotating-wave approximation.
Propagation retains all modulation harmonics.

Writing $\varphi(t)=\int_0^t\delta\omega_1(t')\dd t'$, the coefficient
of $a_1^\dagger a_2$ in the phase frame is
$g\exp\{\ii[\Delta t+\varphi(t)]\}$.
For carrier angular frequency $\Omega$, we write the phase modulation as
\begin{equation}
 \varphi(t)=\operatorname{Im}[z(t)e^{-\ii\Omega t}],
 \qquad z(t)=\rho(t)e^{\ii\chi(t)},\quad 0\leq t\leq T,
 \label{eq:rc:phase}
\end{equation}
where $z$ is the dimensionless complex modulation amplitude,
with magnitude $\rho\geq0$ and phase $\chi$.
Its first harmonic near $\Omega=\Delta$ has complex exchange envelope
\begin{equation}
 G(t)=gJ_1[\rho(t)]e^{\ii\chi(t)}.
 \label{eq:rc:harmonic}
\end{equation}
Here $J_1$ is the first-order Bessel function, and $\ket{mn}$ labels
the bare transmon levels.
The envelope drives $\ket{01}\leftrightarrow\ket{10}$ and couples
$\ket{11}$ to $\ket{20}$ and $\ket{02}$ through
$\sqrt{2}G$ and $\sqrt{2}G^*$, respectively.
These two off-resonant transitions form the qubit double-excitation
leakage class in Fig.~\ref{fig:mechanism}(b).
Using angular frequency $\delta$ and the Fourier convention
\begin{equation}
 \widetilde G(\delta)=\int_0^T G(t)e^{+\ii\delta t}\dd t,
 \label{eq:rc:fourier}
\end{equation}
the first-order spectral-design amplitudes from $\ket{11}$
in the resonant exchange-harmonic model are
\begin{equation}
 c_{20,\rm spec}^{(1)}=-\ii\sqrt{2}\,\widetilde G(\alpha_1),\qquad
 c_{02,\rm spec}^{(1)}=-\ii\sqrt{2}\,\widetilde G(-\alpha_2)^*.
 \label{eq:rc:signed}
\end{equation}
The conjugate matrix element places the second spectral zero at
$-\alpha_2$, giving $(\alpha_1,-\alpha_2)/(2\pi)=(-220,+240)$ MHz.
These two first-order cancellation conditions determine the analytical pulse.
Idle dressing and finite-pulse population return shift the interference
minimum, motivating calibration within the same derivative family.

\subsection{Joint derivative construction with a smooth envelope}

For a nominal full-exchange angle $\Theta=\pi/2$, the base envelope is
\begin{equation}
 A_4(t)=\frac{8\Theta}{3T}\sin^4\!\left(\frac{\pi t}{T}\right),
 \quad 0\leq t\leq T,\qquad \Theta=\frac{\pi}{2},
 \label{eq:rc:base}
\end{equation}
and vanishes outside $0\leq t\leq T$, with area $\Theta$.
The envelope and its first three derivatives vanish at both endpoints.
We combine an in-phase second derivative with a quadrature first derivative,
\begin{equation}
 G(t)=s[A_4(t)+\ii\beta\dot A_4(t)+\gamma\ddot A_4(t)],
 \label{eq:rc:pulse}
\end{equation}
where $s$ is a dimensionless amplitude scale. The real coefficients
$\beta$ and $\gamma$ have units of time and time squared, respectively.
For the Fourier convention in Eq.~\eqref{eq:rc:fourier}, integration
by parts gives
\begin{equation}
 \widetilde G(\delta)=s[1+\beta\delta-\gamma\delta^2]\widetilde A_4(\delta).
 \label{eq:rc:multiplier}
\end{equation}
The construction follows the derivative-control principle of DRAG \cite{Motzoi2009,Motzoi2013}.
Related pulse schemes suppress unwanted transitions in frequency-crowded
devices \cite{Schutjens2013,Theis2016}, superconducting qudits
\cite{Li2025Qudit}, and qubit-defect systems~\cite{Wang2025Balanced}.
The signed roots give
\begin{align}
 \mathcal P_{\rm an}(\delta)&=\left(1-\frac{\delta}{\alpha_1}\right)
                         \left(1+\frac{\delta}{\alpha_2}\right),
 \label{eq:rc:polynomial}\\
 \beta_{\rm an}&=\frac1{\alpha_2}-\frac1{\alpha_1},\qquad
 \gamma_{\rm an}=\frac1{\alpha_1\alpha_2}.
 \label{eq:rc:coefficients}
\end{align}
The quadrature coefficient depends on the anharmonicity mismatch through
$\beta_{\rm an}=(\alpha_1-\alpha_2)\gamma_{\rm an}$.
The second-derivative scale $\gamma_{\rm an}$ is set by the inverse
product of the two anharmonicities.
For this device, $\beta_{\rm an}\simeq0.0603$ ns and
$\gamma_{\rm an}\simeq0.480$ ns$^2$.
The quadrature term supplies the spectral asymmetry needed
to cancel both channels with two derivatives.
A real envelope has a conjugate-symmetric spectrum and requires roots
at both signs of each distinct anharmonicity.
For equal anharmonicities, $\beta_{\rm an}=0$ leaves a real
second-derivative correction.
The Supplemental Material quantifies the contribution of the quadrature term.

The conserved area, $\int_0^T G(t)\,\dd t=s\Theta$, provides the initial
amplitude for full-exchange calibration, which also adjusts the carrier.
Analytical PDRAG fixes the coefficients in Eq.~\eqref{eq:rc:coefficients};
calibrated PDRAG varies both to minimize endpoint leakage
[Fig.~\ref{fig:mechanism}(d),(e)].

\subsection{Bessel inversion and the real frequency command}

On the increasing branch of $J_1$, the designed envelope determines
\begin{equation}
 \rho(t)=J_1^{-1}\!\left(\frac{|G(t)|}{g}\right),\qquad
 z(t)=\rho(t)\frac{G(t)}{|G(t)|},
 \label{eq:rc:bessel}
\end{equation}
with $z=0$ at $G=0$ and the continuous small-amplitude limit
$z\simeq2G/g$. Differentiating Eq.~\eqref{eq:rc:phase} gives
the real control command
\begin{equation}
 \delta\omega_1(t)=\operatorname{Im}
 \left\{[\dot z(t)-\ii\Omega z(t)]e^{-\ii\Omega t}\right\}.
 \label{eq:rc:command}
\end{equation}
The $\dot z$ term incorporates the time dependence of the envelope.
A monotonic transmon frequency--flux relation, fitted to the idle
frequency and anharmonicity, converts this command into applied flux $\Phi$.
We restrict the flux to $0\leq\Phi/\Phi_0\leq0.49$,
where $\Phi_0$ is the superconducting flux quantum.

Because $A_4$ and its first three derivatives vanish at the
endpoints, both $G$ and $\dot G$ approach zero there.
The Bessel-inverted frequency command therefore joins continuously
to idle. For $\delta\omega_1/(2\pi)$, we limit the peak and
root-mean-square (RMS) excursions to $1.0$ and $0.33$ GHz,
the slew to $4.1$ GHz/ns, and the one-sided
$99.9\%$-power bandwidth to $1.0$ GHz.
The Bessel utilization is
$u=\max_t|G|/[gJ_1(\rho_*)]\leq0.98$, where
$\rho_*\simeq1.84$ locates the first positive maximum of $J_1$.
These resources are evaluated for each complete real command,
including its joins to idle (Fig.~S8).

\subsection{Multichannel control through a tunable coupler}
\label{sec:qcq:construction}

The Q--C--Q implementation applies the same spectral polynomial to
the coupler-frequency envelope.
Its three-oscillator Hamiltonian is
\begin{align}
 H_{\rm QCQ}(t)/\hbar={}&\sum_{j=1,c,2}
 \left[\omega_j n_j+\frac{\alpha_j}{2}n_j(n_j-1)\right]
 +\delta\omega_c(t)n_c\notag\\
 &+\sum_{(j,k)}g_{jk}(a_j+a_j^\dagger)(a_k+a_k^\dagger),
 \label{eq:qcq:hamiltonian}
\end{align}
The links are $(1,c)$, $(2,c)$, and $(1,2)$. The parameters are
\[
\begin{aligned}
 (\omega_1,\omega_c,\omega_2)/2\pi&=(5.10,4.80,4.50)\ \mathrm{GHz},\\
 (\alpha_1,\alpha_c,\alpha_2)/2\pi&=(-220,-250,-235)\ \mathrm{MHz},\\
 (g_{1c},g_{2c},g_{12})/2\pi&=(150,142.5,5)\ \mathrm{MHz}.
\end{aligned}
\]
Propagation retains the full transverse couplings, including
counter-rotating terms, with fixed oscillator anharmonicities.
The computational states are the idle dressed eigenstates assigned to
$000,001,100,101$, with labels ordered as $(Q_1,C,Q_2)$.
Although $n_c$ is diagonal in the uncoupled oscillator basis,
qubit--coupler hybridization gives it off-diagonal matrix elements
in the idle eigenbasis. The same frequency modulation therefore
drives both $100\leftrightarrow001$ exchange and leakage.

We classify the seven targeted leakage responses by the bare-state
assignments of their destinations [Fig.~\ref{fig:mechanism}(c)].
Two responses, $101\to200$ and $101\to002$, form the qubit
double-excitation class, analogous to the off-resonant Q--Q channels.
The other five form the coupler-associated class: four responses with
a single coupler excitation, $100\to010$, $001\to010$,
$101\to110$, and $101\to011$, and one with a double coupler
excitation, $101\to020$.
Both classes are driven by the same dressed coupler-number operator $n_c$.
Transitions such as $110\leftrightarrow011$ subsequently redistribute
population within the leakage subspace.

For a response from $a_\ell$ to $b_\ell$, define the signed idle gap
$\Delta_\ell=(E_{b_\ell}-E_{a_\ell})/\hbar$, where $E_a$ is an idle dressed energy.
The real carrier supplies both sidebands; the corresponding root in the
complex envelope is $r_\ell=|\Delta_\ell|-\Omega$.
Seven-root PDRAG assigns one root to each response.
For the two channel pairs
$(100\to010,101\to011)$ and $(001\to010,101\to110)$,
five-root PDRAG uses the arithmetic mean of the two roots in each pair
and retains the other three roots.
Grouping roots reduces the number of adjustable parameters.

We use $A(t)=\sin^{10}(\pi t/T)$ in Eq.~\eqref{eq:general:derivative}
and implement the real frequency command
\begin{equation}
 \delta\omega_c(t)=\operatorname{Re}
       [\mathcal E_m(t)e^{-\ii\Omega t}],\qquad m=0,5,7.
 \label{eq:qcq:command}
\end{equation}
The base pulse has $m=0$.
Analytical pulses retain the idle-spectrum roots; calibrated pulses
vary them as $r_\ell\exp(\eta_\ell)$ with
$-0.35\leq\eta_\ell\leq0.35$.
Here $\eta_\ell$ is a dimensionless shift that preserves each root's sign.
The amplitude scale is normalized to enforce the common envelope bound.
The command and its first two derivatives join continuously to idle.

The coupler frequency is mapped to reduced flux $\varphi=\Phi/\Phi_0$
using the symmetric-SQUID transmon relation
\begin{equation}
 \omega_c(\varphi)=\sqrt{8E_CE_{J\Sigma}|\cos(\pi\varphi)|}/\hbar
                   -E_C/\hbar,
 \label{eq:qcq:flux}
\end{equation}
Here $E_C$ is the charging energy and $E_{J\Sigma}$ is the sum of the
two junction Josephson energies, with $E_C/h=0.25$ GHz and a maximum
transition frequency of 7.5 GHz.
The inverse relation determines the idle bias and the flux waveform
corresponding to each frequency command.
We enforce $0.02\leq\varphi\leq0.47$, a flux slew of at most
$1\,\Phi_0$/ns, and peak frequency excursion and envelope amplitude
of at most $2$ GHz after division by $2\pi$.
Propagation uses the frequency command in Eq.~\eqref{eq:qcq:hamiltonian};
the nonlinear inverse map supplies the associated flux constraints.
The Supplemental Material gives the bandwidth, RMS, and numerical checks.

\subsection{Comparison at fixed exchange action}
\label{sec:comparison}
\label{sec:gate-metrics}

We use unitary propagation to evaluate leakage and gate fidelity
in the coupled device's idle eigenbasis.
Let $\ket{mn}_{\rm d}$ denote an eigenstate of the coupled idle
Hamiltonian assigned to the bare product label $mn$.
The matrix
\begin{equation*}
 B=\big(\ket{00}_{\rm d},\ket{01}_{\rm d},
         \ket{10}_{\rm d},\ket{11}_{\rm d}\big)
\end{equation*}
has these four orthonormal states as columns, so $B^\dagger B=I_4$.
The computational and leakage projectors are $P=BB^\dagger$ and
$Q=I-P$, where $I$ is the identity on the full simulated Hilbert space.
For the full propagator $U(T)$ generated by the device Hamiltonian,
$K=B^\dagger U(T)B$ is the $4\times4$ computational map.
Its row and column indices $0,1,2,3$ follow the order $00,01,10,11$.
The principal exchange angle, $0\leq\theta\leq\pi/2$, is
\begin{equation}
 \theta=\operatorname{atan2}\!\left(
 \sqrt{\frac{|K_{12}|^2+|K_{21}|^2}{2}},
 \sqrt{\frac{|K_{11}|^2+|K_{22}|^2}{2}}\right).
 \label{eq:theta}
\end{equation}
For full exchange, amplitude and carrier are adjusted to null
the complex return amplitude $K_{11}$ for $\ket{01}_{\rm d}$.
The full-exchange target is
$\theta=\pi/2$, with an allowed error of $5$ mrad.

The exchange operations carry a nonlocal conditional phase described
by the fSim family \cite{Foxen2020,Sung2021}.
In the ordered computational basis, $U_{\rm fSim}(\theta,\phi)$ leaves
$\ket{00}$ unchanged and multiplies $\ket{11}$ by $e^{-\ii\phi}$.
Its single-excitation block has diagonal entries $\cos\theta$ and
off-diagonal entries $-\ii\sin\theta$.
The conditional phase extracted from $K$ is
\begin{equation}
 \phi_c=\operatorname{wrap}_{[-\pi,\pi)}
 \arg\!\left(\frac{K_{00}K_{33}}{\det K_{1}}\right),
 \label{eq:phi}
\end{equation}
Here $K_1$ is the single-excitation block of $K$, and
$\operatorname{wrap}_{[-\pi,\pi)}$ reduces angles modulo $2\pi$.
This expression remains defined at zero exchange.
For excitation-preserving unitary evolution at nonzero exchange,
it equals $\arg(K_{00}K_{33}/K_{12}K_{21})-\pi$ modulo $2\pi$.

For $U_{\rm fSim}\ket{11}=e^{-\ii\phi}\ket{11}$, the diagnostic equals
$-\phi$ modulo $2\pi$ and is invariant under local $Z$ rotations.
For a computational target $V$, the average gate fidelity is
\begin{equation}
 F(K,V)=\frac{\Tr(K^\dagger K)+|\Tr(V^\dagger K)|^2}{20}.
 \label{eq:fidelity}
\end{equation}
Here $V$ is a unitary target on the four-dimensional computational subspace,
so the denominator is $4(4+1)$.
The map $K$ remains unnormalized, so computational population loss
reduces fidelity through $\Tr(K^\dagger K)$ \cite{Wood2018,Nielsen2002}.
Equation~\eqref{eq:fidelity} is the uniform pure-state average of the
overlap with the target output.
The figure label $F_{\rm avg}$ denotes this same average.

We optimize pre- and post-gate local $Z$ rotations for both fSim and iSWAP fidelities.
The best-fit fSim fidelity used in the duration scans additionally
optimizes $\theta$ and $\phi$ in Eq.~\eqref{eq:fidelity}.
The two-parameter map in Sec.~\ref{sec:fsim} fixes both angles at their
prescribed targets; strict iSWAP fixes $\theta=\pi/2$ and $\phi=0$.
These rotations are implemented as control-phase updates in the
virtual-$Z$ framework \cite{McKay2017}.

\begin{figure*}[!t]
 \centering
 \includegraphics[width=\textwidth]{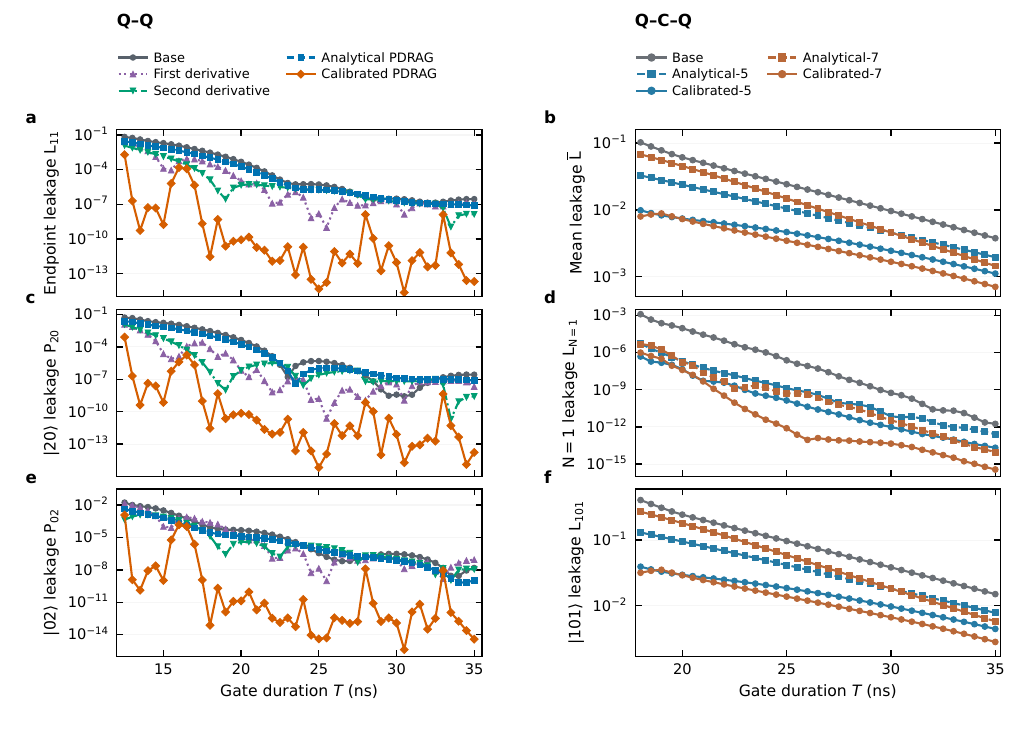}
 \caption{Duration-dependent leakage for direct and coupler-mediated
exchange. (a),(c),(e) Q--Q endpoint leakage $L_{11}$ and its resolved
channels $P_{20}$ and $P_{02}$ for the five derivative families.
The coupling is $g/2\pi=100$ MHz, with full-exchange calibration.
(b),(d),(f) Q--C--Q four-input mean leakage $\overline L$, mean leakage
$L_{N=1}$ from $001$ and $100$, and leakage $L_{101}$ from the doubly
excited computational input. All five Q--C--Q families share a
150-MHz per-link coupling cap.
Markers show 46 Q--Q durations over 12.5--35 ns and 35 Q--C--Q
durations over 18--35 ns, both at 0.5-ns intervals.
Lines connect adjacent samples.
}
 \label{fig:duration}
\end{figure*}

The five Q--Q families are the uncorrected base pulse (Base),
$(\beta,\gamma)=(0,0)$,
first-derivative-only PDRAG $(\gamma=0)$,
second-derivative-only PDRAG $(\beta=0)$, analytical PDRAG with
Eq.~\eqref{eq:rc:coefficients}, and calibrated PDRAG with both
derivative coefficients varied. For these full-exchange duration scans,
the common bounds are
$s\in[0.5,1.3]$, $\beta\in[-4,1.6]$ ns, and
$\gamma\in[-4,3.2]$ ns$^2$. The carrier interval extends
$80$ MHz on each side of the idle one-excitation splitting,
$\sqrt{\Delta^2+4g^2}/(2\pi)\simeq632$ MHz.
Each candidate is evaluated using its duration-specific envelope
and complete Bessel-inverted frequency command.

Leakage is evaluated by orthogonal projection outside the idle
computational subspace,
\begin{equation}
 L_{11}=\big\|QU(T)\ket{11}_{\rm d}\big\|^2.
 \label{eq:leakage}
\end{equation}
The resolved probabilities $P_{20}$ and $P_{02}$ are the endpoint
populations of $\ket{20}_{\rm d}$ and $\ket{02}_{\rm d}$ for this input.
Excitation conservation in the Duffing model gives
$L_{11}=P_{20}+P_{02}$, which is minimized over the derivative coefficients.
The scan covers $46$ durations from $12.5$ to $35$ ns in $0.5$-ns steps.
Calibration combines deterministic derivative grids, multibasin refinement,
bidirectional continuation in duration, and checks of neighboring
parameter branches. Both independently optimized single-derivative
commands enter the joint search at each duration.
The Supplemental Material gives the calibration landscape, selected
coefficients, and independent propagation checks.

\section{Leakage suppression across exchange architectures}
\subsection{Direct qubit--qubit exchange}\label{sec:duffing-results}

Joint derivative control maintains full exchange while suppressing both
leakage channels across the entire 12.5--35 ns scan
[Fig.~\ref{fig:duration}(a),(c),(e)].
The geometric-mean leakage reduction is $5.27\times10^5$ relative
to the base pulse and $1.92\times10^4$ relative to the better
single-derivative control at each duration.
The joint pulse improves on the better single-derivative control
at every sampled duration. All 230 pulses satisfy the exchange
and resource bounds.
Table~S2 gives the complete interval statistics.

At 13 ns, calibration lowers $L_{11}$ from $5.56\times10^{-2}$
to $1.98\times10^{-7}$, a reduction of $2.81\times10^5$.
The first- and second-derivative-only pulses leave
$1.65\times10^{-2}$ and $7.57\times10^{-3}$, respectively.
Joint control therefore suppresses the residual leakage by a further
$3.82\times10^4$ relative to the better single-derivative pulse.
Leakage remains at or below $1.98\times10^{-7}$ at 13, 13.5, and 14 ns.

At 20 ns, the analytical spectral zeros reduce $L_{11}$ from
$4.80\times10^{-4}$ to $1.85\times10^{-4}$; calibration lowers it
further to $8.54\times10^{-11}$.
The additional suppression arises from coherent return of transient
leakage population to the computational subspace [Fig.~\ref{fig:mechanism}(e)].
Section~\ref{sec:floquet_returns} resolves the interfering amplitudes.

At the shortest sampled duration, 12.5 ns, calibrated PDRAG reduces
leakage from $6.82\times10^{-2}$ to $1.98\times10^{-3}$.

\subsection{Coupler-mediated exchange}
\label{sec:qcq:results}

For each Q--C--Q input $b\in\{000,001,100,101\}$, define
$L_b=\|QU(T)\ket b_{\rm d}\|^2$, where $Q$ projects outside the
four idle dressed computational states.
The three plotted measures are
\begin{align}
 \overline L&=\frac{L_{000}+L_{001}+L_{100}+L_{101}}4,\notag\\
 L_{N=1}&=\frac{L_{001}+L_{100}}2.
 \label{eq:qcq:leakage}
\end{align}
The third measure, $L_{101}$, isolates the doubly excited input.
These measures include all population outside the computational
subspace, including residual coupler excitation.

The five Q--C--Q pulse families are compared at the same full-exchange
target and under common waveform constraints.
Each selected pulse has a mean transfer probability of at least 0.99,
directional return probabilities of at most $\sin^2(0.005)$, and
an exchange-angle error of at most 5 mrad.
Both transfer and return probabilities refer to the dressed states
$\ket{001}_{\rm d}$ and $\ket{100}_{\rm d}$ and include leakage loss.
The 10--35 ns search contains 255 family--duration cases.
All five families meet the criteria at the 35 common durations from
18 to 35 ns shown in Fig.~\ref{fig:duration}(b),(d),(f).
Propagation uses six levels per oscillator; comparison with five levels
changes $\overline L$ by at most $1.07\times10^{-7}$ across the complete grid.
The Supplemental Material gives search budgets, selection criteria,
shorter-duration results, and numerical resolution for the smaller $L_{N=1}$ values.

Calibrated five- and seven-root PDRAG reduce $\overline L$ relative
to the base pulse at all 35 common durations.
Their median suppression factors are 4.71 and 6.59, respectively.
At 18 ns, the base leakage of $1.03\times10^{-1}$ falls to
$9.84\times10^{-3}$ with five roots and $7.97\times10^{-3}$
with seven roots, giving 10.4- and
12.9-fold reductions.

At 35 ns, seven-root PDRAG reaches $\overline L=7.03\times10^{-4}$
and 99.93\% best-fit fSim fidelity.
Seven-root PDRAG has lower mean leakage than five-root PDRAG
at 33 of the 35 durations.
At 27 ns, the seven-root pulse leaves $\overline L=2.60\times10^{-3}$
following transient population return to the computational subspace,
with a 6.46-fold reduction from the base pulse (Fig.~S9).

The single-excitation contribution falls more rapidly with gate duration,
leaving the doubly excited input as the dominant leakage source at long
durations [Fig.~\ref{fig:duration}(d),(f)].

\section{Finite-pulse return interference in direct exchange}
\label{sec:floquet_returns}

\begin{figure}[!t]
 \centering
 \setlength{\abovecaptionskip}{4pt}
 \includegraphics[width=\columnwidth,trim=0 2mm 0 0,clip]{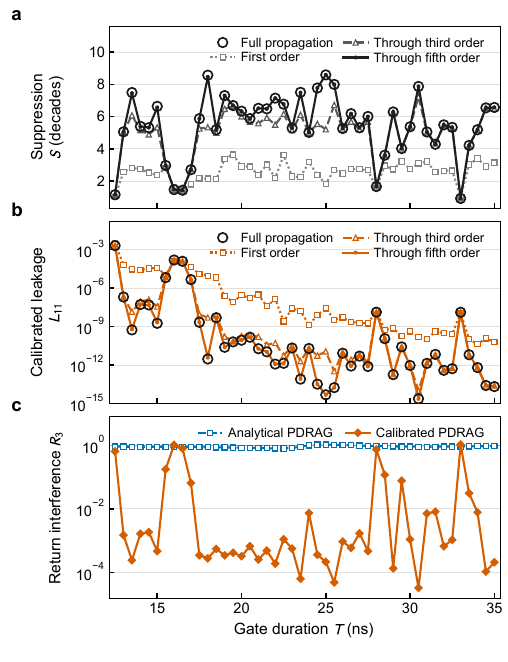}
 \caption{Finite-pulse returns explain the additional suppression from
PDRAG calibration at $g/2\pi=100$ MHz.
(a) Calibration gain $S=\log_{10}(L_{11}^{\rm an}/L_{11}^{\rm cal})$
from full propagation and the Floquet block expansion through first,
third, and fifth order. (b) Calibrated endpoint leakage for the same predictions.
(c) Endpoint interference ratio $R_3$ for analytical and calibrated PDRAG.
The unit line separates destructive and constructive interference between
the first- and third-order amplitudes.
All panels use 46 commands per family over 12.5--35 ns at 0.5-ns intervals,
with commands and Floquet references fixed across expansion orders;
lines connect calculated points.}
 \label{fig:floquet_returns}
\end{figure}

The spectral zeros in Eq.~\eqref{eq:rc:signed} cancel first-order
leakage at the signed bare transition frequencies.
During a finite gate, idle dressing, carrier detuning, and population
transfer modify the amplitudes that interfere at the endpoint.
Calibration adjusts their relative contributions within the same derivative family.
We resolve this interference with a Floquet reference for each complete
frequency command, using a moving basis for the periodically driven
Hamiltonian~\cite{Weinberg2017Floquet}.

Excitation conservation restricts evolution from $\ket{11}_{\rm d}$ to
$\{\ket{11}_{\rm d},\ket{20}_{\rm d},\ket{02}_{\rm d}\}$.
Let $H_2(t)$ be the Duffing Hamiltonian in this idle-dressed sector.
Freezing the complex command envelope gives a periodic Hamiltonian.
We track the Floquet modes of the single-period propagator from
the idle states and assemble them into the matrix $\mathsf F(t)$.
The moving-frame angular-frequency generator is
\begin{equation}
 h_{\rm Fl}=\mathsf F^\dagger
 \left(H_2/\hbar-\omega_{\rm ref}I\right)\mathsf F
 -\ii\mathsf F^\dagger\dot{\mathsf F},
 \label{eq:floquet:frame}
\end{equation}
Here $\omega_{\rm ref}=E_{11}/\hbar$ is the idle dressed energy of
$\ket{11}_{\rm d}$ in angular-frequency units, and $\mathsf F$ is unitary.
The term $-\ii\mathsf F^\dagger\dot{\mathsf F}$ accounts for motion
of the Floquet basis.
In the Floquet coordinates, $\Pi_P=\operatorname{diag}(1,0,0)$ selects
the continued computational branch and $\Pi_Q=I-\Pi_P$ the two leakage
branches. The physical state is
$e^{-\ii\omega_{\rm ref}t}\mathsf F(t)(p,\boldsymbol q)^{\mathsf T}$.
Here $p$ is the scalar computational-branch amplitude and
$\boldsymbol q$ contains the two leakage-branch amplitudes.
The block $h_{{\rm Fl},AB}=\Pi_A h_{\rm Fl}\Pi_B$, restricted to
the corresponding coordinates, couples branches $B$ to $A$,
with $A,B\in\{P,Q\}$.
Intrablock dynamics retain their full time dependence, including
mixing between the two leakage branches.

Expanding only the coupling between these blocks gives
$p=p_0+p_2+p_4+\cdots$ and
$\boldsymbol q=\boldsymbol q_1+\boldsymbol q_3+\boldsymbol q_5+\cdots$.
The contributions follow the recursion
\begin{align}
 \ii\dot p_{2n}&=h_{{\rm Fl},PP}p_{2n}
                  +h_{{\rm Fl},PQ}\boldsymbol q_{2n-1},\notag\\
 \ii\dot{\boldsymbol q}_{2n+1}&=h_{{\rm Fl},QQ}\boldsymbol q_{2n+1}
                  +h_{{\rm Fl},QP}p_{2n},
 \label{eq:floquet:recursion}
\end{align}
for $n=0,1,2$, with $\boldsymbol q_{-1}=0$, $p_0(0)=1$, and all
corrections initially zero. The order counts transitions between blocks.
The contribution $\boldsymbol q_3$ includes one return to the computational
branch, and $\boldsymbol q_5$ includes two.
For these pulses, $\mathsf F(0)=\mathsf F(T)=I$, so the predicted
physical endpoint leakage is
\begin{equation}
 L_{11}^{[2n+1]}=
 \left\|\sum_{j=0}^{n}\boldsymbol q_{2j+1}(T)\right\|^2,
 \qquad n=0,1,2.
 \label{eq:floquet:leakage}
\end{equation}
The Supplemental Material gives the construction and numerical checks.

The return expansion predicts the calibration gain
$S=\log_{10}(L_{11}^{\rm an}/L_{11}^{\rm cal})$ across all $46$ durations
[Fig.~\ref{fig:floquet_returns}(a),(b)].
The median absolute error falls from $3.00$ decades at first order to
$0.0950$ through third order and approximately $5\times10^{-4}$ through fifth order.
The fifth-order expansion predicts every suppression ratio $10^S$ within
a factor of $1.28$ of full propagation. It resolves all eight
third-order errors exceeding one decade.
At $25$ ns, for example, the calibrated leakage changes from the
third-order estimate $1.13\times10^{-11}$ to the fifth-order estimate
$4.71\times10^{-15}$, reproducing the full-propagation value
$4.73\times10^{-15}$.

Calibration drives the leading leakage and return contributions toward
destructive interference [Fig.~\ref{fig:floquet_returns}(c)].
The interference ratio
$R_3=\|\boldsymbol q_1(T)+\boldsymbol q_3(T)\|^2/
(\|\boldsymbol q_1(T)\|^2+\|\boldsymbol q_3(T)\|^2)$
compares the coherent sum with the sum of separate contributions.
Values below unity indicate destructive interference.
Its median falls from $0.907$ for analytical
PDRAG to $6.66\times10^{-4}$ for calibrated PDRAG.
For calibrated pulses, the median endpoint amplitude ratio
$\|\boldsymbol q_5(T)\|/\|\boldsymbol q_3(T)\|$ is $2.1\%$.
This higher-order return accounts for the residual after leading-order
cancellation and reduces the leakage-vector error at every calibrated
duration (Supplemental Material, Floquet return expansion).

\begin{figure}[!t]
 \centering
 \includegraphics[width=\columnwidth]{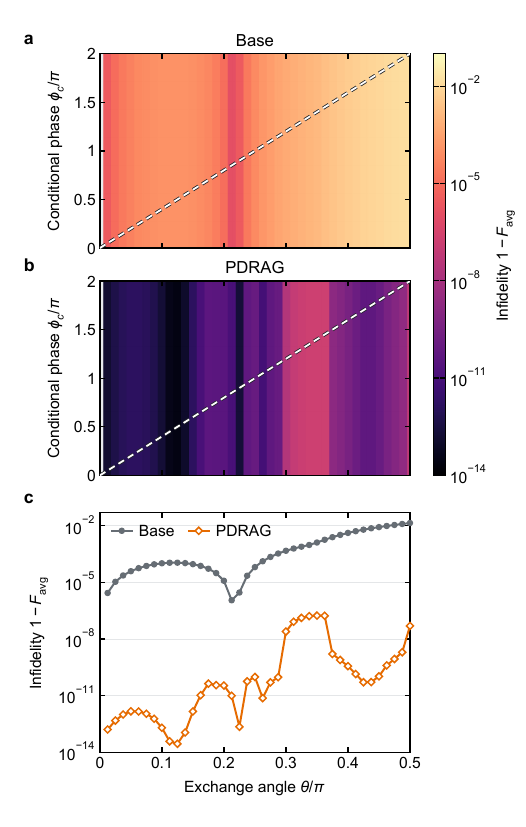}
 \caption{Two-parameter fSim gates at $g/2\pi=100$ MHz.
(a),(b) Fixed-target average gate infidelity for Base and PDRAG,
using a common logarithmic colour scale.
Each target $(\theta,\phi_c=-\phi)$ is realized by a 13-ns exchange
pulse followed by calibrated idle evolution.
(c) Infidelity along the dashed diagonal $\phi_c=4\theta$.
Grey cells at $\theta=0$ denote the exact zero-error idle operation,
which is omitted from the logarithmic cut.}
 \label{fig:fsim}
\end{figure}

\section{Two-parameter fSim gates}
\label{sec:fsim}

We combine leakage-suppressed exchange pulses with calibrated idle
evolution to control excitation exchange and conditional phase independently.
A 13-ns Q--Q pulse is calibrated for each exchange angle
$0\leq\theta\leq\pi/2$ and followed by an idle interval that sets the conditional phase.
Together, these controls realize the two-parameter fSim family
\cite{Foxen2020} while retaining the low leakage of the shaped pulse.

Let $E_{mn}$ be the eigenenergy of $\ket{mn}_{\rm d}$ under the
coupled idle Hamiltonian, and define its conditional frequency shift
\begin{equation}
 \zeta=\frac{E_{11}+E_{00}-E_{01}-E_{10}}{\hbar}.
 \label{eq:fsim:zz}
\end{equation}
Idle evolution is diagonal in this basis.
An idle interval $\tau$ preserves $\theta$ and shifts the conditional phase according to
\begin{equation}
 \phi_c(\tau)=\phi_c(0)-\zeta\tau \pmod{2\pi}.
 \label{eq:fsim:phase}
\end{equation}
For the present Q--Q parameters, $\zeta/2\pi=-24.8636$ MHz.
An idle interval $0\leq\tau<40.22$ ns therefore spans a full
conditional-phase cycle, with total durations from 13 to 53.22 ns.

For the targets in Fig.~\ref{fig:fsim}, both methods calibrate
amplitude and carrier at each of 41 exchange angles within common
resource bounds. PDRAG additionally
optimizes the two derivative coefficients.
Both methods meet a $5\times10^{-6}$-rad exchange tolerance and use
their own waiting interval to reach each of 65 conditional phases.
The Supplemental Material gives the controls, idle calibration,
and independent full-matrix checks.

PDRAG maintains infidelity below $1.76\times10^{-7}$ across the
sampled domain, whereas Base reaches $1.40\times10^{-2}$.
At every sampled nonzero exchange angle, PDRAG lowers the fixed-target
infidelity by at least $5.8\times10^3$.
The diagonal cut in Fig.~\ref{fig:fsim}(c) shows the same advantage
as both target parameters vary simultaneously.
Idle evolution changes diagonal phases in the computational eigenbasis,
preserving gate fidelity and producing the vertical bands in the maps.
The suppression achieved during the exchange pulse therefore extends
across a complete conditional-phase cycle at every calibrated exchange angle.

\section{Transfer to a realistic circuit Hamiltonian}
\label{sec:charge}

\begin{figure}[!t]
 \centering
 \setlength{\abovecaptionskip}{4pt}
 \includegraphics[width=\columnwidth]{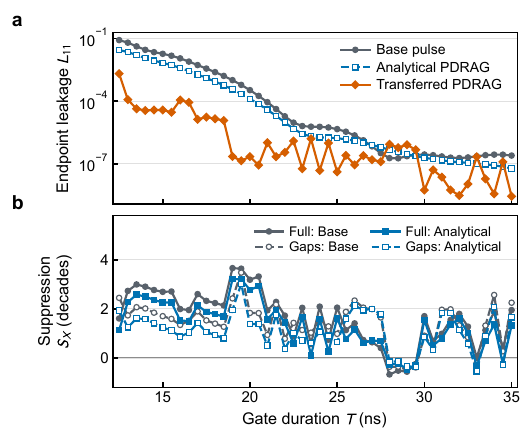}
 \caption{Charge-model transfer and the role of static dressed gaps
at $g/2\pi=100$ MHz.
(a) Endpoint leakage $L_{11}$ for the three pulse families.
(b) Relative suppression $S_X=\log_{10}(L_X/L_{\rm tr})$, where
$L_{\rm tr}$ is transferred-pulse leakage and $L_X$ is Base
($X=B$, grey) or analytical-PDRAG ($X=A$, blue) leakage.
Filled markers on solid lines show full charge propagation; open markers
on dashed lines show the three-state gap reference in Eq.~\eqref{eq:charge:gap-reference}.
Markers show 46 durations over 12.5--35 ns at 0.5-ns intervals;
lines connect samples.}
 \label{fig:charge}
\end{figure}

We next transfer the Q--Q derivative coefficients to a microscopic
charge model with matched idle device parameters [Fig.~\ref{fig:charge}].
The model retains the nonlinear Josephson potential and full capacitive
interaction, extending the multilevel description beyond the Duffing approximation.
The tunable-transmon Hamiltonian is
\begin{align}
 H_q(\Phi)={}&4E_C(\hat n-n_g)^2
 -E_{J\Sigma}\cos(\pi\Phi/\Phi_0)\cos\hat\phi\notag\\
 &-dE_{J\Sigma}\sin(\pi\Phi/\Phi_0)\sin\hat\phi,
 \label{eq:charge}
\end{align}
Here $\hat n$ is the Cooper-pair number operator, $\hat\phi$ its
conjugate phase, $E_C$ the charging energy, and $E_{J\Sigma}$ the sum
of the junction Josephson energies.
The junction asymmetry is
$d=(E_{J1}-E_{J2})/(E_{J1}+E_{J2})=0.15$.
We set the offset charge $n_g=0$ and idle
bias $\Phi/\Phi_0=0.25$.
The fitted $E_C$ and $E_{J\Sigma}$ reproduce the idle frequencies and
anharmonicities in the charge basis $n=-18,\ldots,18$.
The two devices interact through
$H_{\rm int}/\hbar=g_C\hat n_1\hat n_2$, where $g_C$ is the
capacitive-coupling coefficient in angular-frequency units.
Matching the idle exchange matrix element to $g/2\pi=100$ MHz gives
$g_C/2\pi\simeq68.8$ MHz.
Laboratory-frame propagation in the fixed idle transmon basis retains
counter-rotating couplings and flux-dependent spectra and matrix elements.

At each of the same 46 durations, Base and analytical PDRAG retain
their prescribed derivative coefficients. Transferred PDRAG uses the
coefficients selected by Duffing calibration at that duration.
Amplitude and carrier are recalibrated within the common search intervals
to meet the 5-mrad exchange tolerance.
Propagation retains nine idle levels per transmon; all 138 commands
satisfy the exchange and resource constraints.

Transferred PDRAG provides its largest advantage at short gate times.
Across 12.5--22 ns, transferred PDRAG outperforms the base and analytical
pulses at all 20 sampled durations.
The geometric-mean leakage reduction factor relative to the base is 372,
with a minimum reduction of 40.7.
At 13 ns, $L_{11}$ falls from $6.27\times10^{-2}$ to
$1.17\times10^{-4}$; at 20 ns, it falls from $3.33\times10^{-4}$
to $2.17\times10^{-7}$.

Across the full 46-point scan, transferred PDRAG lowers leakage relative
to the base pulse at 42 durations, with a geometric-mean reduction
of 43.8 (Table~S2).
At longer durations, shifted interference minima favor the base at
$28$--$29.5$ ns and analytical PDRAG at those four durations and
at $33$ and $34.5$ ns.

Shifts in the dressed leakage gaps account for this duration dependence.
With matched idle frequencies, anharmonicities, and exchange,
the charge-matrix-element ratios $|n_{21}^{(j)}/n_{10}^{(j)}|$ differ
from the Duffing value $\sqrt{2}$.
Here $n_{rs}^{(j)}=\bra r\hat n_j\ket s$ is evaluated in the isolated
idle eigenbasis of transmon $j$.
The resulting charge-model dressing shifts the two leakage gaps by
$-2.74$ and $+0.53$ MHz relative to Duffing.
We isolate this effect by holding each charge-calibrated command fixed
and replacing the Duffing gaps in the $(11,20,02)$ sector,
\begin{equation}
 h_{\rm G}(t)=\operatorname{diag}(0,\Delta_{20}^{\rm C},\Delta_{02}^{\rm C})
 +\delta\omega_1(t)M_{\rm D}^{(2)},
 \label{eq:charge:gap-reference}
\end{equation}
where $\Delta_{mn}^{\rm C}=(E_{mn}^{\rm C}-E_{11}^{\rm C})/\hbar$,
$E_{mn}^{\rm C}$ is the charge-model idle dressed energy, and
$M_{\rm D}^{(2)}$ is the matrix of $n_1$ in the Duffing idle-dressed
two-excitation sector.
This three-state reference isolates static gap shifts while retaining
the Duffing modulation matrix.
It predicts 91 of 92 pairwise leakage orderings,
including all ten losses of transfer advantage [Fig.~\ref{fig:charge}(b)].
Retaining the Duffing gaps misses all ten reversals, even with the
charge-corrected modulation matrix.
Restoring the flux dependence of exchange recovers the remaining
leakage reduction relative to the base pulse at 33 ns.
Static dressing therefore explains the transfer reversals, while
the driven exchange determines the remaining suppression depth.
The Supplemental Material gives this decomposition, the expanded-basis
check at 33 ns, and the mean leakage over computational inputs.

\subsection{Conditional phase and gate fidelity}

For the 20-ns calibrated Duffing pulse, $\phi_c\simeq2.33$ rad
and the best-fit fSim infidelity is $2.14\times10^{-11}$.
The transferred charge-model pulse has $\phi_c\simeq2.24$ rad
and a best-fit fSim infidelity of $5.44\times10^{-8}$
(Supplemental Material, charge-model gate metrics).
Their strict-iSWAP infidelities are $2.42\times10^{-1}$ and
$2.27\times10^{-1}$, respectively, dominated by the nonzero conditional phase.
The idle interval introduced in Sec.~\ref{sec:fsim} provides independent
control of the conditional phase.

For Q--C--Q exchange, calibrated seven-root PDRAG achieves best-fit
fSim fidelities of 99.20\% at 18 ns and 99.93\% at 35 ns.
Over the same interval, the conditional phase grows from about
0.603 to 2.32 rad, lowering strict-iSWAP fidelity from 97.41\%
to 75.98\% (Fig.~S10).
The Supplemental Material gives both fidelity measures for all five
pulse families across the common duration interval.

\section{Quantum simulation}
\label{sec:simulation}

We test leakage suppression in programmable lattice dynamics using
the SSH model, under both continuous modulation and digital Floquet driving.
Both protocols propagate all $2N-1$ qubit and coupler modes in the
single-excitation sector, where $N$ is the number of qubits.

\begin{figure}[!t]
 \centering
 \setlength{\abovecaptionskip}{4pt}
 \includegraphics[width=\columnwidth]{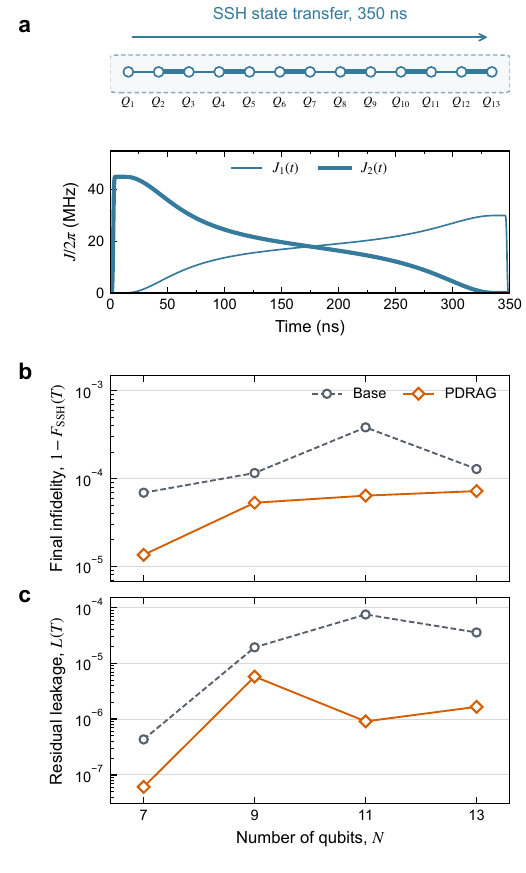}
 \caption{Continuous SSH quantum simulation over $T=350$ ns.
 (a) Thirteen-site chain and target coupling envelopes
 $J_1(t)$ and $J_2(t)$.
 Thin and thick blue curves denote $J_1$ and $J_2$, respectively.
 (b) Final infidelity $1-F_{\rm SSH}(T)$ relative to the finite-time
 ideal SSH state.
 (c) Residual population $L(T)$ in the idle-dressed coupler-like modes.
 Grey open circles and orange open diamonds denote Base and PDRAG.
 Lines guide the eye.}
 \label{fig:ssh}
\end{figure}

\subsection{Continuous SSH dynamics}
\label{sec:ssh}

Continuous modulation of the two SSH bond families enables state
transfer through chains containing $N=7,9,11,$ and $13$ qubits.
Each chain contains $N-1$ parametrically modulated couplers;
the target coupling envelopes for $N=13$ are shown in
Fig.~\ref{fig:ssh}(a).
In the single-excitation logical basis, the target Hamiltonian is
\begin{align}
 \frac{H_{\rm SSH}(t)}{\hbar}
 =\sum_{j=1}^{(N-1)/2}\big[&J_1(t)\ket{2j-1}\bra{2j}
 +J_2(t)\ket{2j}\bra{2j+1}\notag\\
 &+\mathrm{H.c.}\big],
 \label{eq:ssh:target}
\end{align}
where $\ket{j}$ denotes a logical excitation at site $j$,
and H.c. denotes the Hermitian conjugate.
The target bond envelopes are
\begin{align}
 J_1(t)/(2\pi)&=(30\,{\rm MHz})\,w(t)f(t),\notag\\
 J_2(t)/(2\pi)&=(45\,{\rm MHz})\,w(t)[1-f(t)].
 \label{eq:ssh:envelopes}
\end{align}
The common window $w(t)$ switches both couplings on and off over
$5$ ns at each end of the $350$-ns protocol.
Between these ramps, $f(t)$ increases smoothly from zero to one
at a rate adapted to the finite-chain gap.
As $J_1/J_2$ increases, the SSH zero mode,
$\ket{D(t)}\propto\sum_{k=0}^{(N-1)/2}[-J_1(t)/J_2(t)]^k\ket{2k+1}$,
moves from $Q_1$ to $Q_N$ \cite{Mei2018Transfer}.
All four chains use the same schedule rule and duration.

Each PDRAG coupler command contains four independent derivative
coefficients in addition to amplitude, carrier-frequency, phase,
and nonoscillatory frequency-offset (DC) controls.
The ports are calibrated jointly against the prescribed SSH trajectory
under common frequency-excursion and slew limits.
Base varies the non-derivative controls using the same target and
optimization budget.

Both controls are evaluated in a common logical frame that tracks
the idle frequencies and prescribed ac-Stark phase feedforward.
Let $B_{\rm s}$ contain the $N$ idle dressed logical states and
$D_{\rm s}$ be the diagonal matrix of their angular frequencies.
The shared phase reference is
$\Phi_{\rm s}(t)=D_{\rm s}t+\operatorname{diag}[\chi_1(t),\ldots,\chi_N(t)]$,
where $\chi_j(t)$ is the prescribed ac-Stark phase at site $j$.
For an initial excitation at $Q_1$, the logical state is
\begin{equation}
 \ket{\psi_{\rm s}(t)}
 =e^{\ii\Phi_{\rm s}(t)}B_{\rm s}^{\dagger}U(t)B_{\rm s}\ket{1}.
 \label{eq:ssh:state}
\end{equation}
The target SSH Hamiltonian generates the ideal propagator $U_{\rm SSH}(t)$.
With $\ket{\psi_{\rm SSH}(t)}=U_{\rm SSH}(t)\ket{1}$, we evaluate
\begin{align}
 L(t)&=1-\langle\psi_{\rm s}(t)|\psi_{\rm s}(t)\rangle,\notag\\
 F_{\rm SSH}(t)&=
 |\langle\psi_{\rm SSH}(t)|\psi_{\rm s}(t)\rangle|^2.
 \label{eq:ssh:metrics}
\end{align}
The logical-state norm measures the population remaining in the qubit
subspace. Thus, $F_{\rm SSH}$ includes population loss through coupler leakage.
The receiver population is
$P_{Q_N}(t)=|\langle N|\psi_{\rm s}(t)\rangle|^2$.

PDRAG achieves $F_{\rm SSH}(T)>99.99\%$ for all four chain lengths,
reducing the final state error by factors of $1.78$--$5.99$
[Fig.~\ref{fig:ssh}(b)].
For eleven sites, the final error decreases from
$3.83\times10^{-4}$ to $6.39\times10^{-5}$, and the mean trajectory
error decreases by $46.9\%$.
The receiver population increases from $99.897\%$ to $99.975\%$.

Residual coupler leakage is suppressed by factors of $3.35$--$82.5$
across the four chains [Fig.~\ref{fig:ssh}(c)].
The largest reduction occurs for eleven sites, where leakage falls
from $7.52\times10^{-5}$ to $9.12\times10^{-7}$.
The Supplemental Material provides complete trajectories, receiver
populations, calibrated commands, and independent propagation checks.

\begin{figure}[!t]
 \centering
 \setlength{\abovecaptionskip}{4pt}
 \includegraphics[width=\columnwidth]{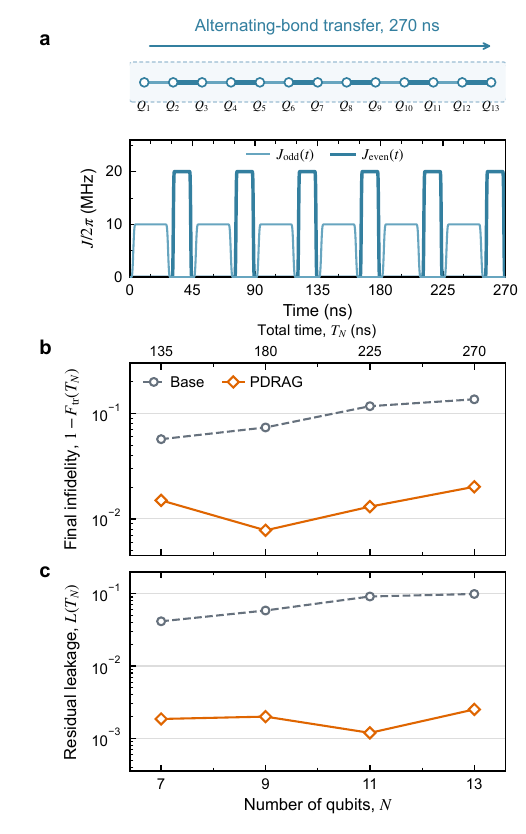}
 \caption{Digital Floquet simulation of SSH dynamics.
(a) Thirteen-site chain and nominal coupling envelopes for the
six-period, $270$-ns sequence. The odd and even bonds have peak couplings
$J_{\rm odd}/(2\pi)=10$ MHz and $J_{\rm even}/(2\pi)=20$ MHz,
respectively.
Light and dark blue denote odd and even bonds.
(b) Final transfer infidelity $1-F_{\rm tr}(T_N)$.
(c) Residual population $L(T_N)$ in the fixed idle-dressed
coupler-like modes.
These panels share qubit number on the lower axis and total duration
on the upper axis.
Grey open circles and orange open diamonds denote Base and PDRAG.
Lines guide the eye.}
 \label{fig:alternating}
\end{figure}

\subsection{Digital Floquet simulation}
\label{sec:ssh:alternating}

Alternating odd- and even-bond interaction layers implement digital
Floquet simulation of SSH dynamics [Fig.~\ref{fig:alternating}(a)]
\cite{Asboth2014DrivenSSH,Zhang2022DigitalFloquet}.
The layer Hamiltonians are the corresponding bond terms of
Eq.~\eqref{eq:ssh:target}.
The odd-bond target envelope reaches $J_{\rm odd}/(2\pi)=10$ MHz
over $30$ ns, whereas the even-bond envelope reaches
$J_{\rm even}/(2\pi)=20$ MHz over $15$ ns.
Doubling the coupling and halving the pulse duration preserves
the exchange angle $\int J(t)\,\dd t=\pi/2$.
Each ideal layer therefore exchanges the excitation across its
active bonds.
The ideal layer propagators define $U_F=U_{\rm even}U_{\rm odd}$
with period $T_F=45$ ns.
After $(N-1)/2$ periods, $U_F^{(N-1)/2}$ transfers the excitation
from $Q_1$ to $Q_N$ in $T_N=22.5(N-1)$ ns.

For each chain length, the odd and even layers are calibrated
independently and then reused with fixed parameters throughout the sequence.
Base and PDRAG optimize amplitude, carrier frequency, phase, and
DC controls under the same frequency-excursion and slew limits.
PDRAG adds four derivative coefficients per active coupler.
We evaluate the transfer fidelity
\begin{equation}
 F_{\rm tr}(T_N)=
 \left|{}_{\rm d}\!\bra{Q_N}U(T_N)\ket{Q_1}_{\rm d}\right|^2,
 \label{eq:ssh:alternating:fidelity}
\end{equation}
Residual leakage $L(T_N)$ is the total population of the
idle-dressed coupler-like modes.

For $N=7,9,11,$ and $13$, PDRAG reduces the final transfer error
by factors of $3.81$--$9.47$ and residual leakage by $22.4$--$77.0$
[Fig.~\ref{fig:alternating}(b,c)].
For eleven sites, the transfer error falls from $11.73\%$ to
$1.306\%$, while leakage decreases from $9.151\%$ to $0.1189\%$.
The thirteen-site sequence completes twelve interaction layers
in $270$ ns, reducing the transfer error from $13.63\%$ to $2.015\%$
and leakage from $9.918\%$ to $0.2509\%$.
These independently calibrated exchange layers retain their advantage
over the repeated Floquet periods required for end-to-end transfer.

\section{Discussion and conclusion}
\label{sec:discussion}
PDRAG combines channel-selective derivative shaping and finite-pulse
calibration to suppress leakage under qubit and coupler frequency modulation.
Two derivative coefficients jointly suppress the conjugate leakage
channels of a Q--Q pair, while five- and seven-root constructions
address qubit and coupler excitations in Q--C--Q circuits.
At strong coupling, the Q--Q construction achieves $13$-ns full
exchange with more than five orders of magnitude of leakage
suppression relative to Base.
Calibrated phase accumulation extends these exchange pulses to
independently specified fSim parameters, preserving infidelity below
$1.8\times10^{-7}$ across the sampled two-parameter domain.
Transferring the derivative coefficients preserves the short-gate
advantage in a realistic circuit Hamiltonian.

Analytical spectral zeros suppress the targeted leakage channels,
and calibration tunes their finite-pulse return amplitudes for destructive interference.
The Floquet expansion resolves this mechanism across the duration
scan, while dressed leakage gaps explain the shifted interference
minima in the realistic circuit.

The SSH simulations demonstrate how the same derivative parametrization
supports both continuously programmed interactions and reusable exchange layers.
It improves trajectory tracking under continuous modulation and suppresses
leakage throughout digital Floquet sequences.
The compact control family also provides a practical basis for
calibration against measured observables \cite{Kelly2014,Werninghaus2021}.

\begin{acknowledgments}
This work was supported by the National Key R\&D Program of China (Grant Nos. 2022YFA1405304 and 2024YFA1409300), the National Natural Science Foundation of China (Grant Nos. 12504588, U21A20436, and 12074179), Quantum Science and Technology-National Science and Technology Major Project (Grant No. 2021ZD0301702), the Natural Science Foundation of Jiangsu Province (Grant Nos. BE2021015-1, BK20232002, and BK20233001), and the Natural Science Foundation of Shandong Province (Grant No. ZR2023LZH002).

\end{acknowledgments}

\begingroup
\makeatletter
\def\section{\@startsection{section}{1}{\z@}%
  {12pt plus 2pt minus 2pt}{8pt}%
  {\normalfont\small\bfseries\centering}}
\makeatother
\section*{Author contributions}
Yiwen Li: Conceptualization, Methodology, Investigation, Data curation,
Writing -- original draft.
Xinsheng Tan: Conceptualization, Supervision, Project administration,
Funding acquisition, Writing -- review and editing.
Yang Yu: Conceptualization, Supervision, Funding acquisition,
Writing -- review and editing.
All authors discussed the results and approved the final manuscript.

\section*{Conflict of interest}
The authors have no conflicts to disclose.

\begin{samepage}
\section*{Data availability}
The data that support the findings of this study are available from the
corresponding authors upon \mbox{reasonable request.}
\par
\end{samepage}
\endgroup

\FloatBarrier
\makeatletter
\let\auto@bib@innerbib\@empty
\makeatother

\onecolumngrid
\clearpage
\setcounter{section}{0}
\setcounter{subsection}{0}
\setcounter{subsubsection}{0}
\setcounter{figure}{0}
\setcounter{table}{0}
\setcounter{equation}{0}
\renewcommand{\thesection}{S\arabic{section}}
\renewcommand{\thesubsection}{\thesection.\arabic{subsection}}
\renewcommand{\thesubsubsection}{\thesubsection.\arabic{subsubsection}}
\makeatletter
\renewcommand{\p@subsection}{}
\renewcommand{\p@subsubsection}{}
\makeatother
\renewcommand{\thefigure}{S\arabic{figure}}
\renewcommand{\thetable}{S\arabic{table}}
\renewcommand{\theequation}{S\arabic{equation}}
\renewcommand{\theHsection}{supp.\arabic{section}}
\renewcommand{\theHsubsection}{\theHsection.\arabic{subsection}}
\renewcommand{\theHsubsubsection}{\theHsubsection.\arabic{subsubsection}}
\renewcommand{\theHfigure}{supp.\arabic{figure}}
\renewcommand{\theHtable}{supp.\arabic{table}}
\renewcommand{\theHequation}{supp.\arabic{equation}}
\setlength{\tabcolsep}{3.5pt}
\setlength{\parskip}{2pt}
\setlength{\textfloatsep}{13pt plus 2pt minus 2pt}
\setlength{\floatsep}{11pt plus 2pt minus 2pt}
\renewcommand{\topfraction}{0.92}
\renewcommand{\bottomfraction}{0.90}
\renewcommand{\textfraction}{0.08}
\renewcommand{\floatpagefraction}{0.78}
\setcounter{topnumber}{3}
\setcounter{bottomnumber}{2}
\setcounter{totalnumber}{4}
\makeatletter
\setlength{\@fptop}{0pt}
\setlength{\@fpsep}{13pt}
\setlength{\@fpbot}{0pt plus 1fil}
\makeatother

\pdfbookmark[0]{Supplemental Material}{supplement-start}
\begin{center}
{\Large\bfseries Supplemental Material}\par\vspace{5pt}
{\large Leakage suppression for high-fidelity exchange gates and quantum simulation in superconducting circuits}\par\vspace{5pt}
Yiwen Li, Xinsheng Tan, and Yang Yu\par\vspace{3pt}
\today
\end{center}

Equation and figure numbers without an S prefix refer to the main text.

\section{Q--Q physical models and common gate metrics}
\label{s4:sec:metrics}
\subsection{Device Hamiltonians and charge-model fit}
\label{sec:supp:dmh:device}

The $\sin^4$ and raised-cosine Q--Q implementations use
$(\omega_1,\omega_2)/(2\pi)=(5.10,4.50)$ GHz,
$(\alpha_1,\alpha_2)/(2\pi)=(-220,-240)$ MHz, and
$g=2\pi\times0.100=0.6283185307$ rad/ns.
The Q--Q Duffing Hamiltonian is
\[
 H_{\rm QQ}(t)/\hbar=\sum_{j=1}^{2}
 \left[\omega_j n_j+\frac{\alpha_j}{2}n_j(n_j-1)\right]
 +\delta\omega_1(t)n_1+g(a_1^\dagger a_2+a_1a_2^\dagger),
\]
Here $a_j$ annihilates an excitation of transmon $j$ and $n_j=a_j^\dagger a_j$.
The angular frequencies $\omega_j$, anharmonicities $\alpha_j$, and coupling $g$
are expressed in rad/ns; $\delta\omega_1(t)$ is the driven frequency shift from idle.
The Duffing calculation retains four levels per transmon and treats static
exchange in the rotating-wave approximation (RWA).
Propagation uses the complete real modulation, including all its harmonics.

The charge model uses the asymmetric-SQUID Hamiltonian
\begin{align*}
 H_q(\Phi)={}&4E_C(\hat n-n_g)^2
 -E_{J\Sigma}\cos(\pi\Phi/\Phi_0)\cos\hat\varphi\\
 &-dE_{J\Sigma}\sin(\pi\Phi/\Phi_0)\sin\hat\varphi
\end{align*}
for each transmon, together with the full capacitive interaction,
\begin{equation}
 H_{\rm C}(\Phi_1,\Phi_2)
 =H_1(\Phi_1)\otimes I+I\otimes H_2(\Phi_2)
   +\hbar g_C\hat n_1\otimes\hat n_2 .
 \label{s4:eq:charge-single}
\end{equation}
Here $\hat n$ and $\hat\varphi$ are the conjugate Cooper-pair number and superconducting phase operators.
The charging energy is $E_C$, $n_g$ is the offset charge in Cooper-pair units,
and $\Phi_0=h/(2e)$ is the flux quantum.
The two junction energies define $E_{J\Sigma}=E_{J1}+E_{J2}$ and
$d=(E_{J1}-E_{J2})/E_{J\Sigma}$, with junction labels chosen for the displayed phase convention.
In Eq.~\eqref{s4:eq:charge-single}, $I$ is the single-transmon identity and
$g_C$ is the capacitive coupling coefficient in angular-frequency units.
The charge operator $\hat n_j$ differs from the Duffing excitation number $n_j$.

Projection onto fixed idle eigenstates retains the flux dependence of the
Hamiltonian and the counter-rotating capacitive interaction.
Table~\ref{s4:tab:device} lists parameters fitted to the idle frequency and anharmonicity of each transmon.
The relative fit residuals are $4.16\times10^{-14}$ and $3.83\times10^{-15}$.

\begin{table}[!hbp]
\centering
\footnotesize
\caption{Charge-Hamiltonian parameters fitted to the idle transition frequencies and anharmonicities. Energies divided by $h$ are in GHz.}\label{s4:tab:device}
\begin{tabular}{lrr}
\toprule
Parameter & Tunable transmon & Fixed transmon\\
\midrule
$E_C/h$ & 0.199475755430 & 0.212526067968\\
$E_{J\Sigma}/h$ & 24.69094569955 & 13.12248658690\\
Junction asymmetry $d$ & 0.15 & 0\\
Offset charge $n_g$ & 0 & 0\\
Idle flux $\Phi/\Phi_0$ & 0.25 & 0\\
\bottomrule
\end{tabular}
\end{table}

The isolated-transmon idle charge matrix elements determine the coupling,
\begin{equation}
 g_C\left|\bra1\hat n_1\ket0\bra0\hat n_2\ket1\right|=g,
 \qquad g_C/(2\pi)=68.787626373\ {\rm MHz}.
 \label{s4:eq:coupling-match}
\end{equation}
This matching fixes the single-excitation exchange strength in both models;
the absolute charge-matrix-element product is $1.45374982788$.
Calculations use charge states $n=-18,\ldots,18$ and
$N=9$ retained idle levels per transmon.
A $1201$-point monotonic frequency-to-flux map covers
$0\leq\Phi_1/\Phi_0\leq0.49$, corresponding to transition frequencies
$2.238215203$--$6.070683627$ GHz.
The nine-level coupled model has a $632.453007$-MHz idle
one-excitation splitting.

The frequency-to-flux map restricts the Duffing search to accessible commands;
the Duffing propagation keeps anharmonicities and coupling fixed.
Charge-model transfer evolves the flux-dependent Hamiltonian
$H_{\rm C}[\Phi_1(t),0]$.

\subsection{Computational projection and gate fidelities}
\label{sec:supp:dmh:metrics}

For each model, $B=(\ket{00}_{\rm d},\ket{01}_{\rm d},
\ket{10}_{\rm d},\ket{11}_{\rm d})$ contains orthonormal eigenstates
of the coupled idle Hamiltonian.
A one-to-one assignment maximizes total squared overlap with the product states;
eigenstate phases make matched components real and positive.
Charge-model product states use isolated-transmon idle eigenstates.
The $N^2\times4$ matrix $B$ satisfies $B^\dagger B=I_4$, with computational
projector $P=BB^\dagger$; its complement is $Q=I-P$.

For the unitary propagator $U$ over the stated evolution window, the columns of $C=UB$ contain the evolved computational inputs.
The projected map $K=B^\dagger C$ retains their amplitudes within the computational subspace.
Logical indices $j=0,1,2,3$ denote $00,01,10,11$.
With $\ket j$ denoting the corresponding four-component unit vector, the leakage probabilities are
\begin{equation}
 L_j=\|(I-BB^\dagger)C\ket j\|^2,\qquad
 \overline L=\frac14\sum_{j=0}^3 L_j,\qquad L_{11}=L_3.
 \label{eq:supp:dmh:leakage}
\end{equation}
Endpoint leakage includes any specified idle guards in the evolution window.
Using the propagator $U$ up to time $t$ gives instantaneous leakage.
Both evaluations use the fixed idle projector to resolve population excursions and return before the endpoint.
For the doubly excited input, the resolved probabilities are
$P_{mn}=|{}_{\rm d}\langle mn|C|3\rangle|^2$ for $mn=20,02$.
Excitation conservation gives $P_{20}+P_{02}=L_{11}$ in the Duffing model.
Charge-model leakage includes every state outside the computational subspace spanned by $B$.
Column norm errors $|(C^\dagger C)_{jj}-1|$ quantify numerical drift.

Exchange calibration suppresses the return amplitude $K_{11}$ of logical input $01$.
The exchange-angle tolerance is $|\theta-\pi/2|\leq5$ mrad, with
\[
 \theta=\operatorname{atan2}\!\left(
 \sqrt{\frac{|K_{12}|^2+|K_{21}|^2}{2}},
 \sqrt{\frac{|K_{11}|^2+|K_{22}|^2}{2}}\right).
\]
The mean exchange and return amplitudes in the one-excitation subspace determine this angle.
The local-$Z$-invariant conditional phase is
\[
 \phi_c=\wrap_{[-\pi,\pi)}
 \arg\!\left(\frac{K_{00}K_{33}}
 {\det K_{\{1,2\},\{1,2\}}}\right).
\]
Here $\wrap_{[-\pi,\pi)}$ selects the equivalent phase in $[-\pi,\pi)$.
The determinant form remains well defined at zero exchange and agrees
with the swap-amplitude phase ratio for excitation-conserving unitary maps.
For the ordered basis $00,01,10,11$, the target family is
\begin{equation}
 U_{\rm fSim}(\theta,\phi)=
 \begin{pmatrix}
 1&0&0&0\\
 0&\cos\theta&-\ii\sin\theta&0\\
 0&-\ii\sin\theta&\cos\theta&0\\
 0&0&0&e^{-\ii\phi}
 \end{pmatrix}.
 \label{eq:supp:dmh:fsim}
\end{equation}
The target exchange angle is $\theta$ and its nonlocal phase is $\phi$,
giving $\phi_c=-\phi$ modulo $2\pi$.
We evaluate the unconditional average fidelity
$F(K,V)=[\Tr(K^\dagger K)+|\Tr(V^\dagger K)|^2]/20$ over
$V=D_Z^{\rm out}U_{\rm fSim}D_Z^{\rm in}$, where
$D_Z(\zeta_1,\zeta_2)=
\operatorname{diag}(1,e^{\ii\zeta_2},e^{\ii\zeta_1},e^{\ii(\zeta_1+\zeta_2)})$.
The phases $\zeta_1$ and $\zeta_2$ specify independent local $Z$ corrections
on the two qubits, applied before and after the gate \cite{McKay2017}.
The best-fit fSim fidelity used in the duration and transfer scans
optimizes $\theta$ and $\phi$;
the strict-iSWAP evaluation fixes them at $\pi/2$ and $0$.
For the two-parameter map in Fig.~4, both target parameters are prescribed,
and only the local $Z$ rotations are optimized
(Sec.~\ref{supp:sec:fsim-map}).
All fidelity evaluations retain the unnormalized map $K$, so leakage reduces the reported fidelity.

Writing $\mathcal Q=\max_V|\Tr(V^\dagger K)|$ gives
\begin{align}
 \mathcal Q_{\rm fSim}&=|K_{00}|+|K_{33}|+\|K_{\{1,2\},\{1,2\}}\|_*,
 \label{eq:supp:dmh:fsimoverlap}\\
 \mathcal Q_{\rm iSWAP}&=\max_{x\in[-\pi,\pi)}
       \bigl(|a+be^{-\ii x}|+|c+de^{-\ii x}|\bigr),
 \quad(a,b,c,d)=(K_{00},\ii K_{12},\ii K_{21},K_{33}).
 \label{eq:supp:dmh:iswapoverlap}
\end{align}
Here $K_{\{1,2\},\{1,2\}}$ is the $2\times2$ exchange subblock,
and its nuclear norm $\|\cdot\|_*$ sums its singular values.
Only the relative local-$Z$ phase $x$ remains; the other phases are optimized analytically.
The iSWAP evaluation samples $128$ phases per period and refines every detected maximum.
The resulting fidelity is $[\Tr(K^\dagger K)+\mathcal Q^2]/20$
\cite{Wood2018,Nielsen2002}.

To evaluate small infidelities without subtracting nearly equal numbers, define
$\mathcal R=\|QC\|_{\rm F}^2=4\overline L$ and
$\mathcal D=\|K-e^{\ii\eta}V\|_{\rm F}^2$,
where $\eta=\arg\Tr(V^\dagger K)$ and $\|\cdot\|_{\rm F}$ is the Frobenius norm.
The isometry condition $C^\dagger C=I_4$ gives the stable identity
\begin{equation}
 1-F=\frac{\mathcal R}{4}+\frac{\mathcal D}{5}
 -\frac{(\mathcal R+\mathcal D)^2}{80}.
 \label{eq:supp:dmh:stableerror}
\end{equation}
Equation~\eqref{eq:supp:dmh:stableerror} evaluates infidelities for
$\sin^4$ Duffing pulses and both raised-cosine models.
For $\sin^4$ charge pulses, direct subtraction gives $1-F>10^{-12}$;
independent norm and consistency checks verify both evaluations.

For an ideal full-exchange fSim operation, local-$Z$ optimization
against strict iSWAP gives $\mathcal Q_{\rm iSWAP}=4\cos(\bar\phi/4)$, where
$\bar\phi=\operatorname{wrap}_{[-\pi,\pi)}\phi$.
Its strict-iSWAP infidelity is
\begin{equation}
 \epsilon_\phi=\frac45\sin^2(\bar\phi/4).
 \label{eq:supp:phase:infidelity}
\end{equation}
Because Eq.~\eqref{eq:supp:phase:infidelity} is even in phase, $\phi_c$ gives the same prediction.
This prediction isolates the strict-iSWAP error caused by the measured conditional phase in an otherwise ideal exchange gate.
Comparison with the simulated infidelity identifies the additional contributions from leakage and imperfect exchange.

\FloatBarrier

\section{Q--Q pulse construction and calibration}
\label{sec:supp:search}
\subsection{Envelopes, signed spectrum, and frequency command}

The Q--Q base envelopes have exchange area $\Theta=\pi/2$ over a gate duration $T$.
Both vanish outside $[0,T]$:
\begin{equation}
 A_4(t)=\frac{8\Theta}{3T}\sin^4(\pi t/T),\qquad
 A_2(t)=\frac{\Theta}{T}[1-\cos(2\pi t/T)]
       =\frac{2\Theta}{T}\sin^2(\pi t/T).
 \label{eq:sin2:base}
\end{equation}
$A_2$ is the raised-cosine envelope \cite{Hyyppa2024}.
For either choice $A$, the complex exchange envelope is
$G=s(A+\ii\beta\dot A+\gamma\ddot A)$.
Here $A$ and $G$ have angular-frequency units, $s$ is dimensionless,
and the real coefficients $\beta$ and $\gamma$ have units of time and time squared.
Dots denote time derivatives.
Since $A=\dot A=0$ at the endpoints, integration by parts gives
\begin{equation}
 \widetilde G(\delta)=s\mathcal P(\delta)\widetilde A(\delta),\qquad
 \mathcal P(\delta)=1+\beta\delta-\gamma\delta^2,\qquad
 \widetilde A(\delta)=\int_0^T A(t)e^{\ii\delta t}\,\dd t.
 \label{s4:eq:filter}
\end{equation}
Here $\delta$ is an angular-frequency offset in the positive-sign Fourier convention.
Both envelopes give $\widetilde G(0)=s\Theta$, preserving the target exchange area up to amplitude calibration.
The leakage channels $11\to20$ and $11\to02$ couple to conjugate exchange envelopes.
Their cancellation therefore requires zeros at $\delta=\alpha_1$ and $\delta=-\alpha_2$, respectively.
Factorization gives the analytical coefficients
\begin{equation}
 \mathcal P_{\rm an}(\delta)=
 \left(1-\frac{\delta}{\alpha_1}\right)\left(1+\frac{\delta}{\alpha_2}\right),
 \quad
 \beta_{\rm an}=0.0602859632924\ {\rm ns},\qquad
 \gamma_{\rm an}=0.479740452852\ {\rm ns}^2 .
 \label{s4:eq:analytic}
\end{equation}
The roots in $\delta/(2\pi)$ are $-220$ and $+240$ MHz;
calibration adjusts $\beta$ and $\gamma$ within this derivative form.

The first-order Bessel function $J_1$ maps modulation depth to exchange strength.
On its increasing branch, define the dimensionless depth
$\rho=J_1^{-1}(|G|/g)$ and complex modulation amplitude $z=\rho G/|G|$.
The continuous zero-amplitude limit is $z=0$ at $G=0$, with $z\simeq2G/g$ nearby.
For a carrier angular frequency $\Omega$, differentiating the modulation phase gives the real frequency command,
\begin{equation}
 \delta\omega_1(t)=
 \operatorname{Im}\{[\dot z(t)-\ii\Omega z(t)]e^{-\ii\Omega t}\}.
 \label{s4:eq:command}
\end{equation}
The phase $\operatorname{Im}[z(t)e^{-\ii\Omega t}]$ is held at its endpoint value after the pulse.
Frame transformations retain both endpoint phases when reconstructing the computational map.
For $A_4$, the first three derivatives also vanish at the joins,
so $G,\dot G$, and the frequency excursion approach zero continuously.
The finite endpoint limits for $A_2$ are given in Sec.~\ref{sec:supp:spec:exact}.

\subsection{Duration scan and derivative search}

Each envelope uses $46$ durations from $12.5$ to $35$ ns in $0.5$-ns steps.
The five Duffing families are Base, first-derivative-only ($\gamma=0$),
second-derivative-only ($\beta=0$), analytical PDRAG, and calibrated PDRAG.
Their common bounds are
\begin{equation}
 s\in[0.5,1.3],\quad
 \beta\in[-4,1.6]\ {\rm ns},\quad
 \gamma\in[-4,3.2]\ {\rm ns}^2,\quad
 \frac{\Omega}{2\pi}=632.455532\pm80\ {\rm MHz}.
 \label{s4:eq:carrier}
\end{equation}
Here $\delta f=(\Omega-\Delta)/(2\pi)$ is the carrier offset from
the bare qubit-frequency detuning $\Delta/(2\pi)=600$ MHz.
Every candidate is constructed and Bessel-inverted at its target duration.
Calibration of the amplitude and carrier suppresses the complex return amplitude $K_{11}$.
Derivative calibration then minimizes $L_{11}$ while maintaining the target exchange angle.

For each envelope, the search scans $721$ points on each single-derivative axis and refines
every detected basin to a bracket width of $2\times10^{-8}$.
The joint grid combines $31\times37$ boundary-inclusive points with
$30\times36$ cell centers, totaling $2227$ candidates.
Bounded local refinement follows Gauss--Newton initialization from
prescribed pulses, grid minima, optimized single-derivative controls,
and neighboring-duration solutions.
Convergence requires coordinate steps no larger than $5\times10^{-6}$
in the respective units and
$\max L-\min L\leq\max(10^{-17},10^{-4}L_{\rm last})$
over the last five iterations, where $L$ denotes $L_{11}$ and
$L_{\rm last}$ is its value in the latest iteration.
Continuation in both duration directions checks adjacent parameter branches.

For $\sin^4$, Newton finite differences use
$(\Delta s,\Delta\delta f)=(-0.0002,0.02\ {\rm MHz})$
for prescribed and single-derivative pulses, and
$(-0.001,0.1\ {\rm MHz})$ for joint calibration.
Iteration limits are $30$ and $24$; updates are bounded by
$0.06$ in $s$ and $4$ MHz in the carrier frequency.
Fine calibration targets $|K_{11}|<2\times10^{-9}$.
The raised-cosine search uses three amplitude--carrier starts for
prescribed pulses and up to $20$ separated starts for joint refinement.
Its Gauss--Newton stage allows $24$ updates before pattern search.
The minimum accepted improvement is $\max(10^{-17},10^{-7}L)$.
Neighboring-duration starts are refined until every competitive branch
falls below this improvement threshold.
Each final joint candidate set includes both optimized single-derivative commands.

Charge-model transfer preserves the derivative coefficients of the
same-duration Duffing envelope and recalibrates $s$ and the carrier.
Seven-level calibration initializes the nine-level calculation.
For $\sin^4$, the Newton increments are $0.001$ and $0.15$ MHz,
with $18$ iterations, update limits $0.06$ and $4$ MHz,
and a return threshold $2\times10^{-5}$.
One-sided differences and backtracking preserve the Bessel and flux branches;
improvements below $1\%$ trigger the stagnation check.
Raised-cosine recalibration starts from the best seven-level
return-amplitude iterate and retains the smallest nine-level exchange error.
Both calibrations target $0.2$ mrad after basis enlargement, within the
common $5$-mrad acceptance tolerance.

\subsection{Complete-command resource definitions}
\label{s4:sec:resources}

Let $x(t)=\delta\omega_1(t)/(2\pi)$, measured in GHz relative to idle.
The shared bounds are
\begin{equation}
 \max|x|\leq1.0\ {\rm GHz},\quad
 x_{\rm RMS}\leq0.33\ {\rm GHz},\quad
 \sup|\dot x|\leq4.1\ {\rm GHz/ns},\quad
 B_{99.9}\leq1.0\ {\rm GHz}.
 \label{eq:shared:resources}
\end{equation}
Flux remains in $[0,0.49]\Phi_0$.
The dimensionless Bessel utilization
$u=\max|G|/[gJ_1(\rho_*)]\leq0.98$ limits the exchange amplitude below the first Bessel maximum,
where $\rho_*=1.8411837813$ and $J_1(\rho_*)=0.5818652243$.
For a uniform record of $N_s$ samples with spacing $\Delta t$, let $X_k$
be the real-input discrete Fourier coefficient at frequency $k/(N_s\Delta t)$.
The bandwidth is the smallest $f_m=m/(N_s\Delta t)$ satisfying
\begin{equation}
 \sum_{k=0}^{m}|X_k|^2\geq0.999
 \sum_{k=0}^{\lfloor N_s/2\rfloor}|X_k|^2.
 \label{eq:supp:dmh:bandwidth}
\end{equation}
Equation~\eqref{eq:supp:dmh:bandwidth} defines the bandwidth from nonnegative-frequency bins with the equal bin weights shown.
The transform includes the zero-frequency (DC) component and uses the full record without mean subtraction or tapering.
For ideal pulses, the root-mean-square (RMS) excursion is $[T^{-1}\int_0^T x^2\,\dd t]^{1/2}$;
the raised-cosine bandwidth record includes $4$ ns of idle on each side.
The $\sin^4$ slew bound includes both joins to idle.
For raised-cosine commands, finite endpoint jumps restrict the slew
criterion to the pulse interior; Sec.~\ref{sec:supp:spec:exact} gives the limits.

\FloatBarrier

\section{Supporting results for the \texorpdfstring{$\sin^4$}{sin4} implementation}
\subsection{Suppression across the duration scan}

The $\sin^4$ comparison measures leakage suppression at equal gate duration (Table~\ref{s4:tab:scan-gains}).
For sampled duration $T_i$, the improvement factor is
$R_i=L_{11}^{\rm ref}(T_i)/L_{11}^{\rm PDRAG}(T_i)$.
With $N_T$ equally weighted durations, the arithmetic and geometric means
are $N_T^{-1}\sum_i R_i$ and $\exp[N_T^{-1}\sum_i\ln R_i]$, respectively.
The geometric mean and median summarize multiplicative gain across
all $46$ durations from $12.5$ to $35$ ns or the $20$ short-duration
points through $22$ ns.
The single-derivative reference uses the lower leakage from the independently
optimized first-derivative-only and second-derivative-only controls at each duration.

\begin{table}[!htbp]
\centering
\footnotesize
\caption{Pointwise leakage-improvement factors $R_i$ for the $\sin^4$ scans.
Statistics include every sampled duration in each interval.}
\label{s4:tab:scan-gains}
\begin{tabular}{llrrrr}
\toprule
Model & Reference & $T$ (ns) & Arithmetic mean & Geometric mean & Median\\
\midrule
Duffing & Base & 12.5--35 & $6.49\times10^7$ & $5.27\times10^5$ & $5.50\times10^5$\\
Duffing & Analytical PDRAG & 12.5--35 & $2.47\times10^7$ & $2.67\times10^5$ & $2.77\times10^5$\\
Duffing & Better single derivative & 12.5--35 & $9.17\times10^5$ & $1.92\times10^4$ & $3.98\times10^4$\\
Charge & Base & 12.5--35 & 405 & 43.8 & 69.6\\
Charge & Analytical PDRAG & 12.5--35 & 153 & 21.4 & 29.3\\
Charge & Base & 12.5--22 & 893 & 372 & 446\\
Charge & Analytical PDRAG & 12.5--22 & 335 & 141 & 160\\
\bottomrule
\end{tabular}
\end{table}

\subsection{Calibration landscape and representative commands}

\begin{figure}[!htbp]
\centering
\includegraphics[width=\linewidth]{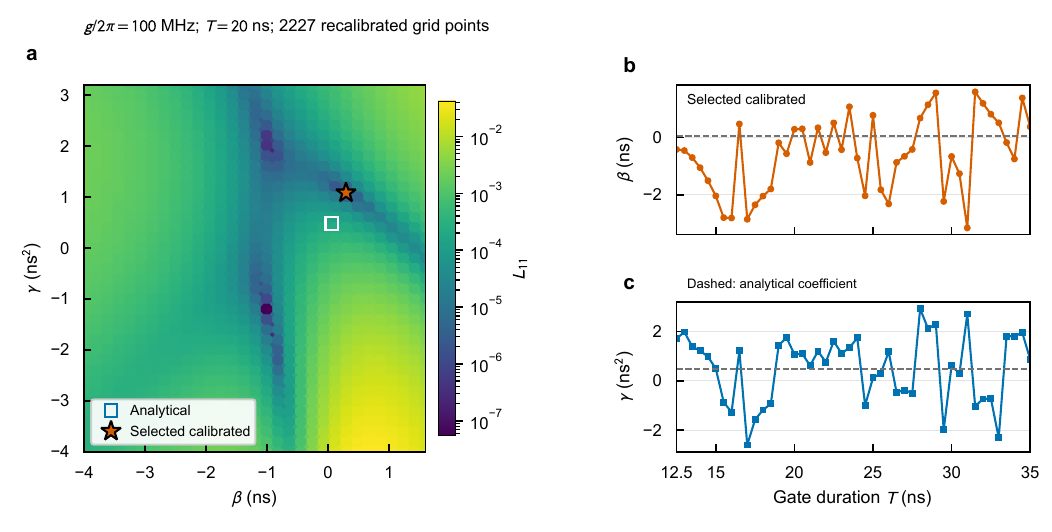}
\caption{$\sin^4$ calibration landscape at $20$ ns. (a) Leakage on the $1147$-point boundary-inclusive grid (cells) and $1080$ cell centers (diamonds), shown on a common logarithmic scale. The square and star mark analytical and calibrated PDRAG. (b),(c) Selected coefficients across the $46$ durations; dashed lines show analytical values.}
\label{s4:fig:landscape}
\end{figure}

The $20$-ns landscape resolves how joint derivative calibration reduces leakage from the analytical starting point (Fig.~\ref{s4:fig:landscape}).
All $2227$ grid points satisfy the exchange and resource constraints.
The best grid value is $5.509\times10^{-8}$ at
$\beta=-1.013333$ ns and $\gamma=-1.200000$ ns$^2$;
local refinement and continuation further suppress the leakage.
Tables~\ref{s4:tab:p1} and \ref{s4:tab:p1metrics} compare the refined
command with the four reference controls.
The tabulated exchange error is $|\theta-\pi/2|$ in milliradians.
All $230$ selected commands satisfy the common constraints across
the duration scan, with a maximum exchange error $\DuffingMaxError$ mrad.

\begin{table}[!htbp]
\centering
\footnotesize
\caption{The five $\sin^4$ Duffing commands at $20$ ns. Carrier offsets are measured from the bare $600$-MHz detuning; all commands meet the exchange and resource criteria.}\label{s4:tab:p1}
\begin{tabular}{lrrrrrr}
\toprule
Family & $\beta$ (ns) & $\gamma$ (ns$^2$) & $s$ & $\delta f$ (MHz) & $L_{11}$ & Error (mrad)\\
\midrule
Base & 0.000000 & 0.000000 & 0.99899507 & 28.719286 & $4.797\times10^{-4}$ & $4.40\times10^{-7}$\\
First only & -1.146007 & 0.000000 & 0.93091817 & 48.690442 & $1.054\times10^{-5}$ & $1.88\times10^{-6}$\\
Second only & 0.000000 & 1.300650 & 0.99921083 & 29.669020 & $4.568\times10^{-6}$ & $1.99\times10^{-6}$\\
Analytical & 0.060286 & 0.479740 & 0.99887641 & 28.051375 & $1.855\times10^{-4}$ & $1.49\times10^{-6}$\\
Calibrated & 0.293213 & 1.085319 & 0.99519655 & 24.830601 & $8.542\times10^{-11}$ & $4.18\times10^{-7}$\\
\bottomrule
\end{tabular}
\end{table}
\begin{table}[!htbp]
\centering
\footnotesize
\caption{Gate metrics for the $\sin^4$ Duffing commands in Table~\ref{s4:tab:p1}, evaluated using Sec.~\ref{sec:supp:dmh:metrics}.}\label{s4:tab:p1metrics}
\begin{tabular}{lrrrr}
\toprule
Family & $\overline L$ & $\phi_{\rm c}$ (rad) & $1-F_{\rm fSim}$ & $1-F_{\rm iSWAP}$\\
\midrule
Base & $1.199\times10^{-4}$ & 2.2328325 & $1.199\times10^{-4}$ & 0.2245309\\
First only & $2.636\times10^{-6}$ & 2.2890927 & $2.636\times10^{-6}$ & 0.2346183\\
Second only & $1.142\times10^{-6}$ & 2.3434280 & $1.142\times10^{-6}$ & 0.2445717\\
Analytical & $4.637\times10^{-5}$ & 2.2767592 & $4.637\times10^{-5}$ & 0.2324091\\
Calibrated & $2.136\times10^{-11}$ & 2.3287565 & $2.136\times10^{-11}$ & 0.2418713\\
\bottomrule
\end{tabular}
\end{table}

\subsection{Anharmonicity asymmetry}

\begin{figure}[!htbp]
\centering
\includegraphics[width=\linewidth]{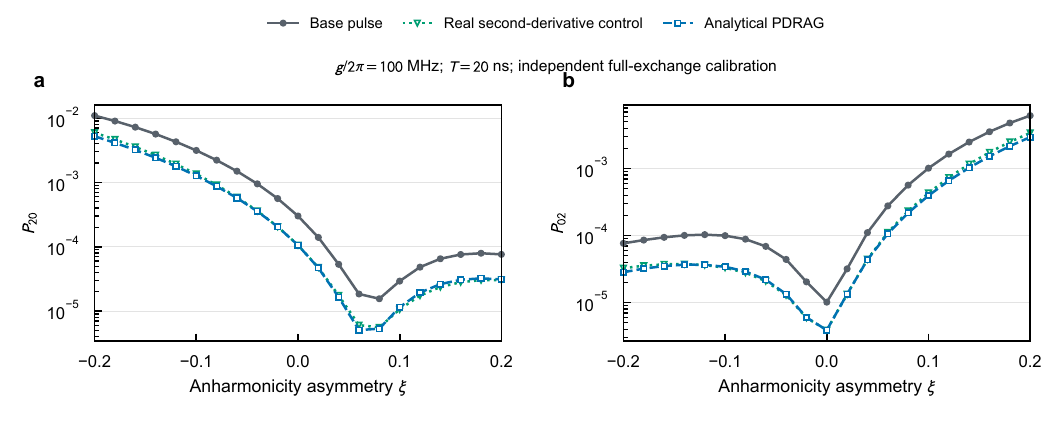}
\caption{$\sin^4$ response to anharmonicity asymmetry at $20$ ns. (a) $P_{20}$ and (b) $P_{02}$ for Base, real second-derivative control, and analytical PDRAG. Derivative coefficients follow Eqs.~\eqref{s4:eq:analytic} and \eqref{s4:eq:asymmetry}; amplitude and carrier are recalibrated at each point.}
\label{s4:fig:asymmetry}
\end{figure}

The quadrature first derivative accounts for the asymmetry between the two signed leakage detunings.
At $T=20$ ns, we vary the dimensionless asymmetry $\xi$ over $21$ equally spaced values:
\begin{equation}
 \alpha_1/(2\pi)=-220(1+\xi)\ {\rm MHz},\qquad
 \alpha_2/(2\pi)=-220(1-\xi)\ {\rm MHz},\qquad
 \xi\in[-0.2,0.2].
 \label{s4:eq:asymmetry}
\end{equation}
At fixed frequencies and coupling, each point uses recomputed idle
states and recalibrated amplitude and carrier.
The real second-derivative control uses $\beta=0$ and
$\gamma=1/(\alpha_1\alpha_2)$; analytical PDRAG adds
$\beta=1/\alpha_2-1/\alpha_1$.
The controls coincide at $\xi=0$.
Analytical PDRAG reduces leakage relative to Base by factors
of $2.099$--$2.844$.
It improves on the second-derivative-only control at every nonzero sampled
asymmetry, by up to a factor of $1.156$.
All $63$ commands satisfy the common exchange and resource constraints.
Independent propagation gives a maximum relative leakage difference of
$2.44\times10^{-8}$ and a map difference of $3.21\times10^{-10}$.
Figure~\ref{s4:fig:asymmetry} resolves the two channels.

\FloatBarrier

\section{Raised-cosine implementation}
\label{sec:supp:sin2}
\subsection{Exact spectrum and endpoint behavior}
\label{sec:supp:spec:exact}

For $A_2$ in Eq.~\eqref{eq:sin2:base}, define the envelope angular frequency
$\kappa=2\pi/T$.
Its derivatives are
\begin{equation}
 \dot A_2=\frac{\kappa\Theta}{T}\sin(\kappa t),\qquad
 \ddot A_2=\frac{\kappa^2\Theta}{T}\cos(\kappa t).
\end{equation}
With $\operatorname{sinc}x=\sin x/x$ and $\operatorname{sinc}0=1$,
\begin{equation}
 \widetilde A_2(\delta)
 =\Theta e^{\ii\delta T/2}
   \frac{\operatorname{sinc}(\delta T/2)}{1-(\delta/\kappa)^2},
 \qquad
 \frac{|\widetilde G(\delta)|^2}{|\widetilde G(0)|^2}
 =\left|\frac{\mathcal P(\delta)\operatorname{sinc}(\delta T/2)}
                   {1-(\delta/\kappa)^2}\right|^2.
 \label{eq:supp:spec:power}
\end{equation}
Here $\mathcal P$ is the derivative polynomial of Eq.~\eqref{s4:eq:filter}.
The apparent singularities at $\delta=\pm\kappa$ are removable,
with $\widetilde A_2(\pm\kappa)=-\Theta/2$.
Stable numerical evaluation near these points uses
$\widetilde A_2=(\Theta/T)[W_T(\delta)
-(W_T(\delta+\kappa)+W_T(\delta-\kappa))/2]$,
where $W_T(\delta)=Te^{\ii\delta T/2}\operatorname{sinc}(\delta T/2)$.
For nonzero $\gamma$, finite endpoint curvature gives the exchange
envelope a power-spectrum tail proportional to $|\delta|^{-2}$.
The occupied control bandwidth is evaluated from the complete frequency
command after Bessel inversion.

The interior endpoint limits are
\begin{equation}
 G(0^+)=G(T^-)=s\gamma C_T,\quad
 \dot G(0^+)=\dot G(T^-)=\ii s\beta C_T,\quad
 C_T=4\pi^2\Theta/T^3.
 \label{eq:sin2:endpoints}
\end{equation}
Here $C_T=\ddot A_2(0)=\ddot A_2(T)$ is the endpoint curvature.
Writing $a_0=s\gamma C_T\ne0$, $b_0=s\beta C_T$, and
$\rho_0=J_1^{-1}(|a_0|/g)$, Eq.~\eqref{s4:eq:command} gives
\begin{equation}
 \delta\omega_1(0^+)=\frac{\rho_0}{|a_0|}(b_0-\Omega a_0),
 \qquad
 \delta\omega_1(T^-)=\delta\omega_1(0^+)\cos(\Omega T).
 \label{eq:supp:dmh:endpoints}
\end{equation}
For $a_0=0$, the first limit is $2b_0/g$.
Both frequency-command endpoint limits vanish when $\beta=\Omega\gamma$.
Propagation uses the stated one-sided command limits and a continuous
integrated modulation phase.
For $A_4$, the vanishing endpoint derivatives remove these frequency-command
jumps, allowing the pulse to join idle continuously.

The idle eigenfrequencies $\varepsilon_{mn}$ in
Eq.~\eqref{supp:floquet:idle} give the signed one-photon references
$\delta_{20}^{\rm d}=\varepsilon_{20}-\varepsilon_{11}-\Omega$ and
$\delta_{02}^{\rm d}=\varepsilon_{11}-\varepsilon_{02}-\Omega$.
Define the negative-carrier coefficient
$q_\omega=(\dot z-\ii\Omega z)/(2\ii)$, so that
$\delta\omega_1=q_\omega e^{-\ii\Omega t}+q_\omega^*e^{\ii\Omega t}$ and
\begin{equation}
 \widetilde q_\omega(\delta)=
 -\frac{\Omega+\delta}{2}\widetilde z(\delta)
 -\frac{\ii}{2}[z(T^-)e^{\ii\delta T}-z(0^+)].
 \label{eq:supp:dr:boundary}
\end{equation}
Near-resonant transitions generated by the dressed modulation matrix
$N_{\rm d}$ of Eq.~\eqref{supp:floquet:idle} sample
$\widetilde q_\omega(\delta_{20}^{\rm d})$ and
$\widetilde q_\omega(\delta_{02}^{\rm d})^*$, with tildes following
the Fourier convention of Eq.~\eqref{s4:eq:filter}.
The boundary term includes the finite endpoint values of the nonlinear
command in the signed spectral response.
Leakage is evaluated by propagating the complete Hamiltonian with that command.

\subsection{Duffing suppression and charge-model transfer}

At $17$ ns, calibrated PDRAG reaches $L_{11}=1.56\times10^{-8}$,
compared with $7.83\times10^{-4}$ for Base and
$1.00\times10^{-4}$ for analytical PDRAG.
The leakage-reduction factors are $5.00\times10^4$ relative to Base and $68.2$ relative to the better
single-derivative control.
The resolved residuals are $P_{20}=2.23\times10^{-10}$
and $P_{02}=1.54\times10^{-8}$.
This endpoint reduction accompanies a much smaller change in peak
transient leakage, from $0.131$ to $0.124$, consistent with coherent return.
The calibrated values $\beta=1.30$ ns, $\gamma=0.0750$ ns$^2$,
and $\Omega/(2\pi)=612$ MHz give spectral roots near
$-118$ MHz and $+2.87$ GHz.

\begin{figure}[!htbp]
\centering
\includegraphics[width=0.85\linewidth]{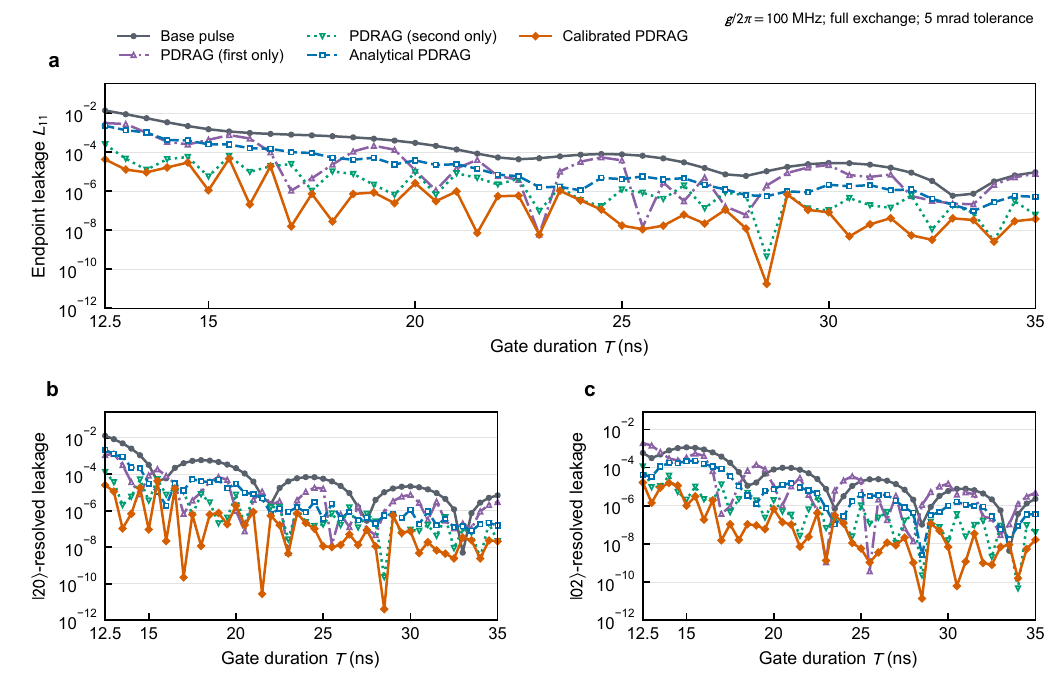}
 \caption{Raised-cosine Duffing pulses across gate durations. (a) Endpoint leakage $L_{11}$ and (b),(c) populations in $\ket{20}_{\rm d}$ and $\ket{02}_{\rm d}$ after evolution from $\ket{11}_{\rm d}$. Each of the five protocols is calibrated for full exchange. Markers show $46$ sampled durations; lines connect adjacent values. The command endpoints follow Sec.~\ref{sec:supp:spec:exact}.}
 \label{fig:sin2:duration}
\end{figure}

Calibrated PDRAG suppresses leakage below Base and both single-derivative controls
at all $46$ durations (Fig.~\ref{fig:sin2:duration}).
Base-to-calibrated gains span $14.3$--$5.91\times10^5$, with a median of $582$;
gains over the better single-derivative control have a median of $3.50$ and a maximum of $669$.
The minimum selected leakage is $1.77\times10^{-11}$ at $28.5$ ns,
against $1.04\times10^{-5}$ for Base and $4.50\times10^{-10}$ for
the better single derivative.
Tables~\ref{tab:duffing:twenty:parameters} and
\ref{tab:duffing:twenty:metrics} give the $20$-ns values.

\begin{table}[!htbp]
\centering
\footnotesize
\caption{Raised-cosine Duffing parameters at $20$ ns. The carrier column gives the full frequency, including the $600$-MHz bare detuning.}
\label{tab:duffing:twenty:parameters}
\begin{tabular}{lrrrr}
\toprule
Protocol & $\beta$ (ns) & $\gamma$ (ns$^2$) & $s$ & $\Omega/2\pi$ (MHz) \\
\midrule
Base & $0.000000$ & $0.000000$ & $0.999345$ & $630.292420$ \\
First derivative only & $-1.396157$ & $0.000000$ & $0.937449$ & $644.727821$ \\
Second derivative only & $0.000000$ & $0.728165$ & $0.999029$ & $630.566815$ \\
Analytical PDRAG & $0.060286$ & $0.479740$ & $0.999023$ & $629.827661$ \\
Calibrated PDRAG & $0.154710$ & $0.695238$ & $0.998330$ & $628.920749$ \\
\bottomrule
\end{tabular}
\end{table}
\begin{table}[!htbp]
\centering
\footnotesize
\caption{Raised-cosine Duffing leakage at $20$ ns. Gains are the reference leakage divided by the listed value; the single-derivative reference is the better optimized first-only or second-only pulse.}
\label{tab:duffing:twenty:metrics}
\begin{tabular}{lrrrrr}
\toprule
Protocol & $L_{11}$ & $P_{20}$ & $P_{02}$ & Gain over Base & Gain over better single \\
\midrule
Base & $2.966\times10^{-4}$ & $2.164\times10^{-4}$ & $8.020\times10^{-5}$ & $1.00$ & $0.03$ \\
First derivative only & $1.152\times10^{-5}$ & $9.932\times10^{-7}$ & $1.053\times10^{-5}$ & $25.74$ & $0.82$ \\
Second derivative only & $9.495\times10^{-6}$ & $7.566\times10^{-6}$ & $1.929\times10^{-6}$ & $31.24$ & $1.00$ \\
Analytical PDRAG & $3.746\times10^{-5}$ & $3.021\times10^{-5}$ & $7.251\times10^{-6}$ & $7.92$ & $0.25$ \\
Calibrated PDRAG & $2.562\times10^{-6}$ & $1.887\times10^{-6}$ & $6.748\times10^{-7}$ & $115.78$ & $3.71$ \\
\bottomrule
\end{tabular}
\end{table}

\begin{figure}[!htbp]
\centering
\includegraphics[width=0.85\linewidth]{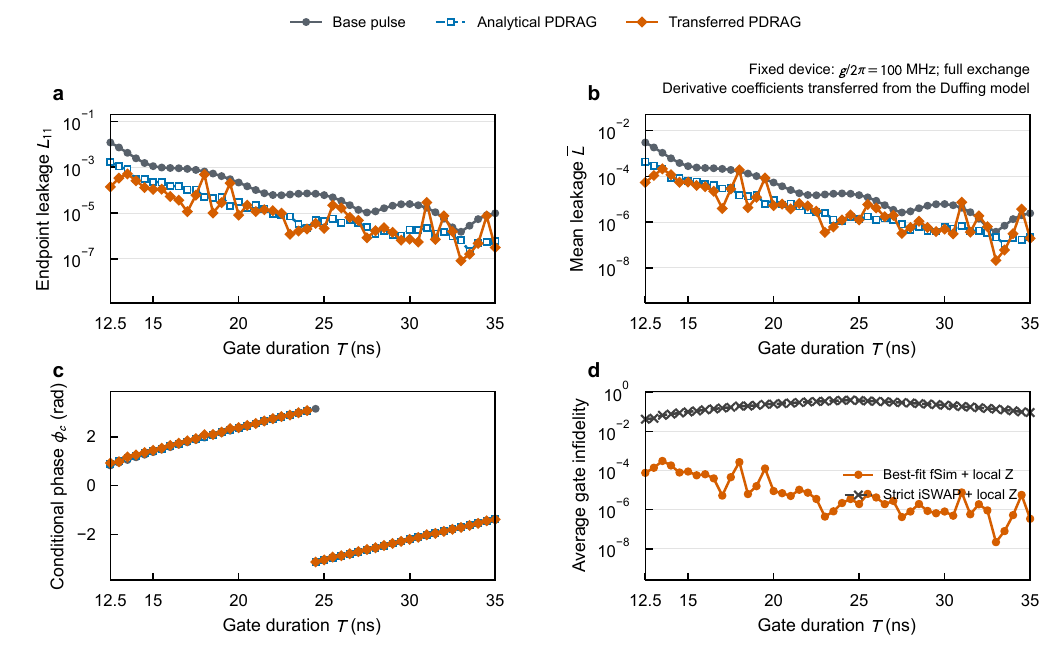}
 \caption{Raised-cosine charge-model transfer. (a) Leakage $L_{11}$ from $\ket{11}_{\rm d}$, (b) four-input mean leakage $\overline L$, (c) conditional phase $\phi_c$, and (d) transferred-PDRAG best-fit fSim and strict-iSWAP infidelities. Derivative coefficients are transferred from Duffing at the same duration; amplitude and carrier are recalibrated for full exchange. Curves use $N=9$ levels per transmon and $n_{\max}=18$; Sec.~\ref{sec:supp:dmh:validation} gives convergence checks.}
 \label{fig:sin2:charge}
\end{figure}

In the charge model (Fig.~\ref{fig:sin2:charge}), transferred raised-cosine
coefficients reduce $20$-ns leakage $L_{11}$ from $2.15\times10^{-4}$ for Base
to $8.09\times10^{-6}$, a factor of $26.6$.
The leakage-reduction factor relative to analytical PDRAG is $3.82$.
Analytical PDRAG improves on Base at all $46$ durations by factors
$2.38$--$34.0$.
Transferred PDRAG improves on Base at $44$ durations, with a median
gain of $12.2$; reversals occur at $31$ and $32$ ns.
Transfer improves on analytical PDRAG at $30$ durations,
with a median gain of $1.58$ across the full scan.

At $20$ ns, transferred PDRAG has
$\overline L=5.27\times10^{-6}$ and fSim gate infidelity
$8.92\times10^{-6}$ (Tables~\ref{tab:charge:twenty:parameters}
and \ref{tab:charge:twenty:metrics}).
Its conditional phase, $\phi_c=2.37$ rad, accounts for nearly all
of the strict-iSWAP infidelity, $0.249$.
The phase-only prediction of Eq.~\eqref{eq:supp:phase:infidelity}
differs from the computed value by $6.47\times10^{-6}$.
For input $\ket{11}_{\rm d}$, leakage beyond the two design channels is
$5.75\times10^{-6}$, compared with $P_{20}+P_{02}=2.34\times10^{-6}$.
The $N=13,n_{\max}=22$ reference gives $L_{11}=8.10\times10^{-6}$,
a $0.20\%$ change that preserves the reduction relative to Base.

\FloatBarrier
\begin{table}[!htbp]
\centering
\footnotesize
\caption{Raised-cosine charge-model parameters at $20$ ns. The carrier column gives the full modulation frequency.}
\label{tab:charge:twenty:parameters}
\begin{tabular}{lrrrr}
\toprule
Protocol & $\beta$ (ns) & $\gamma$ (ns$^2$) & $s$ & $\Omega/2\pi$ (MHz) \\
\midrule
Base & $0.000000$ & $0.000000$ & $1.065919$ & $630.276485$ \\
Analytical PDRAG & $0.060286$ & $0.479740$ & $1.065584$ & $629.812910$ \\
Transferred PDRAG & $0.154710$ & $0.695238$ & $1.064847$ & $628.906517$ \\
\bottomrule
\end{tabular}
\end{table}
\begin{table}[!htbp]
\centering
\footnotesize
\caption{Raised-cosine charge-model metrics at $20$ ns with nine levels per transmon. Gate fidelities follow Sec.~\ref{sec:supp:dmh:metrics}.}
\label{tab:charge:twenty:metrics}
\begin{tabular}{lrrrrr}
\toprule
Protocol & $L_{11}$ & $\overline L$ & $\phi_c$ (rad) & $1-F_{\rm fSim}$ & $1-F_{\rm iSWAP}$ \\
\midrule
Base & $2.152\times10^{-4}$ & $5.379\times10^{-5}$ & $2.338539$ & $5.379\times10^{-5}$ & $2.437\times10^{-1}$ \\
Analytical PDRAG & $3.087\times10^{-5}$ & $9.271\times10^{-6}$ & $2.359054$ & $1.100\times10^{-5}$ & $2.475\times10^{-1}$ \\
Transferred PDRAG & $8.085\times10^{-6}$ & $5.273\times10^{-6}$ & $2.368339$ & $8.921\times10^{-6}$ & $2.492\times10^{-1}$ \\
\bottomrule
\end{tabular}
\end{table}

\FloatBarrier

\section{Sinusoidal exchange envelopes}
\label{supp:simulation}

\begin{figure}[!htbp]
 \centering
 \includegraphics[width=0.90\textwidth]{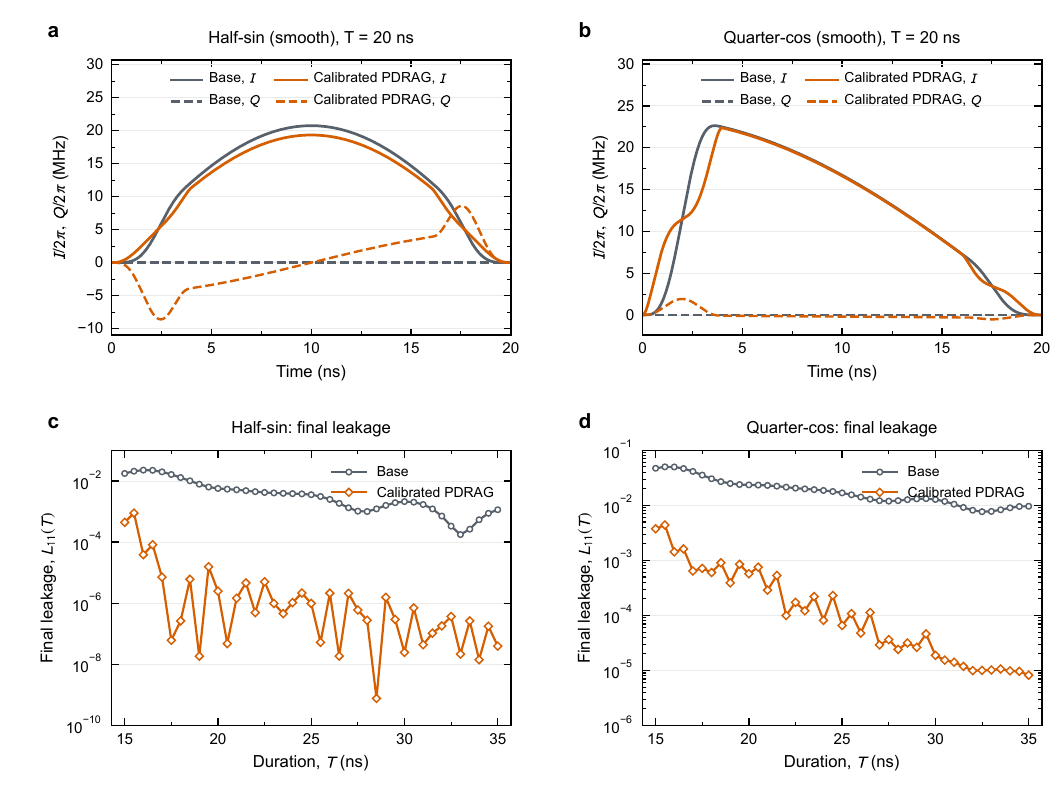}
 \caption{PDRAG for smoothly switched sinusoidal exchange.
 (a),(b) In-phase and quadrature envelopes, $I=\operatorname{Re}G$
 and $Q=\operatorname{Im}G$, for half-sine and quarter-cosine profiles
 at $T=20$ ns.
 Solid and dashed lines show $I/2\pi$ and $Q/2\pi$; grey and orange
 denote the base and calibrated PDRAG controls.
 Both profiles use the same $4$-ns endpoint ramps.
 (c),(d) Endpoint leakage $L_{11}(T)=P_{20}(T)+P_{02}(T)$ from
 $\ket{11}_{\rm d}$ for $T=15$--$35$ ns in $0.5$-ns steps (41 durations).
 Each base--PDRAG pair is calibrated to the same full-exchange
 target and satisfies the common control-resource limits.
 Markers show independently verified calculations; lines connect
 adjacent sampled durations.}
 \label{fig:supp:smooth-sinusoidal}
\end{figure}

The half-sine and quarter-cosine controls in Fig.~\ref{fig:supp:smooth-sinusoidal} use the Q--Q
device, Duffing Hamiltonian, and idle dressed basis defined in
Sec.~\ref{sec:supp:dmh:device}.
Both envelopes are evaluated at 41 durations from $15$ to $35$ ns
in steps of $0.5$ ns, with full-exchange target $\Theta=\pi/2$.

Calibrated PDRAG lowers endpoint leakage at all 41 sampled durations
for both envelopes [Figs.~\ref{fig:supp:smooth-sinusoidal}(c),(d)].
At $20$ ns, PDRAG gives $L_{11}=2.49\times10^{-6}$ for half-sine
and $5.72\times10^{-4}$ for quarter-cosine.
The corresponding Base-to-PDRAG leakage ratios are $2286.43$ and $41.28$.
The lowest sampled leakage values are
$7.65\times10^{-10}$ at $28.5$ ns for half-sine
and $8.22\times10^{-6}$ at $35$ ns for quarter-cosine.

\subsection{Smoothly joined envelopes and derivative controls}
\label{supp:simulation:smooth}

We define the dimensionless sinusoidal profiles
\begin{equation}
 f_{\rm h}(t)=\sin(\pi t/T),\qquad
 f_{\rm q}(t)=\cos(\pi t/2T),
 \qquad 0\leq t\leq T.
 \label{supp:eq:sinusoidal-profiles}
\end{equation}
Their common endpoint window is
\begin{align}
 w(x)&=
 \begin{cases}
  0,&x\leq0,\\
  35x^4-84x^5+70x^6-20x^7,&0<x<1,\\
  1,&x\geq1,
 \end{cases}
 \notag\\
 W_\tau(t)&=w(t/\tau)w[(T-t)/\tau],
 \qquad \tau=4~{\rm ns}.
 \label{supp:eq:sinusoidal-window}
\end{align}
The window is unity on $[\tau,T-\tau]$ and joins continuously
through its third derivative to zero outside $[0,T]$.
Area normalization gives
\begin{equation}
 A_f(t)=
 \frac{\Theta W_\tau(t)f(t)}
 {\displaystyle\int_0^T W_\tau(t')f(t')\,\dd t'},
 \qquad f\in\{f_{\rm h},f_{\rm q}\}.
 \label{supp:eq:sinusoidal-normalization}
\end{equation}
Thus $\int_0^T A_f(t)\,\dd t=\Theta$, and $A_f$ and its first
three derivatives vanish at both joins to idle.
The quarter-cosine profile turns on smoothly within
$[0,\tau]$, then decreases through the interior and turns off smoothly.

Candidate widths $\tau=2,2.5,3,4$ ns are assessed in ascending order
across all 82 calibrated Base envelope--duration combinations.
The first width satisfying the common limits for every Base control is selected.
At $\tau=4$ ns, each Base control satisfies the bounds in
Eq.~\eqref{eq:shared:resources} and the Bessel and flux constraints
of Sec.~\ref{s4:sec:resources}.
This window is held fixed throughout the dense scan and derivative calibration.

Derivative shaping follows Sec.~\ref{sec:supp:search}, with
derivatives evaluated after windowing and area normalization.
The window derivatives contribute to $\ddot A_f$ near the joins,
making $\beta$ and $\gamma$ independent calibration parameters.
The endpoint conditions give $G=\dot G=0$ at $t=0,T$.
Bessel inversion and Eq.~\eqref{s4:eq:command} then produce a
complete real frequency command that connects continuously to idle.
The spectral relation in Eq.~\eqref{s4:eq:filter} applies to the
windowed envelopes, including the quadratic derivative factor.

\subsection{Calibration, resources, and numerical verification}
\label{supp:simulation:calibration}

Base sets $\beta=\gamma=0$ and calibrates $s$ and the carrier
for full exchange.
PDRAG varies both derivative coefficients while recalibrating
the same amplitude and carrier parameters.
Both use the common parameter intervals in Eq.~\eqref{s4:eq:carrier}.
At fixed $T$ and $\tau$, the calibration minimizes the endpoint
leakage $L_{11}=P_{02}+P_{20}$ among controls meeting the exchange
target and the shared resource bounds.
Resources are evaluated from the complete frequency command,
including both joins to idle, with the definitions in
Sec.~\ref{s4:sec:resources}.
Table~\ref{supp:tab:smooth-simulation} gives the selected controls,
and Table~\ref{supp:tab:smooth-resources} gives their resource and
target checks.

\begin{table}[!htbp]
\centering
\footnotesize
\caption{Representative calibrated controls for the
smoothly joined envelopes in Fig.~\ref{fig:supp:smooth-sinusoidal}. The five displayed durations are selected from
the full $15$--$35$ ns scan with $0.5$ ns spacing.
$\delta f=\Omega/(2\pi)-600$ MHz is the carrier offset.
All controls use $\tau=4$ ns, the same full-exchange target, and common resource bounds.}
\label{supp:tab:smooth-simulation}
\begin{tabular}{llrrrrrr}
\toprule
Envelope & Control & $T$ (ns) & $s$ & $\beta$ (ns) & $\gamma$ (ns$^2$) & $\delta f$ (MHz) & $L_{11}(T)$ \\
\midrule
Half-sine & Base & 15 & 0.99896762 & 0.00000000 & 0.00000000 & 29.10479752 & $1.75813\times10^{-2}$ \\
Half-sine & PDRAG & 15 & 0.99956758 & -0.05979492 & 0.53750000 & 30.37467165 & $4.43032\times10^{-4}$ \\
Half-sine & Base & 20 & 0.99947728 & 0.00000000 & 0.00000000 & 30.80353231 & $5.69454\times10^{-3}$ \\
Half-sine & PDRAG & 20 & 0.93893309 & -1.56071777 & 0.32742188 & 43.06631500 & $2.49058\times10^{-6}$ \\
Half-sine & Base & 25 & 0.99969060 & 0.00000000 & 0.00000000 & 31.45811960 & $3.55586\times10^{-3}$ \\
Half-sine & PDRAG & 25 & 0.99927717 & 0.12918457 & 0.67811035 & 30.82871161 & $9.83362\times10^{-7}$ \\
Half-sine & Base & 30 & 0.99978829 & 0.00000000 & 0.00000000 & 31.78432333 & $2.11867\times10^{-3}$ \\
Half-sine & PDRAG & 30 & 0.99967567 & 0.13004069 & 0.67754883 & 31.34641609 & $2.45703\times10^{-8}$ \\
Half-sine & Base & 35 & 0.99984696 & 0.00000000 & 0.00000000 & 31.97184184 & $1.14548\times10^{-3}$ \\
Half-sine & PDRAG & 35 & 0.99964286 & 0.13539551 & 0.68670410 & 31.63744825 & $3.96732\times10^{-8}$ \\
\midrule
Quarter-cosine & Base & 15 & 0.99926671 & 0.00000000 & 0.00000000 & 29.16924251 & $4.73794\times10^{-2}$ \\
Quarter-cosine & PDRAG & 15 & 0.99731940 & 0.17143717 & 0.50000000 & 26.99419986 & $3.75493\times10^{-3}$ \\
Quarter-cosine & Base & 20 & 0.99957759 & 0.00000000 & 0.00000000 & 30.86018218 & $2.36162\times10^{-2}$ \\
Quarter-cosine & PDRAG & 20 & 0.99857793 & 0.15182943 & 0.58085938 & 29.75315661 & $5.72143\times10^{-4}$ \\
Quarter-cosine & Base & 25 & 0.99973485 & 0.00000000 & 0.00000000 & 31.50760575 & $1.68845\times10^{-2}$ \\
Quarter-cosine & PDRAG & 25 & 0.99916193 & 0.14014924 & 0.65224045 & 30.86579147 & $6.57997\times10^{-5}$ \\
Quarter-cosine & Base & 30 & 0.99981598 & 0.00000000 & 0.00000000 & 31.82626225 & $1.28450\times10^{-2}$ \\
Quarter-cosine & PDRAG & 30 & 0.99956854 & 0.13244629 & 0.67101074 & 31.40720001 & $1.88440\times10^{-5}$ \\
Quarter-cosine & Base & 35 & 0.99986667 & 0.00000000 & 0.00000000 & 32.00701193 & $9.66228\times10^{-3}$ \\
Quarter-cosine & PDRAG & 35 & 0.99964026 & 0.13286897 & 0.67580979 & 31.69980470 & $8.21582\times10^{-6}$ \\
\bottomrule
\end{tabular}
\end{table}

At each duration, the derivative search uses a $13\times13$ grid
over the full coefficient intervals and a second $13\times13$ grid
near zero, $\beta\in[-0.8,0.8]$ ns and $\gamma\in[-0.4,0.8]$ ns$^2$.
Each point uses three common amplitude--carrier starts.
Base is calibrated from 20 common starts.
All feasible grid-local minima, the analytic control, and Base
enter bounded local refinement with 50 iterations per pass.
Candidates transferred from adjacent $0.5$-ns durations are recalibrated
and refined at the target duration for 41 half-sine points
and 41 quarter-cosine points.
All competing minima are evaluated at the same final propagation
precision before selection.

For each selected command, the two initial states
$\ket{10}_{\rm d}$ and $\ket{11}_{\rm d}$ are independently propagated
in the full 16-dimensional tensor-product Hamiltonian.
We use the explicit eighth-order Runge--Kutta solver DOP853 with settings
$(\mathrm{rtol},\mathrm{atol},h_{\max})$ of
$(2\times10^{-11},2\times10^{-13},0.025~{\rm ns})$ and
$(2\times10^{-13},2\times10^{-15},0.008~{\rm ns})$.
Here rtol and atol denote relative and absolute tolerances;
$h_{\max}$ is the maximum integration step.
Every physical command parameter is held fixed between these runs.
Across the 164 smooth controls, the largest difference in the evaluated
transfer and leakage probabilities is
$3.38\times10^{-10}$ between the two solver settings
and $2.78\times10^{-11}$ between the excitation-sector
calibration and tight full-space propagation.
The acceptance threshold for the residual return amplitude is
$0.005$; the largest verified residual is $2.07\times10^{-10}$.
Leakage is the sum of absolute dressed-state probabilities
$P_{02}+P_{20}$.

Resources are evaluated at time steps of $0.001$ and $0.0005$ ns.
The bandwidth record includes $4$ ns of zero idle on each side of
the pulse, and the RMS excursion is normalized to the duration $T$.
Every control meets the shared limits on both sampling grids.

\begin{table}[!htbp]
\centering
\footnotesize
\caption{Resource and target checks for the smoothly joined controls.
Each column gives the maximum over all 41 durations and both resource
sampling grids; the flux row gives the combined range.
The return amplitude is evaluated by independent tight-tolerance propagation.}
\label{supp:tab:smooth-resources}
\begin{tabular}{lrrrrr}
\toprule
 & & \multicolumn{2}{c}{Half-sine} & \multicolumn{2}{c}{Quarter-cosine} \\
Quantity & Limit & Base & PDRAG & Base & PDRAG \\
\midrule
Bessel utilization $u$ & .98 & 0.496826 & 0.485404 & 0.542909 & 0.530962 \\
Peak excursion (GHz) & 1.00 & 0.379829 & 0.370644 & 0.414001 & 0.408922 \\
RMS excursion (GHz) & .33 & 0.183367 & 0.177682 & 0.184474 & 0.177037 \\
Peak slew (GHz/ns) & 4.10 & 1.510277 & 1.476088 & 1.686339 & 1.594624 \\
$B_{99.9}$ (GHz) & 1.00 & 0.874982 & 0.999982 & 0.956501 & 0.999988 \\
Flux range ($\Phi_0$) & $[0,.49]$ & $[0.19605,0.29316]$ & $[0.19755,0.29223]$ & $[0.19038,0.29663]$ & $[0.19123,0.29438]$ \\
Return amplitude & $.005$ & $1.95\times10^{-10}$ & $2.07\times10^{-10}$ & $2.01\times10^{-10}$ & $2.03\times10^{-10}$ \\
\bottomrule
\end{tabular}
\end{table}

\FloatBarrier
\subsection{Original finite sinusoidal segments}
\label{supp:simulation:original}

\begin{figure}[!htbp]
\centering
\includegraphics[width=0.90\linewidth]{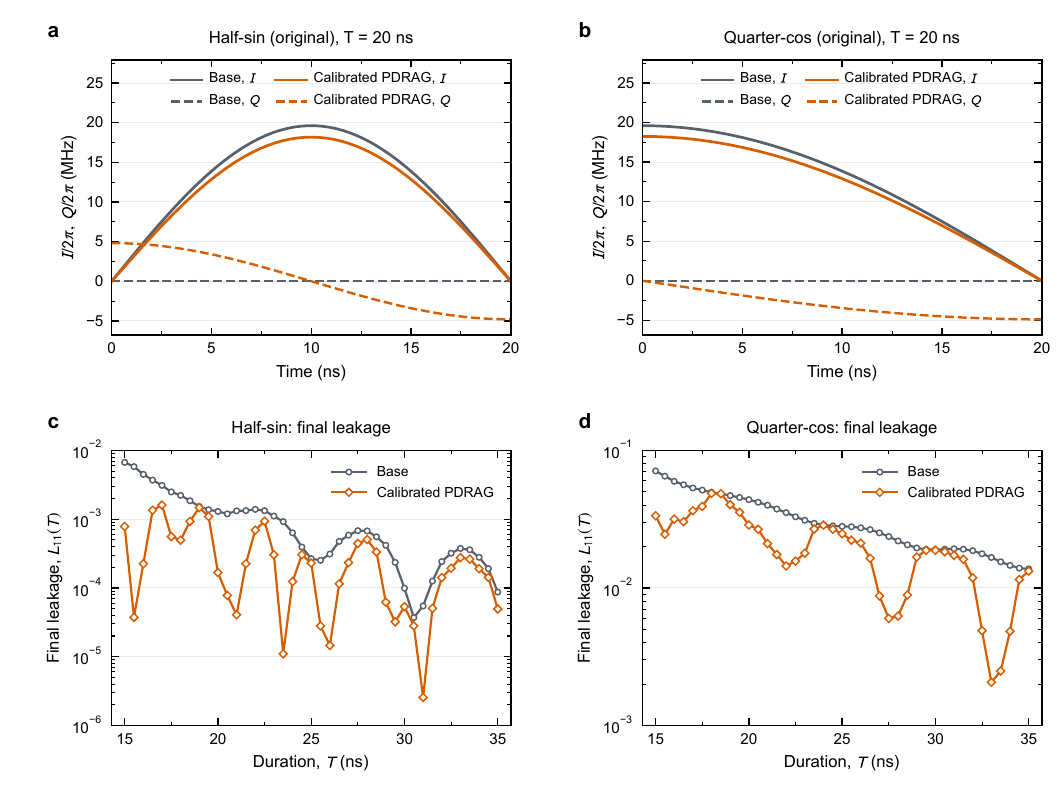}
\caption{Original finite sinusoidal exchange segments.
(a),(b) In-phase and quadrature envelopes,
$I=\operatorname{Re}G$ and $Q=\operatorname{Im}G$, for half-sine
and quarter-cosine controls at $T=20$ ns.
Solid and dashed lines show $I/2\pi$ and $Q/2\pi$; grey and orange
denote Base and calibrated PDRAG.
(c),(d) Endpoint leakage $L_{11}(T)=P_{20}(T)+P_{02}(T)$ from
$\ket{11}_{\rm d}$ for $T=15$--$35$ ns in $0.5$-ns steps (41 durations).
Each pair is calibrated for full exchange using the original
finite segments in Eq.~\eqref{supp:eq:original-segments}.
Propagation uses the Q--Q Duffing model and the complete real
frequency command, with ideal endpoint truncation.
Markers show verified calculations; lines connect adjacent durations.}
\label{fig:supp:original-segments}
\end{figure}

The unwindowed sinusoidal profiles are
\begin{align}
 A_{\rm h}^{(0)}(t)&=
 \frac{\pi\Theta}{2T}\sin(\pi t/T),\notag\\
 A_{\rm q}^{(0)}(t)&=
 \frac{\pi\Theta}{2T}\cos(\pi t/2T),
 \qquad 0\leq t\leq T.
 \label{supp:eq:original-segments}
\end{align}
Both have area $\Theta$.
Figure~\ref{fig:supp:original-segments} shows their calibrated
quadratures and duration-dependent leakage under the same
full-exchange target.
The controls are propagated over $[0,T]$, with ideal truncation
at their original endpoints.

Within each original segment,
$\ddot A^{(0)}=-\kappa^2A^{(0)}$, with $\kappa=\pi/T$ for
half-sine and $\kappa=\pi/(2T)$ for quarter-cosine.
The derivative control therefore reduces to
\begin{align}
 s\big(A^{(0)}+\ii\beta\dot A^{(0)}
                   +\gamma\ddot A^{(0)}\big)
 &=s_{\rm eff}
   \big(A^{(0)}+\ii\beta_{\rm eff}\dot A^{(0)}\big),\notag\\
 s_{\rm eff}&=s(1-\gamma\kappa^2),\qquad
 \beta_{\rm eff}=\frac{\beta}{1-\gamma\kappa^2},
 \label{supp:eq:original-effective}
\end{align}
for $1-\gamma\kappa^2\ne0$.
The implementation uses $s_{\rm eff}$ and $\beta_{\rm eff}$ as
the independent coordinates and sets $\gamma=0$.
The amplitude and carrier intervals are those in
Eq.~\eqref{s4:eq:carrier}.
Both original profiles use the common derivative interval
$\beta_{\rm eff}\in[-16,8]$ ns with $\gamma=0$.
At every duration, a complete one-dimensional coefficient scan
is followed by local refinement of its feasible minima.
Candidate transfer from adjacent $0.5$-ns durations is completed for
41 half-sine points and 41 quarter-cosine points.
Base fixes $\beta_{\rm eff}=0$.

Integration by parts gives the finite-interval spectrum
\begin{equation}
 \widetilde G(\delta)=s_{\rm eff}
 \left[(1+\beta_{\rm eff}\delta)\widetilde A^{(0)}(\delta)
 +\ii\beta_{\rm eff}
 \big(A^{(0)}(T)e^{\ii\delta T}-A^{(0)}(0)\big)\right].
 \label{supp:eq:segment-spectrum}
\end{equation}
For half-sine, $A^{(0)}(0)=A^{(0)}(T)=0$.
For quarter-cosine, $A^{(0)}(0)=\pi\Theta/(2T)$ and
$A^{(0)}(T)=0$, giving a finite boundary contribution.
The original quadratures have finite endpoint values, and the
complete commands have finite jumps at their joins to idle.

Amplitude and carrier are calibrated at each derivative coordinate
with an accepted residual return amplitude of $0.005$.
Selection minimizes $L_{11}$ over target-matched candidates
from the independent and adjacent-duration searches.
Half-sine selection also imposes the finite-window amplitude,
interior-slew, bandwidth, Bessel, and flux bounds.
Quarter-cosine selection uses the exchange and parameter conditions.
Table~\ref{supp:tab:original-simulation} reports the selected
parameters and endpoint leakage.
The 164 original commands use the same fixed-command verification
protocol described in Sec.~\ref{supp:simulation:calibration}.
Their full-space propagation satisfies the exchange target.
The largest verified residual return amplitude is $2.70\times10^{-10}$.
The largest probability differences are
$7.59\times10^{-11}$ between solver settings
and $3.91\times10^{-11}$ between excitation-sector
and tight full-space propagation.

Calibrated PDRAG lowers endpoint leakage at all 41 sampled durations
for both envelopes [Fig.~\ref{fig:supp:original-segments}(c),(d)].
The scan spans $15$--$35$ ns in steps of $0.5$ ns.
At $20$ ns, PDRAG gives $L_{11}=1.68\times10^{-4}$ for half-sine
and $2.87\times10^{-2}$ for quarter-cosine.
The corresponding Base-to-PDRAG leakage ratios are $7.73$ and $1.53$.
The lowest sampled leakages are
$2.56\times10^{-6}$ at $31$ ns for half-sine
and $2.06\times10^{-3}$ at $33$ ns for quarter-cosine.

\FloatBarrier
\noindent\begin{minipage}{\linewidth}
\centering
\footnotesize
\captionof{table}{Representative calibrated controls for the
original finite segments in Fig.~\ref{fig:supp:original-segments}. The five displayed durations are selected from
the full $15$--$35$ ns scan with $0.5$ ns spacing.
$\delta f=\Omega/(2\pi)-600$ MHz is the carrier offset.
Base and PDRAG share the full-exchange target at each duration.}
\label{supp:tab:original-simulation}
\begin{tabular}{llrrrrr}
\toprule
Envelope & Control & $T$ (ns) & $s_{\rm eff}$ & $\beta_{\rm eff}$ (ns) & $\delta f$ (MHz) & $L_{11}(T)$ \\
\midrule
Half-sine & Base & 15 & 0.99858367 & 0.00000000 & 29.88983564 & $6.70797\times10^{-3}$ \\
Half-sine & PDRAG & 15 & 0.97280583 & 0.64070907 & 21.93451028 & $7.83105\times10^{-4}$ \\
Half-sine & Base & 20 & 0.99935528 & 0.00000000 & 31.04707035 & $1.29557\times10^{-3}$ \\
Half-sine & PDRAG & 20 & 0.92572752 & 1.68538772 & 20.22733712 & $1.67532\times10^{-4}$ \\
Half-sine & Base & 25 & 0.99969606 & 0.00000000 & 31.54934034 & $2.66468\times10^{-4}$ \\
Half-sine & PDRAG & 25 & 0.97526319 & 1.12745039 & 26.53555205 & $2.30403\times10^{-4}$ \\
Half-sine & Base & 30 & 0.99963667 & 0.00000000 & 31.82849161 & $9.90255\times10^{-5}$ \\
Half-sine & PDRAG & 30 & 0.99745877 & -0.67836084 & 34.06175542 & $5.34007\times10^{-5}$ \\
Half-sine & Base & 35 & 0.99978354 & 0.00000000 & 31.99873825 & $8.71140\times10^{-5}$ \\
Half-sine & PDRAG & 35 & 0.98304591 & -1.58678044 & 35.77347761 & $4.94245\times10^{-5}$ \\
\midrule
Quarter-cosine & Base & 15 & 0.99808765 & 0.00000000 & 31.76813539 & $7.09087\times10^{-2}$ \\
Quarter-cosine & PDRAG & 15 & 0.81034827 & 5.17972823 & 17.34907596 & $3.34889\times10^{-2}$ \\
Quarter-cosine & Base & 20 & 0.99912118 & 0.00000000 & 31.99218951 & $4.38831\times10^{-2}$ \\
Quarter-cosine & PDRAG & 20 & 0.93006419 & 3.40816952 & 25.00738131 & $2.86912\times10^{-2}$ \\
Quarter-cosine & Base & 25 & 0.99950430 & 0.00000000 & 32.12387123 & $2.80309\times10^{-2}$ \\
Quarter-cosine & PDRAG & 25 & 0.98263557 & 1.99553415 & 29.12825569 & $2.48026\times10^{-2}$ \\
Quarter-cosine & Base & 30 & 0.99959436 & 0.00000000 & 32.21183128 & $1.89842\times10^{-2}$ \\
Quarter-cosine & PDRAG & 30 & 0.99700699 & 0.80136277 & 31.33812111 & $1.87736\times10^{-2}$ \\
Quarter-cosine & Base & 35 & 0.99973906 & 0.00000000 & 32.27049875 & $1.37457\times10^{-2}$ \\
Quarter-cosine & PDRAG & 35 & 0.99423708 & -1.67218932 & 33.59361657 & $1.32575\times10^{-2}$ \\
\bottomrule
\end{tabular}
\end{minipage}
\par\medskip

\makeatletter
\let\auto@bib@innerbib\@empty
\makeatother

\FloatBarrier

\section{Q--Q numerical validation and resource checks}
\label{sec:supp:dmh:validation}
\subsection{Shared propagation checks}

The Duffing calibration propagates the one- and two-excitation sectors,
which contain the exchange dynamics and the two leakage channels.
Fourth-order Runge--Kutta calibration uses maximum steps
$0.010$ ns (coarse) and $0.005$ ns (fine);
leakage evaluation uses $0.004$ and $0.002$ ns, respectively.
For raised-cosine pulses, the coarse and fine tolerances on the residual
single-excitation return amplitude are $2\times10^{-7}$ and
$2\times10^{-9}$, respectively.
Every selected command is propagated in the full 16-state space
with DOP853.
An independent implementation reconstructs the command and integrates its phase.
Restoring the endpoint phases permits comparison between direct and phase-frame propagation.
The map discrepancy is $\max_{jk}|\Delta K_{jk}|$, where
$\Delta K$ is the difference between the two computational maps.

Independent propagation uses three precision tiers,
\begin{equation}
 (\mathrm{rtol},\mathrm{atol},h_{\max})\in
 \left\{
 \begin{aligned}
 &(2\times10^{-13},2\times10^{-15},0.0075\ {\rm ns}),\\
 &(5\times10^{-14},5\times10^{-16},0.00375\ {\rm ns}),\\
 &(3\times10^{-14},3\times10^{-16},0.001875\ {\rm ns})
 \end{aligned}\right\}.
 \label{eq:supp:dmh:independenttiers}
\end{equation}
Let $\boldsymbol\ell_{\rm ref}$ be the reference leakage-amplitude vector
and $\boldsymbol\ell$ the vector being checked, expressed in the same basis.
Their probability difference is bounded by $E_\ell$, where
\begin{equation}
 d_\ell=\|\boldsymbol\ell-\boldsymbol\ell_{\rm ref}\|,\qquad
 E_\ell=2\|\boldsymbol\ell_{\rm ref}\|d_\ell+d_\ell^2.
 \label{eq:supp:dmh:amplitudecheck}
\end{equation}
The bound applies to the orthogonal residual defining $L_{11}$ and
the transition amplitudes defining $P_{20}$ and $P_{02}$.
For $\sin^4$, the probability agreement criterion is $1\%$
between direct and phase-frame propagation and between consecutive
precision tiers. Channels at or below $10^{-11}$ use all three tiers.
For raised-cosine pulses, at least two tiers are used, with
$E_\ell/\|\boldsymbol\ell_{\rm ref}\|^2\leq2\times10^{-3}$,
entrywise map difference at most $10^{-8}$, and norm errors below $10^{-10}$.
All maps and probabilities satisfy these criteria.

Additional raised-cosine checks compare solver settings
$(2\times10^{-11},2\times10^{-13},0.018\ {\rm ns})$
with $(2\times10^{-13},2\times10^{-15},0.006\ {\rm ns})$
at fixed physical commands.
The leakage agreement criterion is $10^{-3}$ relative or
$10^{-20}$ absolute, with the same map and norm bounds.
The reported metrics use the tighter solver settings.

Charge calibration uses seven levels and DOP853 tolerances
$(3\times10^{-8},3\times10^{-10},0.020\ {\rm ns})$.
Nine-level propagation uses $(2\times10^{-11},2\times10^{-13},0.012\ {\rm ns})$;
fixed-command checks use $(2\times10^{-13},2\times10^{-15},0.006\ {\rm ns})$.
An independent tensor-product construction verifies the Hamiltonian.
Factoring out each input's idle phase improves numerical integration;
laboratory-frame maps are restored before evaluating gate metrics.
Raised-cosine ideal-command checks require leakage agreement within
$10^{-3}$ relative or $10^{-15}$ absolute, map difference at most
$10^{-8}$, and norm errors at most $10^{-10}$.

\subsection{Basis convergence and duration-dependent features}

Basis-convergence checks increase $N$ and the cutoff $n_{\max}=\max|n|$
at fixed command and frequency-to-flux map.
All Hamiltonians match the same idle parameters and exchange matrix element.
Tables~\ref{s4:tab:convergence} and \ref{s4:tab:adaptive}
give the $\sin^4$ convergence sequences at $12.5$, $13$, and $20$ ns.
Across $81$ charge-basis calculations, the largest final successive-level
change is $0.0456\%$, and the largest charge-cutoff change is
$3.49\times10^{-6}\%$.
The largest change from the nine-level result is $0.946\%$.
The converged Base-to-transferred leakage ratios are $40.7$, $537$,
and $1.55\times10^3$ at $12.5$, $13$, and $20$ ns, respectively.
The maximum exchange-angle error is $0.189$ mrad.

\begin{table}[!htbp]
\centering
\scriptsize
\caption{Fixed-command $\sin^4$ charge-model convergence. The charge basis is $-n_{\max},\ldots,n_{\max}$; $N$ is the number of retained levels per transmon. Changes are relative to $N=9,n_{\max}=18$. Error denotes $|\theta-\pi/2|$.}\label{s4:tab:convergence}
\begin{tabular}{rlrrrrr}
\toprule
$T$ (ns) & Family & $N$ & $n_{\max}$ & $L_{11}$ & Change (\%) & Error (mrad)\\
\midrule
12.5 & Analytical & 9 & 18 & $2.8052\times10^{-2}$ & 0.000000 & 0.096712\\
12.5 & Analytical & 10 & 18 & $2.8067\times10^{-2}$ & 0.050283 & 0.099188\\
12.5 & Analytical & 11 & 18 & $2.8068\times10^{-2}$ & 0.053989 & 0.108341\\
12.5 & Analytical & 11 & 22 & $2.8068\times10^{-2}$ & 0.053989 & 0.108341\\
12.5 & Base & 9 & 18 & $8.3400\times10^{-2}$ & 0.000000 & 0.160064\\
12.5 & Base & 10 & 18 & $8.3462\times10^{-2}$ & 0.073685 & 0.161949\\
12.5 & Base & 11 & 18 & $8.3460\times10^{-2}$ & 0.072106 & 0.186967\\
12.5 & Base & 11 & 22 & $8.3460\times10^{-2}$ & 0.072106 & 0.186967\\
12.5 & Transferred & 9 & 18 & $2.0497\times10^{-3}$ & 0.000000 & 0.039516\\
12.5 & Transferred & 10 & 18 & $2.0496\times10^{-3}$ & 0.005698 & 0.040927\\
12.5 & Transferred & 11 & 18 & $2.0496\times10^{-3}$ & 0.007251 & 0.042523\\
12.5 & Transferred & 11 & 22 & $2.0496\times10^{-3}$ & 0.007251 & 0.042523\\
13 & Analytical & 9 & 18 & $2.2160\times10^{-2}$ & 0.000000 & 0.088102\\
13 & Analytical & 10 & 18 & $2.2170\times10^{-2}$ & 0.043948 & 0.090524\\
13 & Analytical & 11 & 18 & $2.2171\times10^{-2}$ & 0.048398 & 0.098157\\
13 & Analytical & 11 & 22 & $2.2171\times10^{-2}$ & 0.048398 & 0.098157\\
13 & Base & 9 & 18 & $6.2724\times10^{-2}$ & 0.000000 & 0.133313\\
13 & Base & 10 & 18 & $6.2758\times10^{-2}$ & 0.052994 & 0.135704\\
13 & Base & 11 & 18 & $6.2759\times10^{-2}$ & 0.055207 & 0.152573\\
13 & Base & 11 & 22 & $6.2759\times10^{-2}$ & 0.055207 & 0.152573\\
13 & Transferred & 9 & 18 & $1.1658\times10^{-4}$ & 0.000000 & 0.034472\\
13 & Transferred & 10 & 18 & $1.1679\times10^{-4}$ & 0.178067 & 0.035717\\
13 & Transferred & 11 & 18 & $1.1695\times10^{-4}$ & 0.314333 & 0.036985\\
13 & Transferred & 11 & 22 & $1.1695\times10^{-4}$ & 0.314333 & 0.036985\\
20 & Analytical & 9 & 18 & $1.2821\times10^{-4}$ & 0.000000 & 0.034080\\
20 & Analytical & 10 & 18 & $1.2829\times10^{-4}$ & 0.062832 & 0.035293\\
20 & Analytical & 11 & 18 & $1.2830\times10^{-4}$ & 0.069023 & 0.036624\\
20 & Analytical & 11 & 22 & $1.2830\times10^{-4}$ & 0.069023 & 0.036624\\
20 & Base & 9 & 18 & $3.3304\times10^{-4}$ & 0.000000 & 0.048950\\
20 & Base & 10 & 18 & $3.3322\times10^{-4}$ & 0.053730 & 0.050220\\
20 & Base & 11 & 18 & $3.3323\times10^{-4}$ & 0.058242 & 0.051761\\
20 & Base & 11 & 22 & $3.3323\times10^{-4}$ & 0.058242 & 0.051761\\
20 & Transferred & 9 & 18 & $2.1746\times10^{-7}$ & 0.000005 & 0.029282\\
20 & Transferred & 10 & 18 & $2.1577\times10^{-7}$ & 0.777844 & 0.030327\\
20 & Transferred & 11 & 18 & $2.1550\times10^{-7}$ & 0.901195 & 0.031335\\
20 & Transferred & 11 & 22 & $2.1550\times10^{-7}$ & 0.901195 & 0.031335\\
\bottomrule
\end{tabular}
\end{table}
\begin{table}[!htbp]
\centering\scriptsize
\caption{Extended basis convergence for nine representative $\sin^4$ charge commands. The final cutoff is $n_{\max}=22$. Production change is relative to $N=9$; level change compares the final two level truncations at $n_{\max}=18$. Cutoff change compares $n_{\max}=18$ and $22$ at the final $N$. Error denotes $|\theta-\pi/2|$.}\label{s4:tab:adaptive}
\begin{tabular}{rlrrrrrr}\toprule
$T$ (ns) & Family & $N$ & $L_{11}$ & Production change (\%) & Level change (\%) & Cutoff change (\%) & Error (mrad)\\\midrule
12.5 & Analytical & 12 & $2.8069\times10^{-2}$ & 0.05739 & $3.40\times10^{-3}$ & $2.11\times10^{-9}$ & 0.10840\\
12.5 & Base & 13 & $8.3464\times10^{-2}$ & 0.07692 & $3.55\times10^{-4}$ & $7.54\times10^{-9}$ & 0.18873\\
12.5 & Transferred & 12 & $2.0496\times10^{-3}$ & 0.00760 & $3.45\times10^{-4}$ & $7.82\times10^{-8}$ & 0.04257\\
13 & Analytical & 12 & $2.2172\times10^{-2}$ & 0.05130 & $2.90\times10^{-3}$ & $4.19\times10^{-9}$ & 0.09822\\
13 & Base & 13 & $6.2761\times10^{-2}$ & 0.05872 & $1.17\times10^{-4}$ & $4.01\times10^{-9}$ & 0.15375\\
13 & Transferred & 12 & $1.1696\times10^{-4}$ & 0.32292 & $8.56\times10^{-3}$ & $1.21\times10^{-7}$ & 0.03702\\
20 & Analytical & 12 & $1.2830\times10^{-4}$ & 0.07282 & $3.80\times10^{-3}$ & $6.04\times10^{-8}$ & 0.03666\\
20 & Base & 12 & $3.3324\times10^{-4}$ & 0.06156 & $3.31\times10^{-3}$ & $1.61\times10^{-8}$ & 0.05180\\
20 & Transferred & 12 & $2.1540\times10^{-7}$ & 0.94637 & $4.56\times10^{-2}$ & $3.49\times10^{-6}$ & 0.03137\\
\bottomrule\end{tabular}
\end{table}

\begin{figure}[!htbp]
\centering
\includegraphics[width=\linewidth]{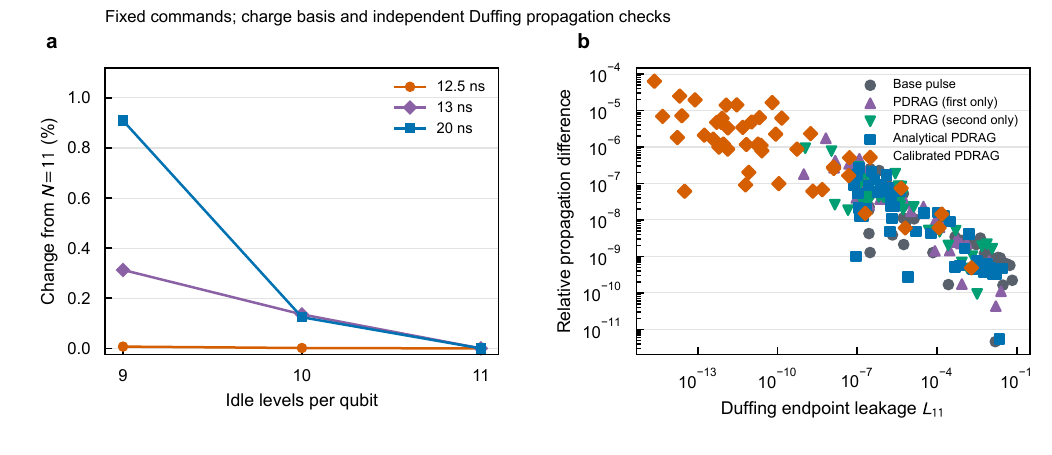}
\caption{Numerical convergence of $\sin^4$ pulses at fixed commands. (a) Charge-model leakage changes for transferred PDRAG at $12.5$, $13$, and $20$ ns, relative to $N=11,n_{\max}=18$. Tables~\ref{s4:tab:convergence} and \ref{s4:tab:adaptive} give the corresponding checks for all three protocols. (b) Direct real-command and phase-frame leakage for all $230$ Duffing commands.}
\label{s4:fig:convergence}
\end{figure}

Independent propagation verifies all $230$ $\sin^4$ Duffing commands,
including $460$ resolved channels, with a maximum map discrepancy of
\DuffingMapDifference{} (Table~\ref{s4:tab:solver}).
A $0.1$-ns duration grid resolves leakage minima near $23$ ns.
Additional calculations at $14$ single-derivative durations and $18$
joint-control points resolve neighboring minima and changes of calibration branch.
Figure~\ref{s4:fig:convergence} summarizes propagation and basis convergence.

\begin{table}[!htbp]
\centering
\scriptsize
\caption{Independent 16-state Duffing propagation of representative $\sin^4$ commands. The final columns compare phase-frame and direct real-command propagation through the relative leakage difference and largest entrywise computational-map difference. The full comparison covers all $230$ commands.}\label{s4:tab:solver}
\begin{tabular}{rlrrrr}
\toprule
$T$ (ns) & Family & Selected $L_{11}$ & Direct $L_{11}$ & Phase/direct (\%) & $\max|\Delta K|$\\
\midrule
12.5 & Analytical & $2.7942\times10^{-2}$ & $2.7942\times10^{-2}$ & 0.000000 & $1.80\times10^{-10}$\\
12.5 & Base & $6.8181\times10^{-2}$ & $6.8181\times10^{-2}$ & 0.000000 & $9.39\times10^{-11}$\\
12.5 & Calibrated & $1.9850\times10^{-3}$ & $1.9850\times10^{-3}$ & 0.000000 & $1.72\times10^{-10}$\\
12.5 & First only & $2.5664\times10^{-2}$ & $2.5664\times10^{-2}$ & 0.000000 & $3.33\times10^{-10}$\\
12.5 & Second only & $1.2080\times10^{-2}$ & $1.2080\times10^{-2}$ & 0.000000 & $1.31\times10^{-10}$\\
13 & Analytical & $2.2360\times10^{-2}$ & $2.2360\times10^{-2}$ & 0.000000 & $1.29\times10^{-10}$\\
13 & Base & $5.5649\times10^{-2}$ & $5.5649\times10^{-2}$ & 0.000000 & $1.89\times10^{-10}$\\
13 & Calibrated & $1.9833\times10^{-7}$ & $1.9833\times10^{-7}$ & 0.000002 & $8.49\times10^{-11}$\\
13 & First only & $1.6470\times10^{-2}$ & $1.6470\times10^{-2}$ & 0.000000 & $3.49\times10^{-10}$\\
13 & Second only & $7.5694\times10^{-3}$ & $7.5694\times10^{-3}$ & 0.000000 & $1.53\times10^{-10}$\\
20 & Analytical & $1.8548\times10^{-4}$ & $1.8548\times10^{-4}$ & 0.000001 & $7.90\times10^{-11}$\\
20 & Base & $4.7974\times10^{-4}$ & $4.7974\times10^{-4}$ & 0.000000 & $9.46\times10^{-11}$\\
20 & Calibrated & $8.5422\times10^{-11}$ & $8.5422\times10^{-11}$ & 0.000226 & $1.06\times10^{-10}$\\
20 & First only & $1.0542\times10^{-5}$ & $1.0542\times10^{-5}$ & 0.000002 & $9.69\times10^{-11}$\\
20 & Second only & $4.5685\times10^{-6}$ & $4.5685\times10^{-6}$ & 0.000008 & $9.72\times10^{-11}$\\
\bottomrule
\end{tabular}
\end{table}

\Needspace{4\baselineskip}
Raised-cosine basis checks cover short pulses, small residuals, representative
longer durations, and neighboring leakages differing by more than a factor of three.
Starting at $N=10$, successive level truncations are compared at
$n_{\max}=18$, followed by a cutoff check at $n_{\max}=22$.
The reference uses at least $N=11$ levels, increasing to at least
$N=13$ for the ideal $12.5$, $13$, and $20$-ns commands.
Further checks extend to $N=14$ as needed.
Convergence tolerances are $10^{-3}$ relative or $10^{-12}$ absolute
for $L_{11},P_{20},P_{02},\overline L$,
$10^{-4}$ rad for $\phi_c$, and $10^{-5}$ rad for $\theta$.
For fSim gate infidelity, they are $10^{-3}$ relative or $10^{-10}$ absolute;
the absolute tolerance for strict-iSWAP infidelity is $10^{-6}$.
Norm errors remain below $10^{-10}$.
All $43$ selected ideal raised-cosine commands reach these tolerances:
$18$ use $N=11$, $16$ use $N=12$, and $9$ use $N=13$.
The nine-level results differ by at most $0.197\%$ in $L_{11}$
and $8.28\times10^{-6}$ in strict-iSWAP infidelity.
At nine levels, $13$ commands meet every comparison tolerance;
the remaining $30$ exceed at least one tolerance.
All $230$ Duffing and $138$ charge commands satisfy the independent
integration and full-exchange criteria.

\subsection{Control resources and reproducibility}

\begin{figure}[!htbp]
\centering
\includegraphics[width=0.84\linewidth]{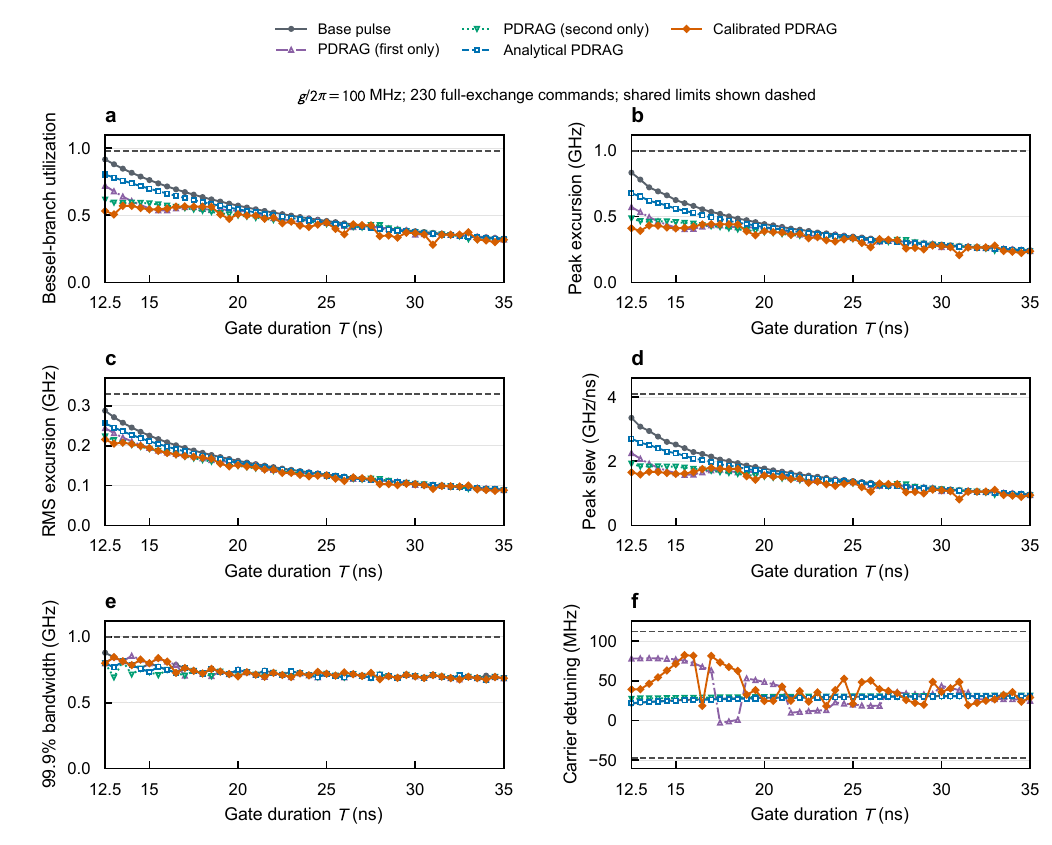}
\caption{Control resources for all $230$ $\sin^4$ Duffing commands. (a) Bessel-branch utilization $u$, (b),(c) peak and RMS frequency excursion, (d) slew, (e) occupied bandwidth $B_{99.9}$, and (f) carrier offset from the $600$-MHz bare detuning. Resources follow Sec.~\ref{s4:sec:resources}; dashed lines mark the common limits.}
\label{s4:fig:resources}
\end{figure}

Independent reconstruction evaluates the real $\sin^4$ command,
slew, and flux map on $24\,001$- and $48\,001$-sample grids.
All $230$ Duffing and $138$ charge commands satisfy the resource
bounds on the denser grid.
Table~\ref{s4:tab:resources} gives the maxima;
the charge-model flux spans $0.0737$--$0.349$ in units of $\Phi_0$.
Figure~\ref{s4:fig:resources} shows the duration dependence.

\begin{table}[!htbp]
\centering
\footnotesize
\caption{Maximum full-command resources for the $230$ $\sin^4$ Duffing and $138$ charge commands on independent $48\,001$-sample grids.}\label{s4:tab:resources}
\begin{tabular}{llrrrl}
\toprule
Model & Resource & Maximum & Limit & $T$ (ns) & Family\\
\midrule
Duffing & $u$ & 0.9162849 & 0.98 & 12.5 & Base\\
Duffing & Peak (GHz) & 0.8345104 & 1 & 12.5 & Base\\
Duffing & RMS (GHz) & 0.2879296 & 0.33 & 12.5 & Base\\
Duffing & Slew (GHz/ns) & 3.3540917 & 4.1 & 12.5 & Base\\
Duffing & $B_{99.9}$ (GHz) & 0.8799817 & 1 & 12.5 & Base\\
Charge & $u$ & 0.9797204 & 0.98 & 12.5 & Base\\
Charge & Peak (GHz) & 0.9797377 & 1 & 12.5 & Base\\
Charge & RMS (GHz) & 0.3238376 & 0.33 & 12.5 & Base\\
Charge & Slew (GHz/ns) & 4.0062199 & 4.1 & 12.5 & Base\\
Charge & $B_{99.9}$ (GHz) & 0.9599800 & 1 & 12.5 & Base\\
\bottomrule
\end{tabular}
\end{table}

The numerical records specify each command and its device,
envelope, solver settings, basis truncation, and source version.
They retain computational maps, evolved states, gate metrics, and convergence checks.
The $\sin^4$ records also contain the $2227$-point landscape,
$63$ asymmetry commands, additional duration points, and figure scripts.
Calculations used Python 3.10.1, NumPy 1.25.0, SciPy 1.10.1,
and \mbox{pandas 1.5.3}.

\FloatBarrier

\section{Spectral-root control of a Q--C--Q exchange gate}
\label{supp:qcq}

\subsection{Three-mode Hamiltonian and frequency-to-flux map}
\label{supp:qcq:model}

A frequency-modulated coupler mediates exchange between fixed-frequency qubits \cite{McKay2016,Roth2017}.
In $Q_1$--$C$--$Q_2$ order, the closed-system Hamiltonian is
\begin{equation}
 \frac{H_{\rm QCQ}(t)}{\hbar}
 =\sum_{j=1,c,2}\left[\omega_j^{(0)}n_j+
       \frac{\alpha_j}{2}n_j(n_j-1)\right]
 +\delta\omega_c(t)n_c
 +\sum_{jk\in\{1c,2c,12\}}g_{jk}(a_j+a_j^\dagger)(a_k+a_k^\dagger).
 \label{supp:qcq:hamiltonian}
\end{equation}
Here $a_j$ and $n_j=a_j^\dagger a_j$ describe mode $j$, and $c$ labels the coupler.
The idle frequencies satisfy $(\omega_1^{(0)},\omega_c^{(0)},\omega_2^{(0)})/(2\pi)=(5.1,4.8,4.5)$ GHz,
with $(\alpha_1,\alpha_c,\alpha_2)/(2\pi)=(-220,-250,-235)$ MHz.
The couplings satisfy $(g_{1c},g_{2c},g_{12})/(2\pi)=(150,142.5,5)$ MHz.
Propagation retains exchange and counter-rotating terms within the exact even- and odd-excitation-parity blocks.
Each block includes every state allowed by the per-mode cutoff.
Device parameters and idle bias remain fixed throughout the duration scan.

The physical command is specified in frequency units,
\begin{equation}
 \delta f_c(t)=\frac{\delta\omega_c(t)}{2\pi}
   =\operatorname{Re}\{E(t)e^{-\ii\Omega t}\},
 \qquad \nu_{\rm car}=\frac{\Omega}{2\pi},
 \label{supp:qcq:command}
\end{equation}
Here $E$ is the complex frequency-modulation envelope in GHz, $\nu_{\rm car}$ is the carrier frequency,
and time is measured in ns.
Thus $2\pi E$ is the angular-frequency envelope used in the general construction of the main text.
A symmetric-SQUID transmon approximation supplies the nonlinear map
\begin{equation}
 f_c(\varphi)=\sqrt{8\mathcal E_C\mathcal E_{J\Sigma}
                    |\cos(\pi\varphi)|}-\mathcal E_C,
 \qquad
 \varphi=\frac{\Phi}{\Phi_0},\quad
 \mathcal E_C=\frac{E_C}{h}=0.25\ {\rm GHz},\quad
 \mathcal E_{J\Sigma}=\frac{(7.5+0.25)^2}{8(0.25)}\ {\rm GHz}.
 \label{supp:qcq:fluxmap}
\end{equation}
The reduced flux is $\varphi$, and $\mathcal E_C$ and $\mathcal E_{J\Sigma}$ are charging and total Josephson energies divided by $h$.
On $0\leq\varphi<1/2$, the map decreases monotonically and has idle bias
$\varphi_0=0.3604149565$.
Its inverse gives the command $\varphi(t)=f_c^{-1}[4.8\ {\rm GHz}+\delta f_c(t)]$.
Each candidate must remain within the accessible frequency and flux ranges and satisfy the flux-slew bound.
Applying the forward map to the reconstructed flux checks frequency--flux consistency.
The Q--C--Q propagation uses Eq.~\eqref{supp:qcq:hamiltonian} with fixed
anharmonicities and coupling matrix elements.
Section~\ref{supp:charge-transfer-mechanism} treats the Q--Q microscopic charge Hamiltonian.

\subsection{Seven signed responses and the five-root reduction}
\label{supp:qcq:roots}

Dressed eigenstates of the full idle Hamiltonian are assigned to product occupations by maximizing the total squared overlap.
The control design resolves seven directed leakage responses,
\begin{align}
 &100\to010,\quad001\to010,\notag\\
 &101\to200,\quad101\to002,\quad101\to110,\quad
   101\to011,\quad101\to020.
 \label{supp:qcq:arrows}
\end{align}
State labels retain the $Q_1$--$C$--$Q_2$ ordering.
The $101\to200$ and $101\to002$ responses populate higher qubit levels.
The remaining five responses populate the coupler, including its second excited state in $101\to020$.
All seven arise from off-diagonal matrix elements of the same operator $n_c$ in the dressed basis.

For response $p\to q$, let $\varepsilon_p$ be the eigenvalue of $H_{\rm QCQ}(0)/\hbar$ for dressed state $p$.
The signed gap $\Delta_{qp}=\varepsilon_q-\varepsilon_p$ and carrier $\Omega$ are angular frequencies;
$s_{qp}=\operatorname{sgn}\Delta_{qp}$ identifies the resonant sideband.
That sideband sets the carrier-relative envelope root
\begin{equation}
 r_{qp}=s_{qp}\Delta_{qp}-\Omega=|\Delta_{qp}|-\Omega.
 \label{supp:qcq:signedroot}
\end{equation}
The positive-sign Fourier convention is
$\widetilde E(\delta)=\int_0^T E(t)e^{\ii\delta t}\,\dd t$.
The real-drive sidebands couple to $E$ and $E^*$.
Their opposite gap signs give the common envelope-root expression in Eq.~\eqref{supp:qcq:signedroot}.
The root retains the sign of $|\Delta_{qp}|-\Omega$, locating the response above or below the carrier.
Modulation matrix elements are evaluated from $n_c$ in the same dressed basis.

The seven-root filter assigns a distinct root to each response.
The five-root filter reduces the parameter count by replacing each pair
$\{100\to010,101\to011\}$ and
$\{001\to010,101\to110\}$ by its arithmetic mean root.
The other three roots remain separate, while propagation retains the
original dressed gaps and all states in Eq.~\eqref{supp:qcq:hamiltonian}.

Using $b_T(t)=\sin^{10}(\pi t/T)$ on $[0,T]$, define
\begin{equation}
 E_m(t)=\frac{A}{\mathcal N_m}
 \prod_{j=1}^{m}\left(1-\frac{\ii}{r'_j}\frac{\dd}{\dd t}\right)b_T(t),
 \qquad r'_j=r_j e^{\eta_j},\qquad
 \widetilde E_m(\delta)=\frac{A}{\mathcal N_m}
 \prod_{j=1}^{m}\left(1-\frac{\delta}{r'_j}\right)\widetilde b_T(\delta).
 \label{supp:qcq:filter}
\end{equation}
Here $m$ is the number of spectral roots, $A$ is an amplitude scale in GHz,
and $\mathcal N_m$ is the dimensionless normalization defined below.
The dimensionless offsets $\eta_j$ move each root multiplicatively,
preserving its sign and the real coefficients of the spectral polynomial.
Base uses $m=0$.
Analytical-5 and Analytical-7 fix $\eta_j=0$;
Calibrated-5 and Calibrated-7 allow $\eta_j\in[-0.35,0.35]$.
The amplitude and carrier provide two calibration parameters for Base and the analytical families.
Independent root offsets increase this count to seven or nine for the calibrated families.
Root design and optimization use four levels per mode.
Larger-cutoff checks keep the selected pulse, carrier and four-level design roots fixed.

The finite Fourier representation evaluates all derivatives analytically,
\begin{equation}
 b_T(t)=\sum_{k=-5}^{5}c_k e^{\ii\omega_k t},\qquad
 c_k=\frac{(-1)^k}{2^{10}}\binom{10}{5-k},\qquad
 \omega_k=\frac{2\pi k}{T}.
 \label{supp:qcq:fourier}
\end{equation}
Each filter factor multiplies $c_k$ by $1+\omega_k/r'_j$,
giving the unnormalized filtered coefficients $d_k=c_k\prod_j(1+\omega_k/r'_j)$.
The base envelope and its derivatives through ninth order vanish at both endpoints.
Consequently, the seven-root envelope and its first two derivatives join continuously to zero.
The five-root envelope has vanishing derivatives through fourth order.
The normalization is
\begin{equation}
 \mathcal N_m=\max_{t\in\mathcal G}|\sum_k d_k e^{\ii\omega_k t}|
       +\frac{(\Delta t)^2}{8}\sum_k |d_k|\omega_k^2,
 \label{supp:qcq:normalization}
\end{equation}
Here $\mathcal G$ contains $2049$ uniformly spaced samples on $[0,T]$, with spacing $\Delta t=T/2048$.
The second term bounds the linear-interpolation remainder, ensuring $\max_t|E_m(t)|\leq A$ between sampled times.
The search uses $A\in[0.001,2]$ GHz and a carrier interval
$\pm0.30$ GHz about the idle dressed qubit splitting, with a positive-frequency lower bound of $0.05$ GHz.
Candidates with $\min_j|r_j|<10^{-7}$ rad/ns are excluded from the polynomial parametrization.

\subsection{Complete-command spectra and resource limits}
\label{supp:qcq:resources}

Equation~\eqref{supp:qcq:fourier} gives the finite-window spectrum analytically using
$\int_0^T e^{\ii xt}\dd t=T e^{\ii xT/2}\operatorname{sinc}(xT/2)$,
where $\operatorname{sinc}(y)=\sin(y)/y$ and $\operatorname{sinc}(0)=1$.
Figure~\ref{fig:supp:qcq-spectrum-dynamics}(a) shows the positive-frequency power spectrum of the complete
real command in Eq.~\eqref{supp:qcq:command}, summing conjugate sidebands before taking the squared modulus.
Each curve is normalized to its positive-frequency peak, located over
$0$--$3000$ MHz and refined on a $0.01$-MHz grid; the display spans $0$--$1200$ MHz.
Amplitude and RMS quantify absolute resources, while this normalization compares spectral shapes.
The roots in Eq.~\eqref{supp:qcq:filter} refer to the carrier-relative envelope.
Figure~1(d) uses the Q--Q normalization
$|\widetilde G(\delta)|^2/|\widetilde G(0)|^2$.

All five families obey the same bounds,
\begin{equation}
\begin{gathered}
 \max|E|\leq2\ {\rm GHz},\quad \max|\delta f_c|\leq2\ {\rm GHz},\quad
 (\delta f_c)_{\rm RMS}\leq1.2\ {\rm GHz},\\
 B_{99.9}\leq2\ {\rm GHz},\quad
 0.02\leq\varphi\leq0.47,\quad
 \max|\dot\varphi|\leq1\ {\rm ns}^{-1}.
\end{gathered}
 \label{supp:qcq:resourcebounds}
\end{equation}
RMS is $[T^{-1}\int_0^T\delta f_c^2\dd t]^{1/2}$.
Resource checks sample at intervals no larger than $0.002$ ns with at least $2049$ points.
Bandwidth uses the complete real command with eightfold zero padding,
retaining DC and applying no mean subtraction or taper.
The one-sided Fourier energy doubles the contributions from interior bins;
the smallest frequency cutoff containing $99.9\%$ of the energy defines $B_{99.9}$.
Flux slew follows by differentiating Eq.~\eqref{supp:qcq:fluxmap}.

Across all $255$ selected pulses on the full duration grid, the maximum envelope,
frequency excursion and RMS are $1.33538$, $1.25994$ and $0.396358$ GHz.
The maximum bandwidth is $1.03750$ GHz and the maximum flux slew is
$0.283239\Phi_0/{\rm ns}$.
The flux range is $0.278874$--$0.423128\Phi_0$.
The maximum frequency--flux round-trip discrepancy is $2.67\times10^{-15}$ GHz.
All selected commands satisfy the resource bounds.

\subsection{Four-input leakage, exchange and gate fidelity}
\label{supp:qcq:metrics}

The fixed idle computational basis is
$B=(\ket{000}_{\rm d},\ket{001}_{\rm d},\ket{100}_{\rm d},\ket{101}_{\rm d})$.
The definitions of $C$, $K$, $L_j$, $\overline L$ and $F(K,V)$ in
Sec.~\ref{sec:supp:dmh:metrics} apply with this ordering.
We additionally report
\begin{equation}
 L_{N=1}=\frac{L_{001}+L_{100}}{2},\qquad
 L_{101}=L_3,\qquad
 \overline L=\frac{L_{000}+2L_{N=1}+L_{101}}4.
 \label{supp:qcq:inputgroups}
\end{equation}
Here $L_{N=1}$ averages leakage from the two single-excitation inputs.
Its index labels the input manifold; propagation conserves excitation parity rather than excitation number.
Leakage includes every output outside $B$ for each input, giving
$4(6^3-4)=848$ input--output probabilities at six levels per mode.
Their sum divided by four yields $\overline L$.
The seven responses in Eq.~\eqref{supp:qcq:arrows} define the resolved design channels;
the remaining-channel contribution sums all other input--output leakage probabilities with the same factor of $1/4$.
The time-dependent curves use the fixed idle dressed projector to resolve population excursions and coherent return.

Numerical exchange acceptance requires
\begin{equation}
 p_{\rm swap}=\frac{|K_{12}|^2+|K_{21}|^2}{2}\geq0.99,\quad
 \max(|K_{11}|^2,|K_{22}|^2)\leq\sin^2(0.005),\quad
 |\theta-\pi/2|\leq0.005,
 \label{supp:qcq:exchange}
\end{equation}
together with the resource bounds and consistent five- and six-level checks.
A separate diagnostic subset uses $p_{\rm swap}\geq0.999$.
All exchange conditions use the unnormalized computational map and retain
both directional return probabilities.

The best-fit fSim fidelity follows the nuclear-norm expression
in Eq.~\eqref{eq:supp:dmh:fsimoverlap}.
For strict iSWAP, Powell's method optimizes the local $Z$ phases
from four fixed starts.
An independent phase-reduced reconstruction checks this result.
The conditional-phase contribution to strict-iSWAP infidelity follows
Eq.~\eqref{eq:supp:phase:infidelity}.

\subsection{Pulse calibration and verification at fixed parameters}
\label{supp:qcq:search}

The grid $T=10,10.5,\ldots,35$ ns contains $255$ family--duration cases.
Five independent initializations and continuation from both shorter and longer durations provide seven searches per case, totaling $1785$ local attempts.
Four initializations use the common amplitude--detuning pairs
$(0.6,-0.10)$, $(1.2,0)$, $(1.8,0.10)$ and $(1.0,-0.28)$ GHz,
where detuning is relative to the dressed exchange carrier.
The fifth initialization uses the nearest preliminary pulse of the same family at $10$, $15$ or $19$ ns.
Ties are resolved in favor of the shorter duration.
Calibrated families initialize from the corresponding same-duration analytical pulse with zero root offsets.
Evaluations of the analytical initializations are counted separately from subsequent calibration.
Each continuation transfers its independently optimized parameters to the adjacent duration and reconstructs the pulse there.

Each attempt starts with trust-region least squares to minimize logical return amplitudes and one-excitation leakage, allowing at most $40$ evaluations.
These evaluations propagate the computational inputs and parameter sensitivities together.
Sequential least-squares programming (SLSQP) then minimizes the four-input
$\overline L$ with at most $80$ optimization evaluations for every family.
All families share the exchange and resource constraints, parameter bounds,
and small numerical margins inside the exchange-acceptance boundary.
Six-level verification evaluates the initial pulse and the output of each optimization stage from all seven attempts, yielding up to $21$ distinct pulses.
We select the feasible command with the smallest $\overline L$.
If none passes exchange acceptance, the diagnostic command minimizes normalized
constraint violation among resource-compliant candidates.

A convergence check flags improvement exceeding $1\%$ over the last $20$ refinement evaluations.
It also flags feasible minima from opposite continuation directions that differ by more than $10\%$.
At the $41$ flagged durations, each family received one additional
$80$-evaluation confirmation refinement.
At $12.5$ and $16$ ns, both directional anchors received this budget in every family.
Identical pulses were evaluated once, reducing $225$ confirmation initializations to $217$ distinct pulses.
Confirmation revisited $205$ family--duration cases, retaining every original candidate for comparison.
The other $50$ cases retained their original selections.

The local-stability criterion permits at most $1\%$ improvement over the last $20$ evaluations relative to an earlier feasible solution.
For shorter histories, it requires successful SLSQP termination and at least one feasible evaluation.
After confirmation, $24$ family-level convergence checks remain unresolved at $16$ durations,
including the opposite-direction Calibrated-7 minima at $16$ ns.
Calibrated-7 searches at $10$, $10.5$, $11$ and $11.5$ ns are incomplete
because finite-difference steps encountered a singular root or inherited that failure during descending continuation.

\subsection{Level, integration and channel checks}
\label{supp:qcq:verification}

Table~\ref{supp:qcq:solver-table} lists the five propagation settings used to verify each pulse at fixed coefficients and carrier.
Reported endpoint metrics use the tight six-level setting; transient trajectories use the tighter six-level setting.
Low-lying labels are matched by embedding the smaller basis in the larger one.
The minimum relevant overlap is $0.999997186$ for four versus six levels
and $0.9999999939$ for five versus six levels.
Matching includes the computational inputs and the targets of the seven responses;
labels for higher excited states retain cutoff-specific assignments.
Cross-cutoff comparisons use total leakage and the remaining-channel sum.

\begin{table}[!htbp]
\centering\footnotesize
\caption{Q--C--Q integration settings for verification at fixed pulse parameters.
All propagations use the eighth-order Dormand--Prince integrator DOP853.
Relative and absolute tolerances are denoted by rtol and atol; $h_{\max}$ is the maximum time step in ns.}
\label{supp:qcq:solver-table}
\begin{tabular}{lrrr}
\toprule
Levels and setting & rtol & atol & $h_{\max}$\\
\midrule
4, standard & $2\times10^{-11}$ & $2\times10^{-13}$ & 0.025\\
4, tight & $2\times10^{-12}$ & $2\times10^{-14}$ & 0.0125\\
5, tight & $2\times10^{-12}$ & $2\times10^{-14}$ & 0.0125\\
6, tight & $2\times10^{-12}$ & $2\times10^{-14}$ & 0.0125\\
6, tighter & $10^{-13}$ & $10^{-15}$ & 0.00625\\
\bottomrule
\end{tabular}
\end{table}

Together, the five settings give $1275$ independent propagations of the $255$ final pulses.
Five- versus six-level changes are at most $1.068\times10^{-7}$ in
$\overline L$, $2.738\times10^{-6}$ rad in conditional phase and
$2.361\times10^{-7}$ in strict-iSWAP fidelity.
Tight versus tighter six-level changes are at most
$1.090\times10^{-13}$, $2.066\times10^{-12}$ rad and
$2.779\times10^{-13}$, respectively.
Maximum column-norm deviations are $3.343\times10^{-12}$;
the maximum Frobenius norm $\|C^\dagger C-I_4\|_{\rm F}$ is $3.394\times10^{-12}$.
Summing the resolved leakage probabilities reproduces the projected mean leakage within $1.67\times10^{-16}$.
Independent reconstruction reproduces the fSim gate fidelity within
$2.23\times10^{-16}$ and the strict local-$Z$ evaluation within $6.35\times10^{-11}$.
All selected commands pass these cutoff, integration and channel checks.

For a leakage metric $L$, its empirical numerical sensitivity is
\begin{equation}
 u_L=\max\{|L_{6,\rm tight}-L_{5,\rm tight}|,
              |L_{6,\rm tighter}-L_{6,\rm tight}|,10^{-12}\}.
 \label{supp:qcq:sensitivity}
\end{equation}
The subscripts specify the level cutoff and integration setting.
We resolve a leakage value when $L>3u_L$, using the larger cutoff or integration discrepancy with an absolute floor of $10^{-12}$.

\subsection{Complete grid and interval statistics}
\label{supp:qcq:grid}

All five families satisfy the exchange conditions at $35$ of the $51$ durations,
giving the common $18$--$35$-ns interval in Fig.~2.
Tables~\ref{supp:qcq:grid-short} and \ref{supp:qcq:grid-long} retain the full $10$--$35$-ns scan.
All $175$ pulses in the common interval pass the propagation and exchange checks.
Of these, $172$ satisfy the initial finite-budget stability check (Table~\ref{supp:qcq:coverage});
Sec.~\ref{supp:qcq:boundary} gives additional stability checks for the other three at their unchanged pulse parameters.

\begin{table}[!htbp]
\centering\footnotesize
\caption{Coverage of the complete $51$-duration Q--C--Q grid.
The last column includes the three additional stability checks described in Sec.~\ref{supp:qcq:boundary}; the preceding column reports the initial finite-budget check.}
\label{supp:qcq:coverage}
\begin{tabular}{lrrrr}
\toprule
Family & Records & Numerical exchange & Original bounded & With separate support\\
\midrule
Base & 51 & 35 & 33 & 35\\
Analytical-5 & 51 & 44 & 44 & 44\\
Calibrated-5 & 51 & 51 & 51 & 51\\
Analytical-7 & 51 & 46 & 46 & 46\\
Calibrated-7 & 51 & 47 & 44 & 45\\
\midrule
Total & 255 & 223 & 218 & 221\\
\bottomrule
\end{tabular}
\end{table}

At each common duration, the leakage-reduction factor is $R_k=\overline L_{\rm Base}/\overline L_k$.
Table~\ref{supp:qcq:gains} weights all $35$ common durations equally,
including the three commands with separate stability support.
All four filtered families improve on Base at every common duration.
Calibrated-5 and Calibrated-7 yield gains spanning $3.3884$--$10.4154$ and $5.3367$--$12.8687$.
Their median gains are $4.7121$ and $6.5919$, with improvement at $100\%$ of the common durations.

\begin{table}[!htbp]
\centering\footnotesize
\caption{Equal-duration Q--C--Q leakage gains over Base across all $35$ points in $18$--$35$ ns.
Arithmetic and geometric means average pointwise ratios; extrema and median use the same denominator.}
\label{supp:qcq:gains}
\begin{tabular}{lrrrrrr}
\toprule
Family & Minimum & Median & Arithmetic mean & Geometric mean & Maximum & Improved\\
\midrule
Analytical-5 & 1.9273 & 2.2564 & 2.2955 & 2.2795 & 3.1300 & 35/35\\
Calibrated-5 & 3.3884 & 4.7121 & 5.4531 & 5.1340 & 10.4154 & 35/35\\
Analytical-7 & 1.3481 & 1.7659 & 1.8430 & 1.8011 & 2.6037 & 35/35\\
Calibrated-7 & 5.3367 & 6.5919 & 6.9476 & 6.7870 & 12.8687 & 35/35\\
\bottomrule
\end{tabular}
\end{table}

Calibrated-7 has lower $\overline L$ than Calibrated-5 at $33$ common durations;
Calibrated-5 has lower leakage at $19$ and $19.5$ ns.
Calibrated-5 has higher strict-iSWAP fidelity at all $35$ common durations.
The $101$ input contributes at least $99.414\%$ of the Base mean leakage,
$99.9976\%$ for Calibrated-5 and $99.9940\%$ for Calibrated-7.

Superscripts identify failed or unconfirmed exchange acceptance (F),
an unresolved initial convergence check (S), incomplete continuation (I), or additional stability support (D).
A D entry retains the pulse and numerical values originally marked S.
Unmarked entries satisfy exchange acceptance and the initial finite-budget stability criterion.

\begin{table}[!htbp]
\centering\footnotesize
\caption{Complete-grid mean leakage, $10^3\overline L$, for $T=10$--$22.5$ ns. A5/C5 and A7/C7 denote the analytical/calibrated five- and seven-root families. Superscripts follow the definitions in the text; each listed duration contributes once to statistics for its stated interval.}
\label{supp:qcq:grid-short}
\begin{tabular}{rrrrrr}
\toprule
$T$ (ns) & Base & A5 & C5 & A7 & C7\\
\midrule
10 & $376.03^{\mathrm{F}}$ & $237.51^{\mathrm{F}}$ & $62.6106$ & $232.625^{\mathrm{F}}$ & $48.4341^{\mathrm{FI}}$\\
10.5 & $309.056^{\mathrm{FS}}$ & $137.233^{\mathrm{FS}}$ & $64.9917$ & $77.9764^{\mathrm{FS}}$ & $37.9071^{\mathrm{FI}}$\\
11 & $265.854^{\mathrm{FS}}$ & $122.976^{\mathrm{FS}}$ & $56.1839$ & $124.3^{\mathrm{FS}}$ & $40.4308^{\mathrm{FI}}$\\
11.5 & $430.834^{\mathrm{F}}$ & $105.752^{\mathrm{F}}$ & $52.4017$ & $148.485^{\mathrm{F}}$ & $35.1708^{\mathrm{FI}}$\\
12 & $448.585^{\mathrm{F}}$ & $91.3738^{\mathrm{F}}$ & $49.7735$ & $184.445^{\mathrm{F}}$ & $30.4707$\\
12.5 & $465.097^{\mathrm{FS}}$ & $83.1812^{\mathrm{FS}}$ & $43.4508$ & $165.198$ & $25.6998^{\mathrm{S}}$\\
13 & $416.904^{\mathrm{FS}}$ & $74.0627^{\mathrm{FS}}$ & $36.6824$ & $167.84$ & $21.3233$\\
13.5 & $195.07^{\mathrm{FS}}$ & $65.6062$ & $32.5103$ & $143.407$ & $16.721$\\
14 & $188.633^{\mathrm{FS}}$ & $60.3141$ & $29.1527$ & $129.817$ & $16.1269$\\
14.5 & $202.587^{\mathrm{FS}}$ & $55.9092$ & $25.8786$ & $122.477$ & $13.9285$\\
15 & $216.118^{\mathrm{FS}}$ & $51.887$ & $22.8314$ & $113.139$ & $11.569$\\
15.5 & $123.011^{\mathrm{FS}}$ & $48.0106$ & $20.0034$ & $104.907$ & $9.73366$\\
16 & $142.279^{\mathrm{FS}}$ & $44.4942$ & $17.4112$ & $96.7753$ & $8.18711^{\mathrm{S}}$\\
16.5 & $150.77^{\mathrm{FS}}$ & $41.2682$ & $15.0291$ & $88.9429$ & $7.15386$\\
17 & $151.105^{\mathrm{FS}}$ & $38.256$ & $12.85$ & $81.538$ & $7.01221$\\
17.5 & $149.031^{\mathrm{FS}}$ & $35.4142$ & $11.081$ & $74.4949$ & $7.39482$\\
18 & $102.5$ & $32.7472$ & $9.84117$ & $67.8162$ & $7.96509$\\
18.5 & $89.1366^{\mathrm{SD}}$ & $30.2623$ & $8.94311$ & $61.5563$ & $8.53017$\\
19 & $77.9147$ & $27.9525$ & $8.28708$ & $55.7423$ & $8.82864$\\
19.5 & $68.4103$ & $25.8029$ & $7.79465$ & $50.3602$ & $7.99225$\\
20 & $61.1903$ & $23.8039$ & $7.40489$ & $45.3905$ & $7.29352^{\mathrm{SD}}$\\
20.5 & $55.5541$ & $21.9484$ & $7.07629$ & $40.8244$ & $6.69581$\\
21 & $50.6339$ & $20.2299$ & $6.78158$ & $36.6534$ & $6.17359$\\
21.5 & $46.0841$ & $18.6404$ & $6.50441$ & $32.8527$ & $5.71083$\\
22 & $41.9343$ & $17.1709$ & $6.23543$ & $29.4029$ & $5.29573$\\
22.5 & $38.2247$ & $15.8133$ & $5.96937$ & $26.282$ & $4.91969$\\
\bottomrule
\end{tabular}
\end{table}

\begin{table}[!htbp]
\centering\footnotesize
\caption{Complete-grid mean leakage, $10^3\overline L$, for $T=23$--$35$ ns. A5/C5 and A7/C7 denote the analytical/calibrated five- and seven-root families. Superscripts follow the definitions in the text; each listed duration contributes once to statistics for its stated interval.}
\label{supp:qcq:grid-long}
\begin{tabular}{rrrrrr}
\toprule
$T$ (ns) & Base & A5 & C5 & A7 & C7\\
\midrule
23 & $34.8973$ & $14.5601$ & $5.70385$ & $23.4662$ & $4.57648$\\
23.5 & $31.8682$ & $13.4042$ & $5.43845$ & $20.9319$ & $4.26106$\\
24 & $29.0897$ & $12.3387$ & $5.17378$ & $18.6559$ & $3.96976$\\
24.5 & $26.5444$ & $11.3569$ & $4.91073$ & $16.616$ & $3.69965$\\
25 & $24.2198$ & $10.4526$ & $4.65081$ & $14.7912$ & $3.44818$\\
25.5 & $22.0973$ & $9.6199$ & $4.39514$ & $13.1611$ & $3.21362$\\
26 & $20.1576$ & $8.85342$ & $4.14498$ & $11.7068$ & $2.9943$\\
26.5 & $18.3845$ & $8.14776$ & $3.90156$ & $10.4109$ & $2.78896$\\
27 & $16.7641$ & $7.49843$ & $3.66554$ & $9.25697$ & $2.59655$\\
27.5 & $15.2844$ & $6.90083$ & $3.43787$ & $8.23066$ & $2.41614$\\
28 & $13.9334$ & $6.35072$ & $3.21891$ & $7.31829$ & $2.24696$\\
28.5 & $12.7007^{\mathrm{SD}}$ & $5.8444$ & $3.0092$ & $6.50754$ & $2.0883$\\
29 & $11.5756$ & $5.37827$ & $2.80886$ & $5.78737$ & $1.93954$\\
29.5 & $10.5493$ & $4.94903$ & $2.61807$ & $5.14785$ & $1.80013$\\
30 & $9.61311$ & $4.55366$ & $2.43688$ & $4.57998$ & $1.66955$\\
30.5 & $8.75923$ & $4.18932$ & $2.26517$ & $4.07571$ & $1.54723$\\
31 & $7.98033$ & $3.85357$ & $2.10282$ & $3.62799$ & $1.43111$\\
31.5 & $7.26987$ & $3.54402$ & $1.94964$ & $3.23038$ & $1.32059$\\
32 & $6.6217$ & $3.25851$ & $1.80537$ & $2.87714$ & $1.21569$\\
32.5 & $6.03023$ & $2.99507$ & $1.66978$ & $2.56332$ & $1.11648$\\
33 & $5.49048$ & $2.75189$ & $1.54248$ & $2.28439$ & $1.02288$\\
33.5 & $4.99775$ & $2.52741$ & $1.42324$ & $2.03638$ & $0.934814$\\
34 & $4.54784$ & $2.3201$ & $1.31165$ & $1.81575$ & $0.852186$\\
34.5 & $4.13692$ & $2.12863$ & $1.20739$ & $1.61945$ & $0.774861$\\
35 & $3.76153$ & $1.95175$ & $1.1101$ & $1.44468$ & $0.702655$\\
\bottomrule
\end{tabular}
\end{table}

\subsection{Short-duration boundary and separate stability diagnostics}
\label{supp:qcq:boundary}

At $10$ ns, Calibrated-5 satisfies the exchange conditions with
$\overline L=0.062611$ and $F_{\rm iSWAP}=0.92354$.
The partial Calibrated-7 search yields $\overline L=0.048434$
and $F_{\rm iSWAP}=0.91635$.
Base fails the exchange conditions at both $10$ and $15$ ns.
At $17.5$ ns, Base reaches a mean swap probability of $0.996664$,
but its maximum return probability is $6.7252\times10^{-5}$ and its angle error is $0.007285$ rad.
Both exceed the respective exchange tolerances, setting the lower boundary of the common comparison interval.

An additional $17.75$-ns calculation probes the $17.5$--$18$-ns transition.
Five independent initializations and continuations from both adjacent endpoints give $35$ attempts across the five families.
All five selected midpoint pulses satisfy the exchange and cutoff checks.
Analytical-7 improves by $1.36848\%$ over its last $20$ evaluations, exceeding the local-stability threshold;
the other four satisfy this criterion.
The five midpoint mean leakages are $0.113928$, $0.0340580$, $0.0104092$,
$0.0711070$ and $0.00767106$ in Base, A5, C5, A7 and C7 order.
This additional point is excluded from statistics on the uniform $0.5$-ns grid.

Initial confirmation stopped with SLSQP line-search status 8 for three pulses:
Base at $18.5$ and $28.5$ ns, and Calibrated-7 at $20$ ns.
Status 8 indicates a positive directional derivative in the line search.
Additional tests started from these pulses, allowing up to $20$ evaluations to restore the exchange constraints,
followed by the original $80$-evaluation leakage refinement.
Each test used three restoration evaluations.
The $18.5$-ns Base and $20$-ns Calibrated-7 refinements converged after five and six evaluations.
The $28.5$-ns Base refinement used all $80$ evaluations, with zero improvement over its final $20$ evaluations.
All three therefore satisfy the additional local-stability check.
The previously selected pulses remain the feasible candidates with the lowest leakage.
Their six-level metrics and plotted values are unchanged.
The table retains their initial S flags and records the additional tests with D.

The $18$-ns point is the earliest common exchange solution on the $0.5$-ns grid that also satisfies the initial finite-budget stability criterion.

\subsection{Representative control and fidelity across the interval}
\label{supp:qcq:representative}

\begin{figure}[!htbp]
\centering
\includegraphics[width=0.94\linewidth]{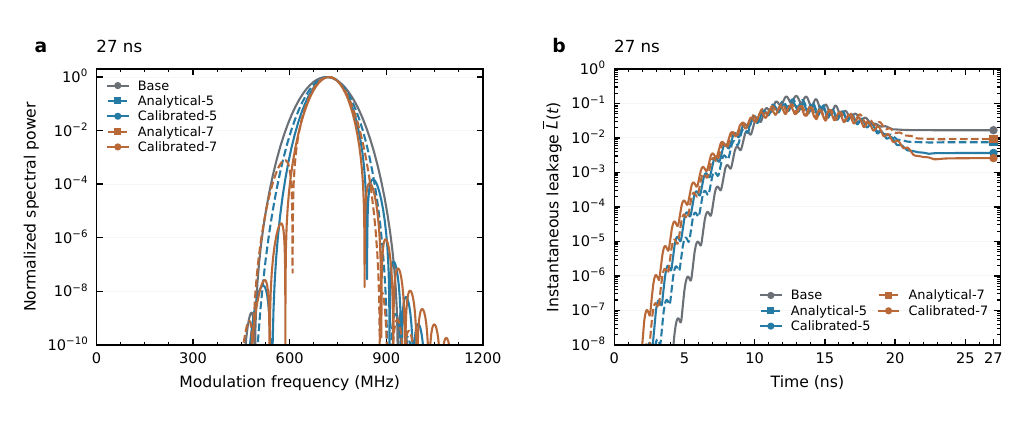}
\caption{Spectrum and leakage dynamics of the $27$-ns Q--C--Q pulses.
(a) Positive-frequency power of the real coupler-frequency modulation, normalized to the peak of each curve.
(b) Four-input mean leakage $\overline L(t)$ in the fixed idle dressed basis; symbols mark endpoint values.
The five pulse families use the device and resource bounds of Secs.~\ref{supp:qcq:model} and \ref{supp:qcq:resources}.}
\label{fig:supp:qcq-spectrum-dynamics}
\end{figure}

At $27$ ns, spectral shaping reduces both transient excursions and residual leakage (Fig.~\ref{fig:supp:qcq-spectrum-dynamics}).
Calibrated-5 and Calibrated-7 reduce the Base endpoint mean leakage
from $0.016764$ to $0.003666$ and $0.002597$, corresponding to suppression factors of $4.57345$ and $6.45631$.
Their fSim fidelities reach 99.63\% and 99.74\%, compared with 98.31\% for Base (Table~\ref{supp:qcq:27ns}).

The transient mean leakage peaks at $0.16516$, $0.13196$, $0.10270$, $0.09761$ and $0.08414$
for Base, A5, C5, A7 and C7, respectively.
Coherent return reduces the population outside the computational subspace before the pulse ends.

\begin{table}[!htbp]
\centering\footnotesize
\caption{The $27$-ns Q--C--Q commands in Fig.~\ref{fig:supp:qcq-spectrum-dynamics}. All metrics are evaluated for the same pulse at six levels per mode. The dagger marks a raw $L_{N=1}$ estimate below its empirical resolution scale $3u_L=3\times10^{-12}$. Conditional phases are in rad; fidelities include leakage.}
\label{supp:qcq:27ns}
\begin{tabular}{lrrrrrr}
\toprule
Family & $\overline L$ & $L_{N=1}$ & $L_{101}$ & $\phi_c$ & $F_{\rm fSim}$ & $F_{\rm iSWAP}$\\
\midrule
Base & $1.67641\times10^{-2}$ & $2.55193\times10^{-8}$ & $6.70565\times10^{-2}$ & 1.226350 & 0.983061 & 0.911436\\
Analytical-5 & $7.49843\times10^{-3}$ & $1.83613\times10^{-10}$ & $2.99937\times10^{-2}$ & 1.340537 & 0.992467 & 0.906577\\
Calibrated-5 & $3.66554\times10^{-3}$ & $1.59754\times10^{-11}$ & $1.46622\times10^{-2}$ & 1.430341 & 0.996326 & 0.898671\\
Analytical-7 & $9.25697\times10^{-3}$ & $1.03462\times10^{-10}$ & $3.70279\times10^{-2}$ & 1.439657 & 0.990691 & 0.892381\\
Calibrated-7 & $2.59655\times10^{-3}$ & $9.61400\times10^{-14}{}^{\dagger}$ & $1.03862\times10^{-2}$ & 1.493864 & 0.997399 & 0.891179\\
\bottomrule
\end{tabular}
\end{table}

\begin{figure}[!htbp]
\centering
\includegraphics[width=\linewidth]{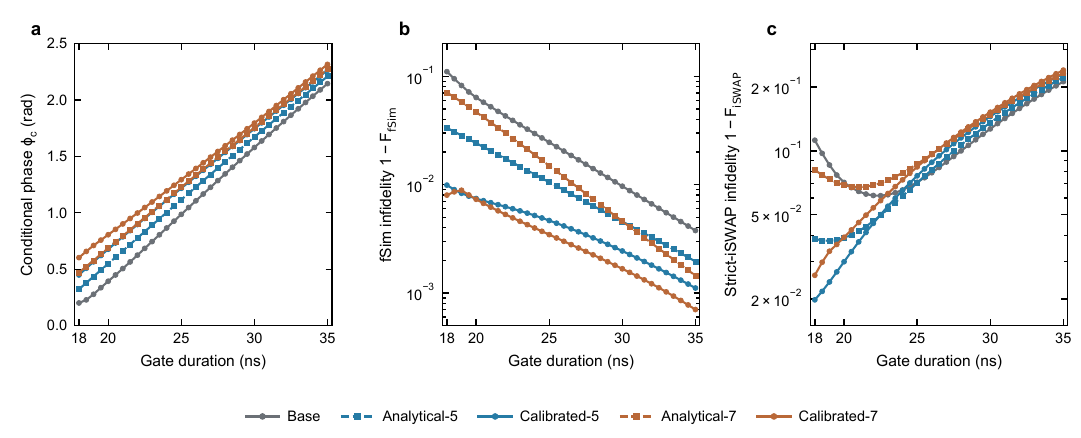}
\caption{Conditional phase and gate infidelity across the Q--C--Q duration scan.
(a) Conditional phase $\phi_c$.
(b) Best-fit fSim infidelity $1-F_{\rm fSim}$.
(c) Strict-iSWAP infidelity $1-F_{\rm iSWAP}$.
The five families contain the same $175$ pulses as Fig.~2, sampled from $18$ to $35$ ns at $0.5$-ns intervals.
Fidelities follow Sec.~\ref{sec:supp:dmh:metrics}.
Symbols mark evaluated pulses; lines connect adjacent durations.
Panels (b) and (c) use logarithmic scales.}
\label{fig:supp:qcq-fidelity-duration}
\end{figure}

Figure~\ref{fig:supp:qcq-fidelity-duration} separates best-fit fSim performance, limited by leakage, from the conditional-phase mismatch to strict iSWAP.
At the shortest common duration of $18$ ns, C5 and C7 achieve mean leakages of $0.00984117$ and $0.00796509$.
These correspond to Base-to-calibrated ratios of $10.4154$ and $12.8687$.
Their fSim fidelities are $0.990100$ and $0.991996$, compared with strict-iSWAP fidelities of $0.980163$ and $0.974107$.

Longer pulses reduce residual leakage but accumulate a larger nonlocal phase.
At $35$ ns, C5 and C7 reach mean leakages of $0.00111010$ and $0.000702655$,
with fSim fidelities of $0.998889$ and $0.999297$.
Their conditional phases are $2.26447$ and $2.31662$ rad, giving strict-iSWAP fidelities of $0.768993$ and $0.759814$.

Across all $35$ common durations, fSim fidelity spans $0.990100$--$0.998889$ for C5
and $0.991124$--$0.999297$ for C7.
The strict-iSWAP ranges are $0.768993$--$0.980163$ and $0.759814$--$0.974107$, respectively.

\FloatBarrier

\section{Finite-pulse Floquet return expansion}
\label{supp:floquet}

The Floquet expansion separates successive leakage excursions and
coherent returns, resolving the interference responsible for the
calibration gain in Fig.~3.
We evaluate Base, analytical, and calibrated $\sin^4$ pulses at $46$
durations from $12.5$ to $35$ ns.
For each of the $138$ commands, $s,\Omega,\beta,\gamma$ remain fixed
as the expansion order increases.

\subsection{Periodic reference for the complete real command}
\label{supp:floquet:reference}

Excitation conservation in the Q--Q Duffing Hamiltonian restricts the
evolution from $\ket{11}_{\rm d}$ to the two-excitation sector.
Let $W$ contain its idle eigenvectors in the order
$(\ket{11}_{\rm d},\ket{20}_{\rm d},\ket{02}_{\rm d})$, expressed in
the bare order $(\ket{20},\ket{11},\ket{02})$.
Overlap assignment and phases follow Sec.~\ref{sec:supp:dmh:metrics}.
Removing the common idle frequency $\omega_{\rm ref}=\varepsilon_{11}$ gives
\begin{equation}
 h(t)=D_{\rm d}+\delta\omega_1(t)N_{\rm d},\quad
 D_{\rm d}=\operatorname{diag}(0,\varepsilon_{20}-\varepsilon_{11},
                         \varepsilon_{02}-\varepsilon_{11}),\quad
 N_{\rm d}=W^\dagger\operatorname{diag}(2,1,0)W.
 \label{supp:floquet:idle}
\end{equation}
Here $h$ has angular-frequency units, $\hbar\varepsilon_{mn}$ are
the idle eigenenergies, and $N_{\rm d}$ is the dressed modulation operator.

The periodic reference freezes the complex command envelope
\begin{equation}
 \mathcal C(t)=\dot z(t)-\ii\Omega z(t)
              =r(t)e^{\ii\chi_C(t)},\qquad
 \delta\omega_1(t)=-r(t)\sin[\Omega t-\chi_C(t)].
 \label{supp:floquet:command}
\end{equation}
Here $r=|\mathcal C|$ is the amplitude of the complete frequency command,
and $\chi_C=\arg\mathcal C$ is its phase.
Freezing $\mathcal C$ retains both the carrier term and the derivative
of $z$; its phase therefore differs from $\chi=\arg z$.
At fixed $r$ and carrier $\Omega$, the periodic propagator obeys
\begin{equation}
 h_r(\tau)=D_{\rm d}-r\sin(\Omega\tau)N_{\rm d},\qquad
 \ii\partial_\tau U_r=h_rU_r,\quad U_r(0)=I,
 \quad \tau_c=2\pi/\Omega.
 \label{supp:floquet:frozen}
\end{equation}
The one-period propagator $U_r(\tau_c)$ has eigenvectors $v_a(r)$
and quasienergies $\varepsilon_a^{\rm F}(r)$ satisfying
$U_r(\tau_c)v_a=e^{-\ii\varepsilon_a^{\rm F}\tau_c}v_a$.
For each command, radial tracking starts at $r=0$ with
$\mathsf V=(v_1,v_2,v_3)=I$ and $\varepsilon_a^{\rm F}=(D_{\rm d})_{aa}$.
Maximizing the total eigenvector overlap matches neighboring radial branches;
integer multiples of $\Omega$ unwrap their quasienergies.
Keeping $(v_a)_a$ real and positive fixes the radial phase convention.
The periodic mode matrix is
\begin{equation}
 \mathsf F(r,\sigma)=U_r(\sigma/\Omega)\mathsf V(r)
    e^{\ii D_{\rm F}(r)\sigma/\Omega},\qquad
 D_{\rm F}=\operatorname{diag}(\varepsilon_a^{\rm F}),
 \quad \mathsf F(r,\sigma+2\pi)=\mathsf F(r,\sigma).
 \label{supp:floquet:modes}
\end{equation}
This Floquet basis separates motion within the carrier period from
envelope variation \cite{Weinberg2017Floquet}.

Radial derivatives are obtained from the variational propagator
$U_{r,r}=\partial_rU_r$, initialized at zero, using
$\ii\partial_\tau U_{r,r}=h_rU_{r,r}
-\sin(\Omega\tau)N_{\rm d}U_r$.
Writing $\mathcal U_c=U_r(\tau_c)$, $\mathcal U_{c,r}=\partial_r\mathcal U_c$, and
$\mu_a=e^{-\ii\varepsilon_a^{\rm F}\tau_c}$ gives
\begin{equation}
 \partial_r\varepsilon_a^{\rm F}
 =\frac{\ii v_a^\dagger\mathcal U_{c,r}v_a}{\tau_c\mu_a},\qquad
 v_b^\dagger\partial_rv_a
 =\frac{v_b^\dagger\mathcal U_{c,r}v_a}{\mu_a-\mu_b}\quad(b\ne a).
 \label{supp:floquet:radial}
\end{equation}
The positive-component phase convention fixes $v_a^\dagger\partial_rv_a$.
Eigenphase separations and neighboring eigenvector overlaps are monitored
to maintain continuous radial branches.

Along the pulse, $\sigma(t)=\Omega t-\chi_C(t)$ and
$\mathsf F(t)=\mathsf F[r(t),\sigma(t)]$.
For $r>0$,
$\dot r=\operatorname{Re}(\mathcal C^*\dot{\mathcal C})/r$ and
$\dot\chi_C=\operatorname{Im}(\mathcal C^*\dot{\mathcal C})/r^2$.
With the shifted idle-basis state written as
$\boldsymbol\psi=\mathsf F\boldsymbol c$, its exact moving-frame
angular-frequency generator is
\begin{align}
 h_{\rm Fl}
 &=\mathsf F^\dagger h\mathsf F-\ii\mathsf F^\dagger\dot{\mathsf F}
   =D_{\rm F}+\mathcal A,\notag\\
 \mathcal A
 &=-\ii\mathsf F^\dagger
     (\mathsf F_r\dot r-\mathsf F_\sigma\dot\chi_C),\qquad
 \Omega\mathsf F_\sigma=-\ii h\mathsf F+\ii\mathsf F D_{\rm F}.
 \label{supp:floquet:connection}
\end{align}
Subscripts $r$ and $\sigma$ denote partial derivatives of $\mathsf F$.
The frozen Floquet equation accounts for motion within the carrier period,
while $\mathcal A$ contains radial and envelope-phase variation.
Both its diagonal and off-diagonal blocks enter the return expansion.

\subsection{Successive departures and returns}
\label{supp:floquet:returns}

Let $\Pi_P=\operatorname{diag}(1,0,0)$ and $\Pi_Q=I-\Pi_P$
select the branch connected to $\ket{11}_{\rm d}$ and its two complements,
defining the expansion's computational and leakage blocks.
The physical computational projector in this sector is
$P_{\rm d}=\ket{11}_{\rm d}{}_{\rm d}\!\bra{11}$;
in the moving basis it is $\mathsf F^\dagger P_{\rm d}\mathsf F$,
which generally differs from $\Pi_P$ during the pulse.
The full Hilbert-space computational projector remains $BB^\dagger$.

Write the moving-frame state as
$\boldsymbol c=(p,\boldsymbol q^{\mathsf T})^{\mathsf T}$,
where $p$ is the amplitude in $\Pi_P$ and $\boldsymbol q$ contains
the two amplitudes in $\Pi_Q$.
Define $h_P=(h_{\rm Fl})_{PP}$, the two-dimensional block
$h_Q=(h_{\rm Fl})_{QQ}$, and the coupling vector
$\boldsymbol v=(h_{\rm Fl})_{QP}$.
A dimensionless expansion parameter $\lambda$ counts transfers between these blocks:
\begin{equation}
 h_{\rm Fl}(\lambda)=
 \begin{pmatrix}h_P&\lambda\boldsymbol v^\dagger\\
 \lambda\boldsymbol v&h_Q\end{pmatrix},\qquad
 p=\sum_{j\geq0}\lambda^{2j}p_{2j},\quad
 \boldsymbol q=\sum_{j\geq0}\lambda^{2j+1}\boldsymbol q_{2j+1}.
 \label{supp:floquet:counting}
\end{equation}
Setting $\lambda=1$ recovers the physical coupling.
Expansion order counts transfers between Floquet blocks; derivative order
specifies the PDRAG envelope. Through fifth order,
\begin{align}
 \ii\dot p_0&=h_Pp_0,&
 \ii\dot{\boldsymbol q}_1&=h_Q\boldsymbol q_1+\boldsymbol v p_0,\notag\\
 \ii\dot p_2&=h_Pp_2+\boldsymbol v^\dagger\boldsymbol q_1,&
 \ii\dot{\boldsymbol q}_3&=h_Q\boldsymbol q_3+\boldsymbol v p_2,\notag\\
 \ii\dot p_4&=h_Pp_4+\boldsymbol v^\dagger\boldsymbol q_3,&
 \ii\dot{\boldsymbol q}_5&=h_Q\boldsymbol q_5+\boldsymbol v p_4.
 \label{supp:floquet:recursion}
\end{align}
Initially $p_0=1$ and all higher-order amplitudes vanish.
The $\sin^4$ command starts at idle, where $\mathsf F=I$.
The first departure $\boldsymbol q_1$ feeds the return $p_2$ and
second departure $\boldsymbol q_3$; $\boldsymbol q_3\to p_4\to\boldsymbol q_5$
adds the second return cycle.

The full $Q$-block propagator obeys
$\ii\dot U_Q=h_QU_Q$, $U_Q(0)=I_2$.
Since $p_0(t)=\exp[-\ii\int_0^t h_P(s)\dd s]$, variation of constants gives
\begin{align}
 \boldsymbol q_{2j+1}(T)
 &=-\ii U_Q(T)\int_0^T U_Q^\dagger(t)
                  \boldsymbol v(t)p_{2j}(t)\dd t,\quad j=0,1,2,\notag\\
 p_{2j}(T)
 &=-\ii p_0(T)\int_0^T
      \frac{\boldsymbol v^\dagger(t)\boldsymbol q_{2j-1}(t)}{p_0(t)}\dd t,
      \quad j=1,2.
 \label{supp:floquet:integrals}
\end{align}
The propagator $U_Q(T)$ retains the relative phase and mixing of the
two leakage channels throughout each excursion.

Let $J_Q$ be the $2\times3$ matrix that selects the
$\ket{20}_{\rm d},\ket{02}_{\rm d}$ coordinates from the fixed idle basis.
The fifth-order endpoint leakage vector and probability are
\begin{equation}
 \boldsymbol\ell^{[5]}=e^{-\ii\omega_{\rm ref}T}J_Q\mathsf F(T)
 \begin{pmatrix}p_0+p_2+p_4\\
 \boldsymbol q_1+\boldsymbol q_3+\boldsymbol q_5\end{pmatrix}_{t=T},
 \qquad L_{11}^{[5]}=\|\boldsymbol\ell^{[5]}\|^2.
 \label{supp:floquet:endpoint}
\end{equation}
Third order retains $p_0,p_2,\boldsymbol q_1,\boldsymbol q_3$;
first order retains $p_0,\boldsymbol q_1$.
The endpoint transformation and restored laboratory phase express all
orders in the same physical basis.
Because $\mathsf F(T)=I$, leakage is the squared norm of the coherent
sum of the retained leakage amplitudes.

\subsection{Accuracy across the duration scan and the fifth-order contribution}
\label{supp:floquet:statistics}

For a fixed command, define the logarithmic leakage error
$e_L^{[n]}=\log_{10}(L_{11}^{[n]}/L_{11})$, where $n=1,3,5$ and
$L_{11}$ is obtained by direct propagation.
The calibration gain is
$S=\log_{10}(L_{11}^{\rm an}/L_{11}^{\rm cal})$;
$S^{[n]}$ uses the corresponding order-$n$ predictions, with error
$e_S^{[n]}=S^{[n]}-S$.
One decade corresponds to a factor of ten in the leakage ratio.
Table~\ref{supp:floquet:tab:statistics} weights all $46$ durations equally.
The fifth-order expansion reduces $|e_S|$ at $44$ durations and keeps every
prediction within $0.3$ decade.
Table~\ref{supp:floquet:tab:eight} lists the eight third-order errors
exceeding one decade.

\begin{table}[!htbp]
\centering\footnotesize
\caption{Absolute logarithmic prediction errors over all $46$ durations.
The final column counts points within $0.3$, $0.5$, and $1.0$ decade,
respectively. Base, analytical, and calibrated rows use $|e_L|$;
the suppression rows use $|e_S|$.}
\label{supp:floquet:tab:statistics}
\begin{tabular}{llrrrr}
\toprule
Quantity & Order & Median & Mean & Maximum & Counts\\
\midrule
Base & 1 & 0.038901 & 0.040435 & 0.081121 & 46/46/46\\
 & 3 & 0.000550 & 0.000615 & 0.002054 & 46/46/46\\
 & 5 & $5.48\times10^{-6}$ & $9.59\times10^{-6}$ & $4.35\times10^{-5}$ & 46/46/46\\
Analytical & 1 & 0.040770 & 0.042982 & 0.089007 & 46/46/46\\
 & 3 & 0.000393 & 0.000540 & 0.001351 & 46/46/46\\
 & 5 & $4.27\times10^{-6}$ & $6.96\times10^{-6}$ & $3.51\times10^{-5}$ & 46/46/46\\
Calibrated & 1 & 3.069830 & 2.906898 & 6.736111 & 6/6/9\\
 & 3 & 0.095192 & 0.498419 & 3.378674 & 28/31/38\\
 & 5 & 0.000525 & 0.006137 & 0.104553 & 46/46/46\\
Suppression $S$ & 1 & 2.996793 & 2.871664 & 6.762651 & 6/6/9\\
 & 3 & 0.094966 & 0.498532 & 3.377407 & 28/32/38\\
 & 5 & 0.000522 & 0.006140 & 0.104556 & 46/46/46\\
\bottomrule
\end{tabular}
\end{table}

The fifth-order contribution corrects both the magnitude and direction
of the third-order residual.
For the directly propagated endpoint $\boldsymbol\ell$ in the same
idle basis, define
$\boldsymbol r_n=\boldsymbol\ell-\boldsymbol\ell^{[n]}$ and
$\delta\boldsymbol\ell_5=
 e^{-\ii\omega_{\rm ref}T}J_Q\mathsf F(T)(p_4,\boldsymbol q_5)^{\mathsf T}$.
For all calibrated commands, the alignment
\begin{equation}
 A_5=\frac{\operatorname{Re}
  (\boldsymbol r_3^\dagger\delta\boldsymbol\ell_5)}
 {\|\boldsymbol r_3\|\|\delta\boldsymbol\ell_5\|}
 \label{supp:floquet:alignment}
\end{equation}
is at least $0.998958$ in the fixed idle-basis phase convention.
Values near unity show that the fifth-order correction aligns with
the missing complex amplitude.
The norm ratio $\|\delta\boldsymbol\ell_5\|/\|\boldsymbol r_3\|$
lies between $0.991$ and $1.02$.
The vector error decreases at all $46$ durations by factors
$21.9$--$1.35\times10^3$ (median $115$); the median
$\|\boldsymbol r_5\|/\|\boldsymbol r_3\|$ is $8.72\times10^{-3}$.
The endpoint ratio $\|\boldsymbol q_5\|/\|\boldsymbol q_3\|$
ranges from $2.66\times10^{-3}$ to $4.14\times10^{-2}$ (median $2.11\times10^{-2}$).
Thus, a small fifth-order amplitude controls the residual left by
cancellation of the lower orders.

\begin{table}[!htbp]
\centering\footnotesize
\caption{The eight durations with $|e_S^{[3]}|>1$ decade.
The last two columns quantify the complex endpoint error for calibrated pulses.}
\label{supp:floquet:tab:eight}
\begin{tabular}{rrrrr}
\toprule
$T$ (ns) & $e_S^{[3]}$ & $e_S^{[5]}$ & $\|\boldsymbol r_5\|/\|\boldsymbol r_3\|$ & $A_5$\\
\midrule
13.5 & $-1.417249$ & $-0.041510$ & 0.012271 & 0.999933\\
15.0 & $-1.306905$ & $-0.068476$ & 0.024762 & 0.999773\\
18.0 & $-3.235382$ & $-0.104556$ & 0.004488 & 0.999993\\
22.0 & $-1.644820$ & $-0.000325$ & 0.005727 & 0.999999\\
23.5 & $-1.399373$ & $-0.019207$ & 0.016470 & 0.999982\\
24.5 & $-2.225078$ & $+0.002011$ & 0.016339 & 0.999997\\
25.0 & $-3.377407$ & $+0.001682$ & 0.006453 & 0.999998\\
25.5 & $-1.276893$ & $-0.003011$ & 0.020747 & 0.999962\\
\bottomrule
\end{tabular}
\end{table}

Vector and total-leakage errors probe different aspects of convergence,
because probability errors in the two channels can cancel.
For
$\boldsymbol e=\boldsymbol\ell^{[n]}-\boldsymbol\ell$, the leakage
error is
$\Delta L=2\operatorname{Re}(\boldsymbol\ell^\dagger\boldsymbol e)
+\|\boldsymbol e\|^2$.
At $13$ ns, fifth order reduces the calibrated vector error
from $2.26\times10^{-4}$ to $2.68\times10^{-6}$, a factor of $84.2$.
The relative total-leakage error changes from $-0.308\%$ to $+0.684\%$.
The third-order channel-probability errors, $-3.10\times10^{-8}$ and
$+3.04\times10^{-8}$, nearly cancel.
Fifth order reduces them to $+1.45\times10^{-9}$ and
$-8.83\times10^{-11}$, changing $e_S$ from
$+1.61\times10^{-3}$ to $-2.99\times10^{-3}$ decade.
At $16.5$ ns the calibrated vector error decreases by a factor of $338$, whereas
$e_S$ changes from $+3.56\times10^{-6}$ to $-4.65\times10^{-6}$ decade.
Base has one analogous total-leakage change at $15.5$ ns:
$e_L^{[3]}=-3.56\times10^{-6}$ and
$e_L^{[5]}=-2.11\times10^{-5}$ decade, while its vector error decreases
from $1.85\times10^{-5}$ to $3.25\times10^{-6}$.
Analytical PDRAG improves in total leakage at all $46$ points;
Base and calibrated PDRAG improve at $45$ each.

\subsection{Numerical construction and precision}
\label{supp:floquet:numerics}

Each periodic reference is sampled on uniform radial and carrier-phase
grids spanning $0\leq r\leq1.025\max_t r(t)$.
Quintic splines interpolate the reference and $\mathcal C(t)$;
$\dot{\mathcal C}$ is obtained by differentiating the same spline.
Table~\ref{supp:floquet:tab:precision} lists the two precision settings.
Direct propagation of $h(t)$ checks the endpoint reconstructed from
the moving frame.
All expansion orders use the same $h_{\rm Fl}$ and $\mathsf F$.

\begin{table}[!htbp]
\centering\footnotesize
\caption{Two precision settings for all $138$ fixed commands.
The frozen-period and return equations use DOP853 with the listed
relative and absolute tolerances.}
\label{supp:floquet:tab:precision}
\begin{tabular}{lrr}
\toprule
Setting & Coarse & Refined\\
\midrule
Radial samples & 41 & 81\\
Carrier-phase samples & 257 & 513\\
Envelope time spacing (ns) & 0.010 & 0.005\\
Maximum frozen-period step & $\tau_c/128$ & $\tau_c/256$\\
Maximum return-equation step (ns) & 0.010 & 0.005\\
Relative tolerance & $2\times10^{-12}$ & $2\times10^{-13}$\\
Absolute tolerance & $2\times10^{-14}$ & $2\times10^{-15}$\\
\bottomrule
\end{tabular}
\end{table}

At the joins to idle, $r=0$ and $\mathsf F=I$.
For long Base pulses, $\mathcal C$ vanishes cubically near each join.
The polar representation retains nonzero interior values down to
$r\sim10^{-11}$ rad/ns, preserving continuity of the command phase.

Order-by-order norm conservation gives
\begin{align}
 \mathcal I_2&=2\operatorname{Re}(p_0^*p_2)+\|\boldsymbol q_1\|^2=0,\notag\\
 \mathcal I_4&=|p_2|^2+2\operatorname{Re}(p_0^*p_4)
          +2\operatorname{Re}(\boldsymbol q_1^\dagger\boldsymbol q_3)=0.
 \label{supp:floquet:norm}
\end{align}
Across all refined calculations, their maximum residuals are
$5.14\times10^{-16}$ and $3.73\times10^{-17}$.
Independent Simpson integration of Eq.~\eqref{supp:floquet:integrals}
on the full grid and a grid containing every other point verifies the terminal sources.
For $p_4,\boldsymbol q_5$, the full-grid discrepancies are at most
$1.27\times10^{-13}$ and $4.71\times10^{-14}$; the alternate-grid
values are $3.74\times10^{-12}$ and $8.66\times10^{-13}$.
Recomputed lower orders agree with the saved predictions within
$2.87\times10^{-17}$ in amplitude. The combined fifth-order state has
a maximum norm defect of $1.01\times10^{-4}$; the leakage prediction
uses its directly computed amplitudes.

For the logarithmic metrics, $\delta_{\rm num}$ is the largest of three
amplitude discrepancies: coarse versus refined endpoints, summed integral
checks for $\boldsymbol q_1,\boldsymbol q_3,\boldsymbol q_5$, and full
moving-frame versus direct propagation after projection onto $Q$.
All three comparisons use the fixed idle basis and the same laboratory phase.
All evaluated commands satisfy
$\delta_{\rm num}<\sqrt{L_{11}^{[5]}}$, giving the empirical numerical scales
\begin{equation}
 \nu_L=\max_{\pm}\left|2\log_{10}
       \left(1\pm\frac{\delta_{\rm num}}{\sqrt{L_{11}^{[5]}}}\right)\right|,
 \qquad \nu_S=\nu_L^{\rm an}+\nu_L^{\rm cal}.
 \label{supp:floquet:numscale}
\end{equation}
The two precision settings give $276$ calculations, from which the
numerical scales for leakage and suppression remain below $0.05$ decade.
The largest $\nu_S$ is $0.0128$ decade at $13.5$ ns.
These empirical scales quantify numerical convergence; $e_L$, $e_S$,
and the vector residuals measure truncation error in the return expansion.

The residual $\boldsymbol r_5$ resolves amplitude and phase errors alongside
the probability metrics in Tables~\ref{supp:floquet:tab:statistics} and \ref{supp:floquet:tab:eight}.
At $25$ ns, its norm is $31.5\%$ of the exact leakage amplitude,
while $e_S^{[5]}=0.001682$ decade.
At $13.5$ and $33.5$--$35$ ns, some residuals lie within a factor
of ten of the numerical vector-comparison scale.

\FloatBarrier

\section{Two-parameter fSim construction with a fixed exchange pulse}
\label{supp:sec:fsim-map}

Figure~4 combines variable-angle exchange pulses with calibrated idle
evolution to cover the fSim parameter domain.
The pulse duration is fixed at $T_{\rm p}=13$ ns for both Base and PDRAG.
We calibrate $41$ exchange targets $\theta_j=j\pi/80$, $j=0,\ldots,40$,
with angle tolerance $5\times10^{-6}$ rad.
The $\sin^4$ envelope and derivative form are unchanged; the amplitude
interval extends to $s\in[0,1.3]$ to include zero exchange.
The derivative, carrier, and physical waveform bounds remain those
specified above.
At each angle, the selected Base or PDRAG pulse remains fixed throughout the phase scan.

Let $H_0$ be the idle Duffing Hamiltonian and
$\varepsilon_b=E_b/\hbar$ its angular-frequency eigenvalues for the four
computational states $b=00,01,10,11$.
For the pulse map $K_{\rm p}=B^\dagger U(T_{\rm p})B$, a subsequent
idle interval $\tau$ gives
\begin{align}
 K(\tau)&=D(\tau)K_{\rm p},\qquad
 D(\tau)=\operatorname{diag}_b e^{-\ii\varepsilon_b\tau},\notag\\
 \zeta&=\varepsilon_{11}-\varepsilon_{10}
             -\varepsilon_{01}+\varepsilon_{00},\notag\\
 \phi_c(\tau)&=\phi_c^{({\rm p})}-\zeta\tau\pmod{2\pi}.
 \label{supp:eq:fsim-idle}
\end{align}
Here $\phi_c^{({\rm p})}$ is the conditional phase of the selected pulse.
For the stated device, $\zeta/(2\pi)=-24.8635846524$ MHz,
so the conditional phase increases at $-\zeta=0.156222509772$ rad/ns.
The idle Hamiltonian commutes with the computational projector.
Consequently, idle evolution preserves the exchange angle and every
input leakage probability, including $\overline L=L_{11}/4$ in this
excitation-conserving Duffing calculation.

For a prescribed conditional phase $\phi_c^{\rm tar}$, we choose
\begin{equation}
 \tau=\frac{\operatorname{mod}_{2\pi}
       [\phi_c^{\rm tar}-\phi_c^{({\rm p})}]}{-\zeta},
 \qquad 0\leq\tau<T_\phi=\frac{2\pi}{|\zeta|}.
 \label{supp:eq:fsim-wait}
\end{equation}
The function $\operatorname{mod}_{2\pi}$ returns an angle in $[0,2\pi)$.
Here $T_\phi=40.219462076$ ns, giving total durations
$13\leq T_{\rm p}+\tau<53.219462076$ ns.
Base and PDRAG reach the same prescribed gate with separately calibrated
waiting times and generally different total durations.
During this interval, the command remains at the original idle bias.

For every prescribed target pair, Eq.~\eqref{eq:supp:dmh:stableerror}
evaluates the fidelity against
$V=D_Z^{\rm out}U_{\rm fSim}(\theta,-\phi_c^{\rm tar})D_Z^{\rm in}$,
optimizing the local $Z$ phases as defined in Sec.~\ref{sec:supp:dmh:metrics}.
Simultaneous multiplication of $K$ and $V$ by $D(\tau)$ preserves
both terms in the fidelity expression.
Each exchange pulse therefore has constant fidelity along the
conditional-phase direction.
This construction extends the pulse's leakage suppression across the
full conditional-phase range.

The displayed grid contains $41$ exchange angles and $65$ phases
$\phi_{c,k}=k\pi/32$, $k=0,\ldots,64$, per pulse family.
The two phase endpoints represent the same physical gate.
The dense phase grid follows the exact idle relation in
Eq.~\eqref{supp:eq:fsim-idle}, applied to independently validated
four-level-per-transmon pulse maps.
Full $16$-dimensional idle matrix exponentials verify nine phases
per exchange angle and family, giving $738$ validation maps.
An additional $82$ evaluations verify every point on the comparison path
$\phi_c^{\rm tar}=4\theta$.
The maximum phase discrepancy is $3.70\times10^{-12}$ rad, and the
maximum absolute fidelity discrepancy from the analytic construction is
$4.81\times10^{-15}$.
At zero drive, the exact analytical fidelity is unity; numerical
residuals are retained in the validation data.
The source data include both waiting times, total durations, fixed-target
metrics, pulse parameters, and independent validation records.

\FloatBarrier

\section{Charge-ladder corrections and finite-pulse transfer}
\label{supp:charge-transfer-mechanism}

\subsection{Transferred controls}

All $138$ charge-model commands satisfy the exchange and resource
constraints, with a maximum exchange error $\ChargeMaxError$ mrad.
Tables~\ref{s4:tab:chargeparameters} and \ref{s4:tab:charge} report
the $12.5$, $13$, and $20$-ns commands; Sec.~\ref{sec:supp:dmh:validation}
gives their basis-convergence checks.

\begin{table}[!htbp]
\centering
\footnotesize
\caption{$\sin^4$ charge-model commands at the short-duration verification points and $20$ ns. Transferred PDRAG retains the same-duration Duffing derivative coefficients.}\label{s4:tab:chargeparameters}
\begin{tabular}{rlrrrr}
\toprule
$T$ (ns) & Family & $\beta$ (ns) & $\gamma$ (ns$^2$) & $s$ & $\delta f$ (MHz)\\
\midrule
12.5 & Analytical & 0.060286 & 0.479740 & 1.06457707 & 21.931332\\
12.5 & Base & 0.000000 & 0.000000 & 1.06887229 & 20.299256\\
12.5 & Transferred & -0.411871 & 1.715234 & 1.05158633 & 39.271396\\
13 & Analytical & 0.060286 & 0.479740 & 1.06453252 & 22.646983\\
13 & Base & 0.000000 & 0.000000 & 1.06651160 & 21.863677\\
13 & Transferred & -0.450000 & 1.986031 & 1.05118722 & 39.459629\\
20 & Analytical & 0.060286 & 0.479740 & 1.06534327 & 28.018300\\
20 & Base & 0.000000 & 0.000000 & 1.06543328 & 28.679176\\
20 & Transferred & 0.293213 & 1.085319 & 1.06144756 & 24.804618\\
\bottomrule
\end{tabular}
\end{table}
\begin{table}[!htbp]
\centering
\scriptsize
\caption{Nine-level charge-model metrics for the $\sin^4$ commands in Table~\ref{s4:tab:chargeparameters}. Gate fidelities follow Sec.~\ref{sec:supp:dmh:metrics}.}\label{s4:tab:charge}
\begin{tabular}{rlrrrrrr}
\toprule
$T$ (ns) & Family & $L_{11}$ & $\overline L$ & $\phi_{\rm c}$ (rad) & $1-F_{\rm fSim}$ & $1-F_{\rm iSWAP}$ & Error (mrad)\\
\midrule
12.5 & Analytical & $2.805\times10^{-2}$ & $7.013\times10^{-3}$ & 0.70432 & $7.043\times10^{-3}$ & 0.031417 & 0.09671\\
12.5 & Base & $8.340\times10^{-2}$ & $2.085\times10^{-2}$ & 0.57246 & $2.113\times10^{-2}$ & 0.037048 & 0.16006\\
12.5 & Transferred & $2.050\times10^{-3}$ & $5.124\times10^{-4}$ & 0.91423 & $5.126\times10^{-4}$ & 0.041560 & 0.03952\\
13 & Analytical & $2.216\times10^{-2}$ & $5.540\times10^{-3}$ & 0.81020 & $5.559\times10^{-3}$ & 0.037753 & 0.08810\\
13 & Base & $6.272\times10^{-2}$ & $1.568\times10^{-2}$ & 0.68551 & $1.583\times10^{-2}$ & 0.038727 & 0.13331\\
13 & Transferred & $1.166\times10^{-4}$ & $2.916\times10^{-5}$ & 1.03594 & $2.917\times10^{-5}$ & 0.052497 & 0.03447\\
20 & Analytical & $1.282\times10^{-4}$ & $3.205\times10^{-5}$ & 2.19334 & $3.205\times10^{-5}$ & 0.217401 & 0.03408\\
20 & Base & $3.330\times10^{-4}$ & $8.326\times10^{-5}$ & 2.15019 & $8.326\times10^{-5}$ & 0.209806 & 0.04895\\
20 & Transferred & $2.175\times10^{-7}$ & $5.438\times10^{-8}$ & 2.24446 & $5.439\times10^{-8}$ & 0.226530 & 0.02928\\
\bottomrule
\end{tabular}
\end{table}

\subsection{Fixed commands and common dressed labels}

Let $f\in\{B,A,\mathrm{tr}\}$ label Base,
analytical, and transferred PDRAG, with parameters
$\vartheta_{f,a}=(\beta_{f,a},\gamma_{f,a},s_{f,a},\Omega_{f,a})$.
Here $a=\mathrm D,\mathrm C$ labels the Duffing and charge commands.
Derivative coefficients transfer at equal duration; amplitude and carrier
are recalibrated for full exchange.
To isolate the device response, each model comparison uses the same
complete charge command $\delta\omega_{1,f}(t)$.
The family index $f$ is omitted from the following Hamiltonian definitions.

The overlap assignment of Sec.~\ref{sec:supp:dmh:metrics} defines
common idle-dressed labels. Let $W_a$ map these labels to the product
coordinates of model $a$, and let $\varepsilon_{a,mn}=E_{mn}^a/\hbar$.
Static comparisons use the same $N^2$-dimensional label space
with $N=9$ levels per transmon.
All generators $h$ below have angular-frequency units.
Let $e_{mn}$ denote the unit vector for dressed label $mn$.
The static references start in $e_{11}$ and use the computational projector
$P_\ell=\sum_{mn\in\{00,01,10,11\}}e_{mn}e_{mn}^\dagger$.
With $N_1=\operatorname{diag}(m)\otimes I$, define
\begin{equation}
 M_{\rm D}=W_{\rm D}^\dagger N_1W_{\rm D},\qquad
 M_{\rm C}=W_{\rm C}^\dagger\mathsf R_0N_1
             \mathsf R_0^\dagger W_{\rm C}.
 \label{supp:charge:modulation-matrices}
\end{equation}
Here $N_1$ counts isolated transmon levels, whereas $\hat n_1$ is the
Cooper-pair number operator. The matrix $\mathsf R_0$ maps the instantaneous
product eigenbasis at idle to the fixed idle basis; here $\mathsf R_0=I$.
The matrix $M_{\rm C}$ captures static dressing of the level-index operator;
moving-basis references retain the full flux response.

\subsection{Microscopic origin of the gap shifts}

In the isolated idle eigenbasis, write
$n^{(j)}_{rs}=\bra r\hat n_j\ket s$.
Choose constant level phases such that
$n^{(j)}_{m,m+1}=-\ii|n^{(j)}_{m,m+1}|$.
Matching $g=g_C|n^{(1)}_{10}n^{(2)}_{01}|$ fixes the leakage couplings
\begin{equation}
 J_{20}=g\frac{|n^{(1)}_{21}|}{|n^{(1)}_{10}|},\qquad
 J_{02}=g\frac{|n^{(2)}_{21}|}{|n^{(2)}_{10}|}.
 \label{supp:charge:ladder}
\end{equation}
The charge ratios are $1.383$ and $1.375$, below the Duffing value
$\sqrt{2}$ by about $2.2\%$ and $2.7\%$, respectively.
They give $J_{20}/(2\pi)=138.3$ MHz and
$J_{02}/(2\pi)=137.5$ MHz, compared with $141.4$ MHz for both Duffing channels.
These weaker leakage couplings modify the idle dressed gaps even when
the isolated transition energies and computational exchange are matched.

The adjacent-exchange RWA isolates the resulting change in idle dressing.
Subtracting the isolated $11$ energy gives, in product order $(20,11,02)$,
\begin{equation}
 h_{2,\mathrm{RWA}}^{\mathrm{idle}}=
 \begin{pmatrix}a&J_{20}&0\\J_{20}&0&J_{02}\\0&J_{02}&b\end{pmatrix},
 \qquad a=\Delta+\alpha_1,\quad b=-\Delta+\alpha_2.
 \label{supp:charge:microscopic-three-state}
\end{equation}
Here $\Delta=\omega_1-\omega_2$, and $a,b$ are the bare angular-frequency
offsets of $\ket{20}$ and $\ket{02}$ from $\ket{11}$.
The eigenvalues $\lambda$ obey
\begin{equation}
 \lambda(\lambda-a)(\lambda-b)
 -J_{20}^2(\lambda-b)-J_{02}^2(\lambda-a)=0.
 \label{supp:charge:secular}
\end{equation}
Overlap assignment gives the dressed gaps
$\Delta_{mn}=\lambda_{mn}-\lambda_{11}$ for $mn=20,02$.
Matching the first three isolated energies makes the charge-ladder
ratios the sole difference between the two RWA blocks.

\begin{table}[!htbp]
\centering\footnotesize
\caption{Idle dressed gaps in MHz. The gap reference uses the full
charge idle Hamiltonian; the adjacent-exchange RWA isolates the ladder contribution.}
\label{supp:charge:gap-table}
\begin{tabular}{lrr}
\toprule
Idle model & $\Delta_{20}/(2\pi)$ & $\Delta_{02}/(2\pi)$\\
\midrule
Duffing, static exchange RWA & 453.325544 & $-838.734790$\\
Charge, adjacent-exchange RWA & 450.622427 & $-838.196902$\\
Charge, full capacitive interaction & 450.584317 & $-838.207338$\\
\bottomrule
\end{tabular}
\end{table}

The ladder correction shifts $\Delta_{20}/(2\pi)$ and
$\Delta_{02}/(2\pi)$ by $-2.70$ and $+0.538$ MHz, respectively.
The remaining static capacitive terms, including nonadjacent and
excitation-nonconserving couplings, contribute $-0.0381$ and $-0.0104$ MHz.
The full shifts are $-2.74$ and $+0.527$ MHz,
so adjacent charge matrix elements account for most of the static gap change.

\subsection{Three-state gap reference and leakage response}

Four static references separate the idle spectrum from the dressed
modulation matrix,
\begin{equation}
 h_{ab}(t)=\operatorname{diag}(\boldsymbol\varepsilon_a-\varepsilon_{a,11})
       +\delta\omega_1(t)M_b,\qquad a,b\in\{\mathrm D,\mathrm C\}.
 \label{supp:charge:static-bridge}
\end{equation}
The vector $\boldsymbol\varepsilon_a$ collects the assigned idle
eigenfrequencies $\varepsilon_{a,mn}$.
The four references are D (Duffing spectrum and modulation), G
(charge gaps with Duffing modulation), M (Duffing gaps with charge modulation),
and $C_s$ (charge gaps and static charge modulation).
They correspond to $(a,b)=(\mathrm D,\mathrm D),(\mathrm C,\mathrm D),
(\mathrm D,\mathrm C),(\mathrm C,\mathrm C)$, respectively.
The Duffing idle Hamiltonian and $N_1$ conserve total excitation.
Both $M_{\rm D}$ and the diagonal energy replacement preserve the
two-excitation subspace, so G closes exactly on $(11,20,02)$ for arbitrary
$\delta\omega_1(t)$, giving
\begin{equation}
 \ii\dot{\boldsymbol c}=h_{\rm G}(t)\boldsymbol c,\qquad
 \boldsymbol c(0)=(1,0,0)^{\mathsf T},\qquad
 L_{11}^{\rm G}=|c_{20}(T)|^2+|c_{02}(T)|^2.
 \label{supp:charge:gap-leakage}
\end{equation}
In the same label order and phase convention,
$M_{\rm D}^{(2)}=N_{\rm d}$ of the Floquet construction.
The largest matrix element connecting this block to the remaining
states is $2.21\times10^{-15}$.

Let $h_\eta=h_0+\eta\delta h$ interpolate between generators in common
coordinates, with dimensionless parameter $\eta$ and propagator $U_\eta$.
Define
$\boldsymbol\ell_\eta=Q_\ell U_\eta(T,0)e_{11}$, where $Q_\ell=I-P_\ell$.
Then
\begin{align}
 \partial_\eta\boldsymbol\ell_\eta
 &=-\ii Q_\ell\int_0^T U_\eta(T,t)\delta h(t)U_\eta(t,0)e_{11}\,\dd t,\notag\\
 \partial_\eta\ln L_\eta
 &=\frac{2\operatorname{Re}[\boldsymbol\ell_\eta^\dagger
                  \partial_\eta\boldsymbol\ell_\eta]}
          {\|\boldsymbol\ell_\eta\|^2}.
 \label{supp:charge:response-identity}
\end{align}
Here $L_\eta=\|\boldsymbol\ell_\eta\|^2$, and the logarithmic derivative
is evaluated where $L_\eta>0$.
The overlap between the residual amplitude and its correction determines
whether leakage increases or decreases.
Near a deep minimum, the small denominator makes relative leakage
sensitive to corrections that are small compared with the Hamiltonian scale.

\subsection{Driven exchange in the full charge space}

Diagonalizing the tunable transmon in its fixed idle basis gives
$\mathsf R(t)=R_1(t)\otimes I$, with $\mathsf R_0=\mathsf R(0)$.
The complete moving-basis generator is
\begin{equation}
 h_{\rm C}^{\rm inst}(t)=\mathsf E_{\rm C}(t)+V_{\rm C}(t)+\mathcal A_{\rm C}(t),
 \quad V_{\rm C}(t)=g_C\mathsf n_1(t)\otimes\mathsf n_2,
 \quad\mathcal A_{\rm C}=-\ii\mathsf R^\dagger\dot{\mathsf R}.
 \label{supp:charge:moving-generator}
\end{equation}
Here $\mathsf E_{\rm C}$ is the product spectrum divided by $\hbar$,
and $\mathcal A_{\rm C}$ accounts for motion of the instantaneous basis.
The charge operators are represented by $\mathsf n_1(t)=R_1^\dagger(t)
\mathsf n_{1,\mathrm{idle}}R_1(t)$ and the time-independent matrix $\mathsf n_2$.
Instantaneous eigenvectors are phased to have positive overlap with
their assigned idle states.
The adjacent excitation-conserving interaction is
\begin{equation}
 V_{\rm ex}(t)=\sum_{m=0}^{N-2}\sum_{n=1}^{N-1}
 \left\{g_C[\mathsf n_1(t)]_{m+1,m}[\mathsf n_2]_{n-1,n}
 \ket{m+1,n-1}\bra{m,n}+\mathrm{H.c.}\right\},
 \label{supp:charge:adjacent-exchange}
\end{equation}
Define $\Delta V_{\rm ex}(t)=V_{\rm ex}(t)-V_{\rm ex}(0)$ and
\begin{equation}
 h_{\rm F}(t)=\mathsf E_{\rm C}(t)+V_{\rm C}(0)+\mathcal A_{\rm C}(t),
 \qquad h_{\rm R}(t)=h_{\rm F}(t)+\Delta V_{\rm ex}(t).
 \label{supp:charge:physical-reference}
\end{equation}
Reference F freezes the interaction at idle while retaining the
81-state spectrum and basis connection.
Reference R restores the time-dependent adjacent exchange, isolating
its contribution on the same background.
The full charge model C additionally includes
$V_{\rm C}(t)-V_{\rm C}(0)-\Delta V_{\rm ex}(t)$,
the remaining driven interaction terms.
The physical initial state and endpoint projection are
\begin{equation}
 \psi_{\rm inst}(0)=\mathsf R_0^\dagger\ket{11}_{\rm d},\qquad
 L_{11}=\|Q\mathsf R(T)\psi_{\rm inst}(T)\|^2.
 \label{supp:charge:reference-projection}
\end{equation}
The common physical projector is $Q=I-BB^\dagger$.
The endpoint vectors include the restored basis transformation and any removed scalar phase.
The frequency-to-flux map acts on the complete fixed command.

\subsection{Ordering, suppression depth, and command transfer}

\begin{figure}[!tp]
 \centering
 \includegraphics[width=\textwidth]{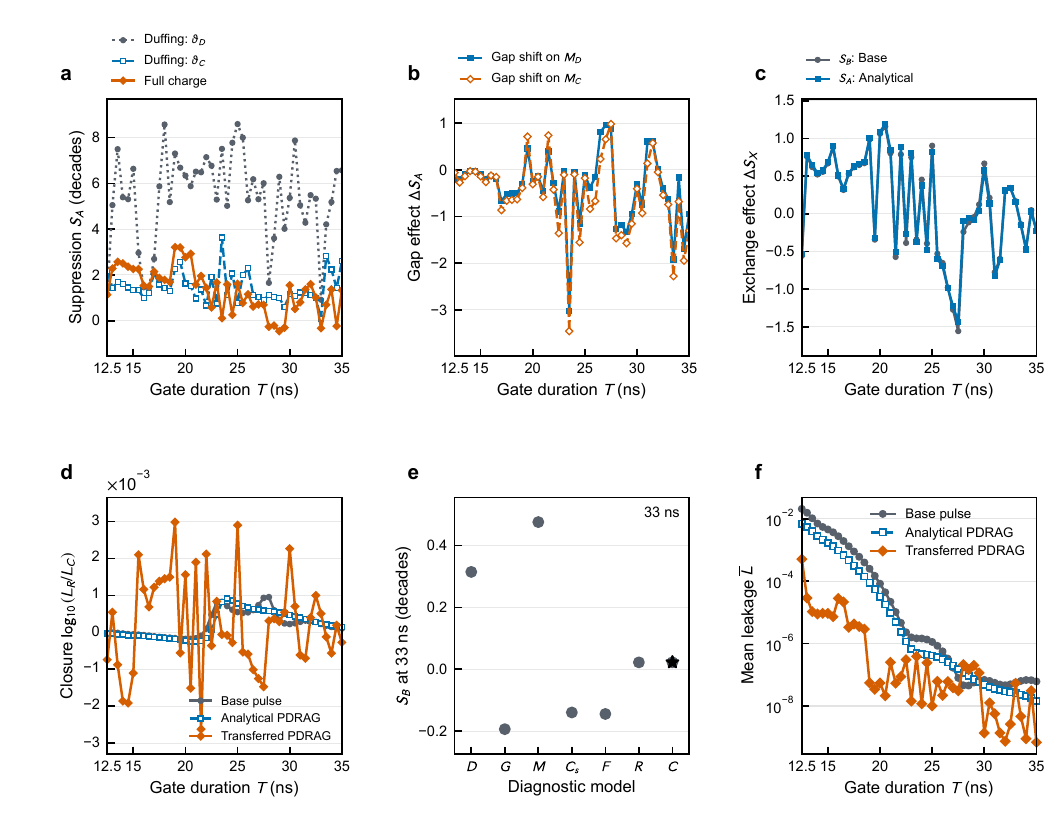}
 \caption{Contributions to charge-model transfer for $138$ commands.
 The transfer gain is $S_X=\log_{10}(L_X/L_{\rm tr})$, with
 $X=A$ (analytical) or $B$ (Base); positive values favor transferred PDRAG.
 (a) $S_A$ before and after amplitude--carrier recalibration in Duffing,
 followed by the full charge response. The command sets are
 $\vartheta_{\rm D}$ and $\vartheta_{\rm C}$, respectively.
 (b) Static-gap contributions on Duffing and charge-modulation backgrounds.
 (c) Contribution of driven adjacent exchange, $S_X^{\rm R}-S_X^{\rm F}$.
 (d) Leakage error of reference R relative to the full charge model C.
 (e) $S_B$ at $33$ ns for the seven reference Hamiltonians defined in
 Sec.~\ref{supp:charge-transfer-mechanism}; F and R differ only by driven adjacent exchange.
 (f) Full-charge mean computational leakage $\overline L$.
 Markers show sampled commands and lines connect adjacent durations.}
 \label{fig:supp:charge_mechanism}
\end{figure}

For model $Y$, define the transfer gain
$S_X^Y=\log_{10}(L_X^Y/L_{\rm tr}^Y)$, where $X=B,A$ labels Base
and analytical PDRAG.
Here $L_f^Y$ is the endpoint $L_{11}$ for family $f$ in model $Y$;
positive $S_X^Y$ means that transferred PDRAG has lower leakage.
Figure~\ref{fig:supp:charge_mechanism} separates command adjustment,
static-gap corrections, and driven exchange.

In the full charge model, leakage under transferred PDRAG exceeds that under Base at
four durations from $28$ to $29.5$ ns.
It also exceeds the leakage under analytical PDRAG at those durations and at $33$ and $34.5$ ns.
D and M predict lower transferred leakage throughout the scan.
Replacing the static gaps reproduces all ten reversals on both
modulation backgrounds, G and $C_s$.
G reproduces the sign of $S_X$ in $91$ of $92$ comparisons and
the complete three-family ordering at $44$ of $46$ durations.
Its median and maximum absolute errors in $S_X$ are $0.767$ and
$1.53$ decades, respectively.
The gap correction captures where the ordering changes, while the
full driven interaction is needed to reproduce the depth of suppression.

At $33$ ns, $S_B$ is $+0.315$ in D, $-0.193$ in G,
and $-0.144$ in F.
Restoring driven adjacent exchange changes the sign to
$S_B^{\rm R}=+0.0224$, close to the full-charge value
$S_B^{\rm C}=+0.0227$.
Its contribution on the F background is
\begin{equation}
 \Delta S_{X,\rm ex}=S_X^{\rm R}-S_X^{\rm F},
 \label{supp:charge:relative-exchange}
\end{equation}
which increases both transfer gains at $27$ of $46$ durations,
including $17$ of $20$ points between $12.5$ and $22$ ns.
Within that short-gate interval, the median increments are $0.633$
and $0.652$ decades relative to Base and analytical PDRAG.
Driven adjacent exchange reduces the gains at $12.5$, $19.5$, and
$21.5$ ns, and at several longer durations.

The four static references separate the gap and modulation responses
and their nonadditive contribution,
\begin{equation}
 \log_{10}\frac{L_f^{C_s}}{L_f^{\rm D}}
 =\log_{10}\frac{L_f^{\rm G}}{L_f^{\rm D}}
 +\log_{10}\frac{L_f^{\rm M}}{L_f^{\rm D}}
 +\log_{10}\frac{L_f^{C_s}L_f^{\rm D}}{L_f^{\rm G}L_f^{\rm M}}.
 \label{supp:charge:static-interaction}
\end{equation}
The first two terms isolate changes in the gaps and modulation matrix.
The final term measures the nonadditive contribution when both change together.
The modulation-matrix correction is substantial at $19.5$ and $21.5$ ns.

Let $\boldsymbol\vartheta_a=\{\vartheta_{f,a}\}_{f=B,A,\mathrm{tr}}$
collect the calibrated commands for all three families.
The total change in suppression separates into command adjustment
and device response,
\begin{align}
 S_X^{\rm C}(\boldsymbol\vartheta_{\rm C})-S_X^{\rm D}(\boldsymbol\vartheta_{\rm D})
 ={}&[S_X^{\rm D}(\boldsymbol\vartheta_{\rm C})-S_X^{\rm D}(\boldsymbol\vartheta_{\rm D})]
 \notag\\
 &+[S_X^{\rm C}(\boldsymbol\vartheta_{\rm C})-S_X^{\rm D}(\boldsymbol\vartheta_{\rm C})].
 \label{supp:charge:command-device-bridge}
\end{align}
The first bracket compares the charge-calibrated and Duffing-calibrated
commands in the Duffing model. The second changes the device while
holding the charge-calibrated commands fixed.
At $20$ ns, amplitude--carrier recalibration reduces $S_A$ from
$6.34$ to $1.63$ decades.
The charge response then raises it to $2.77$ decades, giving contributions
of $-4.71$ and $+1.14$ decades, respectively.
Across the scan, command adjustment reduces both transfer gains at
$45$ durations.
The fixed-command device response increases both gains at $26$ durations,
including $19$ of the $20$ points between $12.5$ and $22$ ns.

\subsection{Numerical verification}

All 138 commands use two integration profiles, giving 276 calculations.
The three-state reference retains the complete real modulation;
the Duffing operator uses the original static-exchange RWA.
DOP853 tolerances $(\mathrm{rtol},\mathrm{atol})$ are
$(2\times10^{-10},2\times10^{-12})$ and
$(2\times10^{-12},2\times10^{-14})$, with maximum steps 0.010 and 0.005 ns.
Command interpolation spacings are 0.001 and 0.0005 ns; moving-component
spacings are 0.004 and 0.002 ns.
Refinement changes logarithmic leakage by at most
$1.15\times10^{-6}$ decade and the complex endpoint amplitude in the $Q$ leakage subspace by
$1.54\times10^{-8}$.
Independent full-charge propagation agrees within $6.68\times10^{-9}$
in the leakage endpoint vector; the largest norm error is $2.80\times10^{-14}$.

Reference R reproduces all $138$ leakages and $92$ transfer gains with
maximum errors of $2.98\times10^{-3}$ and $3.16\times10^{-3}$ decade, respectively.
It also reproduces every pairwise ordering and the ranking of all three families.
For the fixed Base and transferred commands at $33$ ns, full-charge
$S_B$ remains positive as the basis increases.
Its values are $0.0227286$, $0.0227469$, and $0.0227520$ for
$N=9,11,12$, respectively, at charge cutoff $n_{\max}=18$.
At $N=11$, restoring driven adjacent exchange changes $S_B$ from
$-0.14428$ to $+0.02247$.
Independent laboratory-frame propagation agrees within $7.57\times10^{-10}$
in the leakage endpoint vector under the same phase convention.

The mean leakage in Fig.~\ref{fig:supp:charge_mechanism}(f) averages
all four computational inputs, using Eq.~\eqref{eq:supp:dmh:leakage}.
At 20 ns, transfer reduces it from $8.33\times10^{-5}$ for Base
to $5.44\times10^{-8}$.

\FloatBarrier

\section{PDRAG for continuous state transfer in SSH chains of seven to thirteen sites}
\label{supp:sec:ssh}

\subsection{Microscopic chain and two-coupling transport schedule}
\label{supp:ssh:model}

Figure~6 compares continuous transport from $Q_1$ to $Q_N$ in open
Su--Schrieffer--Heeger (SSH) chains with $N=7,9,11,$ and $13$.
All chains share the device-frequency rule, nominal coupling scales,
$350$-ns duration, target weights, physical resource limits, and optimization budget.
Numerical evaluation uses ns and GHz, with angular frequencies in rad/ns.
The microscopic Hamiltonian is
\begin{align}
 \frac{H(t)}{\hbar}={}&\sum_{j=1}^{N}\omega_{Q_j}n_{Q_j}
 +\sum_{m=1}^{N-1}[\omega_{C_m}+u_m(t)]n_{C_m}\notag\\
 &+g\sum_{m=1}^{N-1}
 \left[(a_{Q_m}^{\dagger}+a_{Q_{m+1}}^{\dagger})a_{C_m}
 +\mathrm{h.c.}\right]\notag\\
 &+g_{\rm d}\sum_{m=1}^{N-1}
 (a_{Q_m}^{\dagger}a_{Q_{m+1}}+\mathrm{h.c.}),
 \label{supp:eq:ssh:device}
\end{align}
with $g/(2\pi)=150$ MHz and $g_{\rm d}/(2\pi)=5$ MHz.
The bare frequencies, in GHz, are
\begin{align}
 \omega_{Q_{2k+1}}/(2\pi)&=5.10+0.10k,&&k=0,\ldots,(N-1)/2,\notag\\
 \omega_{Q_{2k+2}}/(2\pi)&=4.50+0.10k,&&k=0,\ldots,(N-3)/2,\notag\\
 \omega_{C_m}/(2\pi)&=4.80+0.05(m-1),&&m=1,\ldots,N-1.
 \label{supp:eq:ssh:frequencies}
\end{align}
We propagate the $(2N-1)$-dimensional single-excitation sector of this
excitation-conserving Hamiltonian coherently.
Diagonalizing the idle Hamiltonian gives $H_0/\hbar=BEB^{\dagger}$.
A one-to-one assignment maximizes the summed squared overlaps between
eigenvectors and bare modes ordered as
$(Q_1,C_1,Q_2,\ldots,C_{N-1},Q_N)$; each assigned bare-mode component
is chosen positive.
The resulting states $\ket{Q_j}_{\rm d}$ and $\ket{C_m}_{\rm d}$ define
the logical and coupler-like subspaces.

The target Hamiltonian is
\begin{align}
 \frac{H_{\rm SSH}(t)}{\hbar}={}&
 J_1(t)\sum_{k=1}^{(N-1)/2}
 \left(\ket{Q_{2k-1}}_{\rm d}\!\bra{Q_{2k}}_{\rm d}
 +\mathrm{h.c.}\right)\notag\\
 &+J_2(t)\sum_{k=1}^{(N-1)/2}
 \left(\ket{Q_{2k}}_{\rm d}\!\bra{Q_{2k+1}}_{\rm d}
 +\mathrm{h.c.}\right).
 \label{supp:eq:ssh:target}
\end{align}
For positive couplings, its zero mode is
\begin{equation}
 \ket{D(t)}=\mathcal N(t)
 \sum_{k=0}^{(N-1)/2}\left[-\frac{J_1(t)}{J_2(t)}\right]^k
 \ket{Q_{2k+1}}_{\rm d},
 \label{supp:eq:ssh:dark}
\end{equation}
where $\mathcal N$ normalizes the state.
Increasing $J_1/J_2$ moves this mode from the left endpoint to the
right endpoint. The protocol lasts $T=350$ ns, with switch duration
$\tau=5$ ns and peak ordinary frequencies
$\mathcal J_1=30$ MHz and $\mathcal J_2=45$ MHz.

We define a smooth switch $W(z)=0$ for $z\leq0$, $W(z)=1$ for
$z\geq1$, and, between these limits,
\begin{align}
 W(z)&=z+\sum_{k=1}^{4}\frac{a_k}{2\pi k}\sin(2\pi kz),\notag\\
 (a_1,a_2,a_3,a_4)&=(-56,28,-8,1)/35.
 \label{supp:eq:ssh:window}
\end{align}
Its derivative is $W'(z)=128\sin^8(\pi z)/35$.
The common switch window is
$w(t)=W(t/\tau)W[(T-t)/\tau]$.
To specify the transport progress, set
$x=\operatorname{clip}[(t-\tau)/(T-2\tau),0,1]$,
$e=0.05$, and $I(z)=\int_0^z W(r)\,\dd r$.
The progress function used by the \texttt{ss5} schedule is
\begin{align}
 G_e(x)&=\begin{cases}
 eI(x/e)/(1-e),&0\leq x<e,\\
 (x-e/2)/(1-e),&e\leq x\leq1-e,\\
 1-eI[(1-x)/e]/(1-e),&1-e<x\leq1,
 \end{cases}\notag\\
 s(x)&=\sin^2[\pi G_e(x)/2].
 \label{supp:eq:ssh:progress}
\end{align}
The inverse-gap schedule is obtained from
\begin{align}
 c&=\cos[2\pi/(N+1)],\qquad
 a=\mathcal J_1^2+\mathcal J_2^2+2\mathcal J_1\mathcal J_2c,\notag\\
 \mu&=\frac{\mathcal J_2^2+\mathcal J_1\mathcal J_2c}{a},
 \qquad b=\sqrt{\mathcal J_2^2/a-\mu^2},\notag\\
 \vartheta_0&=\arctan(-\mu/b),\qquad
 \vartheta_1=\arctan[(1-\mu)/b],\notag\\
 q(t)&=\mu+b\tan\{\vartheta_0+s(x)(\vartheta_1-\vartheta_0)\}.
 \label{supp:eq:ssh:gap_schedule}
\end{align}
Thus the two dimensionless envelopes and target couplings are
\begin{align}
 f_1(t)&=w(t)q(t),& J_1(t)/(2\pi)&=\mathcal J_1f_1(t),\notag\\
 f_2(t)&=w(t)[1-q(t)],& J_2(t)/(2\pi)&=\mathcal J_2f_2(t).
 \label{supp:eq:ssh:envelopes}
\end{align}
Equation~\eqref{supp:eq:ssh:gap_schedule} evaluates the inverse cumulative
integral of the reciprocal squared core SSH gap,
$\Delta_{\rm SSH}^2=J_1^2+J_2^2-2J_1J_2c$, at the smooth progress
$s(x)$ in Eq.~\eqref{supp:eq:ssh:progress}.
During switch-on $J_1=0$, and during switch-off $J_2=0$.
In each ideal limit, the populated endpoint is decoupled from the chain.

\subsection{Fixed Stark feedforward and the comparison frame}
\label{supp:ssh:frame}

Let $f_a^{\rm d}=E_a/(2\pi)$ denote an idle dressed frequency and
$N_m=B^{\dagger}\ket{C_m}\bra{C_m}B$ the coupler-frequency operator
in the dressed basis. For port $m$, define
\begin{align}
 f_m^{(0)}&=|f_{Q_m}^{\rm d}-f_{Q_{m+1}}^{\rm d}|,\notag\\
 \eta_m&=\operatorname{sgn}(f_{Q_m}^{\rm d}-f_{Q_{m+1}}^{\rm d}),\notag\\
 A_m^{(0)}&=\frac{2\mathcal J_{p(m)}}
 {|(N_m)_{Q_m,Q_{m+1}}|},
 \qquad
 p(m)=\begin{cases}1,&m\ \mathrm{odd},\\2,&m\ \mathrm{even}.
 \end{cases}
 \label{supp:eq:ssh:seed}
\end{align}
All ordinary frequencies in the Stark calculation are expressed in GHz.
The common second-order coefficients are
\begin{equation}
 S_{j\alpha}=2\pi
 \sum_{m:p(m)=\alpha}\frac{[A_m^{(0)}]^2}{4}
 \sum_{b\ne Q_j}\sum_{\sigma=\pm1}^{\prime}
 \frac{|(N_m)_{Q_jb}|^2}
 {f_{Q_j}^{\rm d}-f_b^{\rm d}+\sigma f_m^{(0)}}.
 \label{supp:eq:ssh:stark}
\end{equation}
The sum includes all other logical and coupler-like modes; the prime
retains denominators with magnitude at least $0.020$ GHz.
The accumulated phases and port feedforward are
\begin{align}
 \chi_j(t)&=\int_0^t
 [S_{j1}f_1^2(t')+S_{j2}f_2^2(t')]\,\dd t',\notag\\
 \Phi_m^{\rm S}(t)&=\eta_m[\chi_m(t)-\chi_{m+1}(t)].
 \label{supp:eq:ssh:chirp}
\end{align}
These coefficients are computed once from the nominal $30/45$-MHz
target and held fixed for both pulse families.
Table~\ref{supp:tab:ssh:stark} gives the $N=11$ values.
The accompanying numerical source files provide the values for every chain length.

\begin{table}[!htbp]
\centering\footnotesize
\caption{Idle carrier frequencies and fixed Stark coefficients for $N=11$.
The row index denotes port $m=j$ in the carrier column and logical
site $j$ in the Stark columns. All entries are in MHz.}
\label{supp:tab:ssh:stark}
\begin{tabular}{rrrr}
\toprule
$j$ & $f_j^{(0)}$ & $S_{j1}/(2\pi)$ & $S_{j2}/(2\pi)$ \\
\midrule
1 & 778.986871739 & -8.630201272 & -0.294380793 \\
2 & 923.064594648 & 6.126326485 & 25.341326874 \\
3 & 823.243916632 & -9.116840804 & -25.874765865 \\
4 & 923.265059143 & 8.802783619 & 25.436902799 \\
5 & 823.266143108 & -9.107542123 & -26.029515018 \\
6 & 923.265231681 & 8.825223337 & 25.438954803 \\
7 & 823.265223974 & -9.108251352 & -26.029241013 \\
8 & 922.905687383 & 8.825725799 & 25.307406114 \\
9 & 822.748851749 & -8.994116209 & -25.960459931 \\
10 & 870.910768523 & 8.573955655 & 17.448349239 \\
11 & \textemdash & 0.177700490 & -24.565293634 \\
\bottomrule
\end{tabular}
\end{table}

For the bare-basis state $\ket{\psi_b(t)}$ propagated after subtracting
the common angular frequency $E_{Q_1}$ from the Hamiltonian, introduce
\begin{align}
 \ket{\psi_I(t)}&=e^{\ii(E-E_{Q_1})t}B^{\dagger}\ket{\psi_b(t)},\notag\\
 \ket{\widetilde\psi(t)}&=
 \left[\sum_j e^{\ii\chi_j(t)}\ket{Q_j}_{\rm d}\!\bra{Q_j}_{\rm d}
 +\sum_m\ket{C_m}_{\rm d}\!\bra{C_m}_{\rm d}\right]
 \ket{\psi_I(t)}.
 \label{supp:eq:ssh:tracked_frame}
\end{align}
The target state $\ket{\psi_{\rm SSH}(t)}$ solves
$\ii\partial_t\ket{\psi_{\rm SSH}}=H_{\rm SSH}\ket{\psi_{\rm SSH}}/\hbar$
from $\ket{Q_1}_{\rm d}$ and is compared with
$\ket{\widetilde\psi(t)}$ in this common Stark-tracked frame.

\subsection{Continuous port commands and numerical parameters}
\label{supp:ssh:pulses}

The envelope grid is $t_k=0,0.05,\ldots,350$ ns and includes
$\tau$ and $T-\tau$.
For each envelope in Eq.~\eqref{supp:eq:ssh:envelopes}, a degree-nine
interpolating B-spline with the default not-a-knot boundary condition
provides its zeroth through fourth derivatives on this grid.
The values of all five arrays at $t=0,T$ are set to zero, and each array
is then represented by a cubic spline, denoted $D_{n\alpha}(t)$.
The target and commands use $f_\alpha=D_{0\alpha}$.
The derivative normalization is
$M_{n\alpha}=\max_k|D_{n\alpha}(t_k)|$ for $n=1,\ldots,4$.
For the accumulated phase, cumulative trapezoidal integration of
$f_\alpha^2(t_k)$ followed by cubic interpolation implements
Eq.~\eqref{supp:eq:ssh:chirp}.
The target-state integration uses DOP853 with relative tolerance
$2\times10^{-12}$, absolute tolerance $2\times10^{-14}$, and maximum
step $2$ ns.

\begin{table}[!htbp]
\centering\footnotesize
\caption{Derivative normalization for the two SSH envelopes at $N=11$.
The entries in row $n$ have units ns$^{-n}$.}
\label{supp:tab:ssh:normalization}
\begin{tabular}{rrr}
\toprule
$n$ & $M_{n1}$ & $M_{n2}$ \\
\midrule
1 & 0.731428571429 & 0.731428571429 \\
2 & 0.81299237288 & 0.81299237288 \\
3 & 2.31005140724 & 2.31005140724 \\
4 & 5.62773131278 & 5.62773131234 \\
\bottomrule
\end{tabular}
\end{table}

For each of the $N-1$ ports, the continuous command is
\begin{align}
 \mathcal E_m(t)&=f_{p(m)}(t)+\sum_{n=1}^{4}
 c_{mn}(-\ii)^n\frac{D_{n,p(m)}(t)}{M_{n,p(m)}},\notag\\
 \varphi_m(t)&=2\pi[f_m^{(0)}+\delta f_m]t
 +\phi_m+\Phi_m^{\rm S}(t),\notag\\
 \frac{u_m(t)}{2\pi}&=A_m\operatorname{Re}
 [\mathcal E_m(t)e^{-\ii\varphi_m(t)}]
 +d_{m1}f_1^2(t)+d_{m2}f_2^2(t).
 \label{supp:eq:ssh:command}
\end{align}
The amplitude, carrier correction, phase, and both DC coefficients are
port-specific.
The four real coefficients $c_{mn}$ set the first through fourth derivative components,
with odd orders in the imaginary quadrature and even orders in the real
quadrature. Base fixes these four coefficients to zero.
Tables~\ref{supp:tab:ssh:controls} and~\ref{supp:tab:ssh:derivatives}
list the final $N=11$ controls used for the endpoint comparison in Fig.~6.
Machine-readable controls for all four chain lengths accompany the source data.

\begin{table}[!htbp]
\centering\footnotesize
\caption{Final non-derivative controls for continuous eleven-site transfer.
$A_m$, $\delta f_m$, $d_{m1}$ and $d_{m2}$ are in MHz;
$\phi_m$ is in radians. The total carrier is
$f_m^{(0)}+\delta f_m$, with $f_m^{(0)}$ from
Table~\ref{supp:tab:ssh:stark}.}
\label{supp:tab:ssh:controls}
\begin{tabular}{lrrrrrr}
\toprule
Control & $m$ & $A_m$ & $\delta f_m$ & $\phi_m$ & $d_{m1}$ & $d_{m2}$ \\
\midrule
Base & 1 & 499.356870973 & 0.729575457 & 0.893280180 & 0.721912653 & 3.634924206 \\
Base & 2 & 908.360393229 & 0.700516529 & 0.875247296 & 25.408126165 & -2.200635732 \\
Base & 3 & 323.104448609 & 0.085174957 & 3.439167945 & -1.866080585 & 30.436548031 \\
Base & 4 & 858.102033073 & -0.540238565 & 4.066421927 & -3.453590845 & -8.347867580 \\
Base & 5 & 503.092690522 & -0.213302587 & 2.863877249 & -10.011271067 & -3.668036201 \\
Base & 6 & 909.663054130 & -0.228471229 & 2.840810995 & 31.731385823 & -25.121604338 \\
Base & 7 & 475.602536159 & 0.547749374 & 3.019163716 & 25.967547361 & -11.540750837 \\
Base & 8 & 782.060469625 & 1.145679655 & 2.106088624 & -8.449096585 & 32.761678288 \\
Base & 9 & 417.872241012 & 0.536677445 & 0.740236736 & -14.856156138 & -29.494207617 \\
Base & 10 & 894.871650749 & -1.917721784 & 4.080416637 & 5.408497466 & 4.999699121 \\
\midrule
PDRAG & 1 & 480.086306606 & 0.730725188 & 1.976266210 & 0.387064400 & 0.086883417 \\
PDRAG & 2 & 908.181148892 & 0.910927125 & 1.777917522 & 14.299311023 & 27.389277931 \\
PDRAG & 3 & 296.799019030 & 0.305518501 & 3.050312039 & -6.155854459 & 15.002762383 \\
PDRAG & 4 & 800.010421386 & -0.062848609 & 3.457166035 & -7.382315635 & -17.413621145 \\
PDRAG & 5 & 545.302703176 & -0.237144323 & 3.384612836 & -8.499043736 & 26.775252215 \\
PDRAG & 6 & 911.169339124 & 0.030322953 & 3.006395293 & 30.424585508 & -25.291967315 \\
PDRAG & 7 & 425.504822096 & -0.331870999 & 3.521205569 & 19.499521234 & -30.615285366 \\
PDRAG & 8 & 756.930245515 & 0.074695827 & 2.991254417 & -5.292591390 & 25.842984183 \\
PDRAG & 9 & 408.483882096 & 1.658074294 & 1.224484133 & -22.011571663 & 2.104219373 \\
PDRAG & 10 & 809.997544032 & 0.665709890 & 2.544672009 & 2.216735635 & 4.376911364 \\
\bottomrule
\end{tabular}
\end{table}

\begin{table}[!htbp]
\centering\footnotesize
\caption{Dimensionless PDRAG coefficients for $N=11$ in
Eq.~\eqref{supp:eq:ssh:command}; all corresponding Base coefficients
are zero.}
\label{supp:tab:ssh:derivatives}
\begin{tabular}{rrrrr}
\toprule
$m$ & $c_{m1}$ & $c_{m2}$ & $c_{m3}$ & $c_{m4}$ \\
\midrule
1 & -0.228750918425 & 0.158355030748 & 0.096067549800 & 0.106378044679 \\
2 & 0.432742250720 & -0.390686031047 & -0.067607739022 & 0.153857676907 \\
3 & -0.189700236330 & 0.129642720418 & 0.129932869162 & 0.083399145140 \\
4 & 0.152294886487 & -0.193647880408 & 0.087986276626 & -0.344608640089 \\
5 & -0.152887262982 & 0.116820533031 & 0.237343248516 & 0.074010721863 \\
6 & -0.244827070538 & -0.348227185029 & -0.400047729938 & -0.101080100463 \\
7 & -0.195090115398 & 0.121147888243 & 0.138916132574 & 0.248637795676 \\
8 & 0.011764924496 & -0.052469835101 & -0.064499790528 & -0.088016419515 \\
9 & 1.999804173388 & -1.775812309395 & -0.198989749477 & 1.356997289346 \\
10 & -0.280200898600 & 0.157704340493 & 0.603505601638 & 0.376411846675 \\
\bottomrule
\end{tabular}
\end{table}

\subsection{Joint trajectory calibration and resource enforcement}
\label{supp:ssh:calibration}

All $N-1$ ports are optimized jointly for the initial state
$\ket{Q_1}_{\rm d}$.
Let $P_C=\sum_m\ket{C_m}_{\rm d}\!\bra{C_m}_{\rm d}$ and define
\begin{align}
 \epsilon(t)&=\min_\theta
 \left\|e^{-\ii\theta}\ket{\widetilde\psi(t)}
 -\ket{\psi_{\rm SSH}(t)}\right\|^2,\notag\\
 \mathcal C_{\rm state}&=\frac{0.03}{25}\sum_{k=1}^{25}\epsilon(t_k)
 +\epsilon(T)+9\langle\widetilde\psi(T)|P_C|\widetilde\psi(T)\rangle,\notag\\
 t_k&=\tau+\frac{k-1}{24}(T-2\tau).
 \label{supp:eq:ssh:cost}
\end{align}
The endpoint has unit weight, and its coupler-state residual has total
weight ten after including the contribution already present in
$\epsilon(T)$.

With $v_m=u_m/(2\pi)$ in GHz and
$\sigma_m=\dot u_m/(2\pi)$ in GHz/ns, the physical limits are
$\max_t|v_m(t)|\leq1.9$ GHz and
$\max_t|\sigma_m(t)|\leq5$ GHz/ns for every port.
The slew is the analytic time derivative of the complete command,
including the carrier, fixed Stark phase, and all spline components.
During optimization, a conservative resource penalty is
\begin{align}
 \mathcal C_{\rm res}={}&36\sum_{m=1}^{N-1}
 [\max_k|v_m(t_k^{\rm r})|-1.89]_+^2\notag\\
 &+36\sum_{m=1}^{N-1}
 [\max_k|\sigma_m(t_k^{\rm r})|-4.98]_+^2,
 \label{supp:eq:ssh:penalty}
\end{align}
where $[z]_+=\max(z,0)$ and $t_k^{\rm r}=0,0.025,\ldots,T$ ns.
Optimization minimizes $\mathcal C_{\rm state}+\mathcal C_{\rm res}$.
Final candidate ranking minimizes the physical objective
$\mathcal C_{\rm state}$ among controls satisfying the physical limits.
The candidate pool retains every feasible seed and recorded optimization iterate,
including each earlier run's best control.
Each method uses one selected control for all reported observables.

Base has $5(N-1)$ free parameters and PDRAG has $9(N-1)$.
Their common bounds are $A_m\in[1,1900]$ MHz,
$\delta f_m\in[-8,8]$ MHz, $\phi_m\in[-2\pi,2\pi]$, and
$d_{m1},d_{m2}\in[-300,300]$ MHz, with $c_{mn}\in[-2,2]$ for PDRAG.
Calibration uses the bounded limited-memory Broyden--Fletcher--Goldfarb--Shanno
algorithm (L-BFGS-B) with a fourth-order symmetric split-operator adjoint
at a $0.025$-ns outer step.
Each Strang stage propagates the idle Hamiltonian exactly and applies
the coupler-frequency kick in the bare basis.
The three composition coefficients are $(a_Y,1-2a_Y,a_Y)$ with
$a_Y=(2-2^{1/3})^{-1}$; commands are evaluated at each stage midpoint.
The optimizer uses function tolerance $10^{-12}$, gradient tolerance
$10^{-8}$, memory length $30$, and at most $25$ line-search steps.
The fixed parameter scales are $70$ MHz for $A_m$, $0.15$ MHz for
$\delta f_m$, $1$ rad for $\phi_m$, $4$ MHz for each DC coefficient,
and $0.15$ for each derivative coefficient.

Each family receives four starting controls, each with a budget
of $400$ objective-and-gradient evaluations, followed by a $400$-evaluation
pass initialized from the other family's best available control.
Base sets all derivative coefficients to zero when importing a PDRAG seed.
The starts cover a nominal physical guess, mapped eleven-site Base and
PDRAG controls, and the best feasible distinct control available for the
same chain length or a nearby duration.
Seed lists are fixed before each job, and both families are evaluated
against the same target and resource constraints.
Initial seed assessments are counted separately from the equal
$2000$-evaluation optimization cap per family.
Table~\ref{supp:tab:ssh:optimization} records the actual optimization counts.
The counts include interrupted evaluations, which count toward the fixed budget.
Final resource checks use a $0.0025$-ns grid.

\begin{table}[!htbp]
\centering\footnotesize
\caption{Objective-and-gradient evaluation counts and final physical
resources. Excursion and analytic slew are maximized over every port and
the full command at $0.0025$-ns resolution.}
\label{supp:tab:ssh:optimization}
\begin{tabular}{rlrrr}
\toprule
$N$ & Control & Evaluations & Excursion (GHz) & Slew (GHz/ns) \\
\midrule
7 & Base & 1692 & 0.907479130 & 4.842171876 \\
7 & PDRAG & 1866 & 0.877165874 & 4.995020064 \\
9 & Base & 2000 & 0.840690588 & 4.573341527 \\
9 & PDRAG & 1874 & 0.901216194 & 4.995538052 \\
11 & Base & 1998 & 0.934777244 & 4.981519270 \\
11 & PDRAG & 1454 & 1.100403411 & 4.991248478 \\
13 & Base & 2000 & 0.958157275 & 4.985731745 \\
13 & PDRAG & 1700 & 0.964246438 & 4.998762596 \\
\bottomrule
\end{tabular}
\end{table}

\subsection{Observables and independent propagation}
\label{supp:ssh:validation}

\begin{figure}[!p]
\begingroup
\centering\footnotesize
\captionof{table}{Independent propagation checks. The ODE discrepancy is the larger
of the maximum complex-state difference and endpoint $F_{\rm SSH}$
difference between bare- and dressed-basis integrations.
The split diagnostic is the larger of the endpoint differences in receiver
population and leakage between the $0.025$-ns split propagator and tight-tolerance DOP853.}
\label{supp:tab:ssh:validation}
\begin{tabular}{rrrr}
\toprule
$N$ & Independent ODE discrepancy & Norm error & Split diagnostic \\
\midrule
7 & $3.4010\times10^{-10}$ & $2.1110\times10^{-12}$ & $5.1781\times10^{-8}$ \\
9 & $3.3022\times10^{-10}$ & $1.2410\times10^{-12}$ & $4.3866\times10^{-8}$ \\
11 & $8.0277\times10^{-10}$ & $4.6601\times10^{-12}$ & $1.2612\times10^{-7}$ \\
13 & $8.3232\times10^{-10}$ & $3.9184\times10^{-12}$ & $3.6257\times10^{-8}$ \\
\bottomrule
\end{tabular}
\endgroup

\par\bigskip

 \centering
 \includegraphics[width=0.90\textwidth]{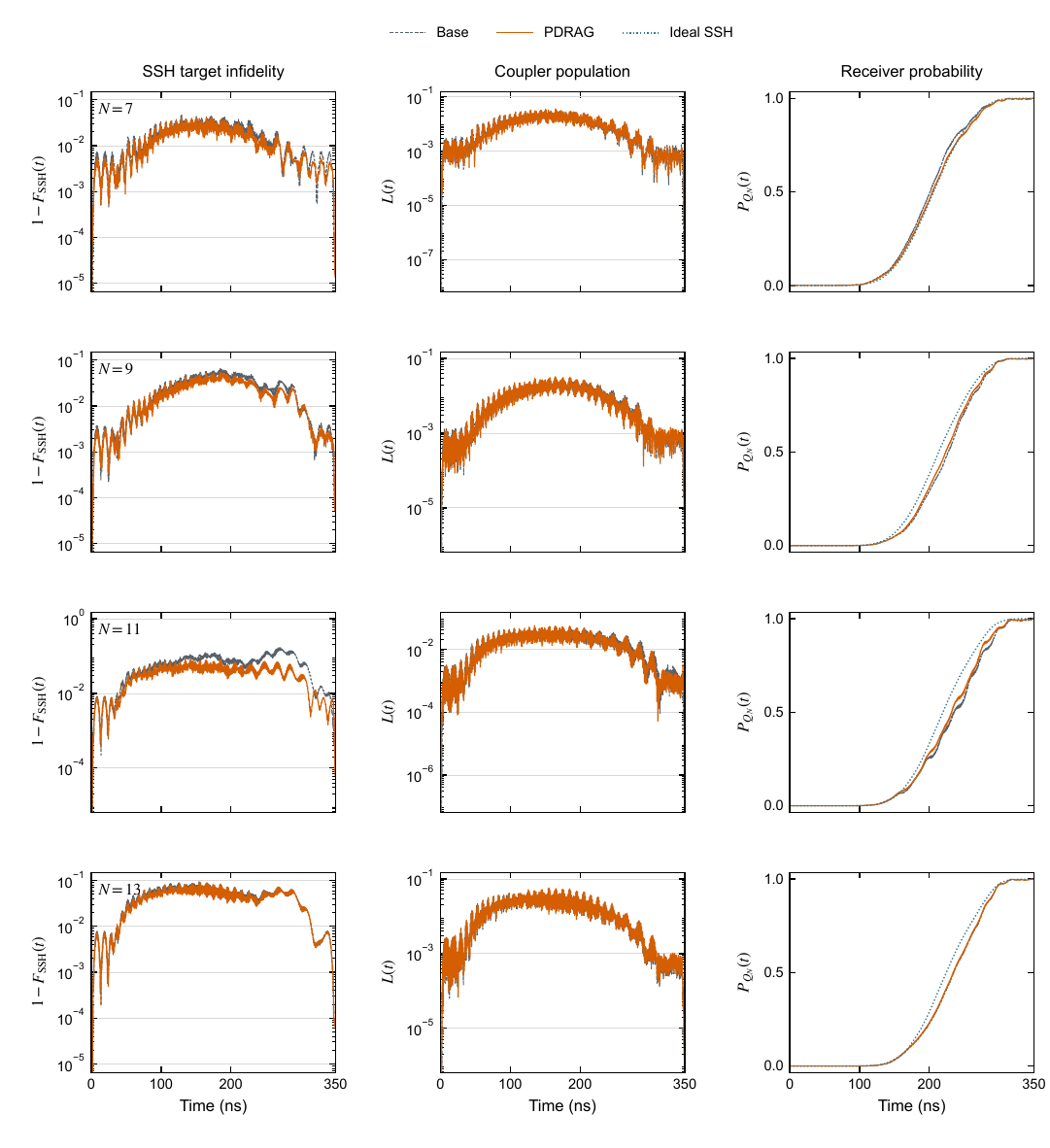}
 \caption{Complete $0.05$-ns trajectories at $T=350$ ns for
 $N=7,9,11,$ and $13$ (rows).
 The columns show infidelity relative to the finite-time ideal SSH state,
 idle dressed coupler population, and unconditional receiver probability.
 Grey dashed and orange solid curves denote Base and PDRAG; the blue
 dotted curves show the ideal SSH receiver probability.
 Logarithmic panels display positive samples.}
 \label{supp:fig:ssh:trajectories}
\end{figure}

The endpoint quantities in Fig.~6 and the complete trajectories in
Fig.~\ref{supp:fig:ssh:trajectories} are evaluated as
\begin{align}
 F_{\rm SSH}(t)&=|\langle\psi_{\rm SSH}(t)|\widetilde\psi(t)\rangle|^2,\notag\\
 L(t)&=\sum_{m=1}^{N-1}
 |{}_{\rm d}\langle C_m|\widetilde\psi(t)\rangle|^2,\notag\\
 P_{Q_N}(t)&=|{}_{\rm d}\langle Q_N|\widetilde\psi(t)\rangle|^2.
 \label{supp:eq:ssh:observables}
\end{align}
All three observables use the full $(2N-1)$-component state norm.
$F_{\rm SSH}$ is the unconditional overlap with the finite-time ideal
SSH state in the common tracked frame, and $P_{Q_N}$ is the population
at the receiving site.
Table~\ref{supp:tab:ssh:results} gives both metrics and residual leakage
for every chain length.
The ideal receiver probabilities are $0.999970435$, $0.999944496$,
$0.999677722$, and $0.998587674$ for $N=7,9,11,$ and $13$, respectively.
At $N=13$, the receiver probabilities are $0.998457603$ for Base and
$0.998454161$ for PDRAG, while $F_{\rm SSH}(T)$ increases from
$0.999871486$ to $0.999927977$.

\begin{table}[!htbp]
\centering\footnotesize
\caption{Independently validated endpoint observables at the common
duration $T=350$ ns. The same selected control determines every entry in
each row and the corresponding point in Fig.~6.}
\label{supp:tab:ssh:results}
\begin{tabular}{rlrrrr}
\toprule
$N$ & Control & $F_{\rm SSH}(T)$ & $1-F_{\rm SSH}(T)$ & $L(T)$ & $P_{Q_N}(T)$ \\
\midrule
7 & Base & 0.999930973468 & $6.90265\times10^{-5}$ & $4.33619\times10^{-7}$ & 0.999847568811 \\
7 & PDRAG & 0.999986453151 & $1.35468\times10^{-5}$ & $6.08331\times10^{-8}$ & 0.999962866202 \\
9 & Base & 0.999884307869 & $1.15692\times10^{-4}$ & $1.93797\times10^{-5}$ & 0.999797718581 \\
9 & PDRAG & 0.999946942953 & $5.30570\times10^{-5}$ & $5.79090\times10^{-6}$ & 0.999894908545 \\
11 & Base & 0.999617024291 & $3.82976\times10^{-4}$ & $7.52411\times10^{-5}$ & 0.998973596996 \\
11 & PDRAG & 0.999936066820 & $6.39332\times10^{-5}$ & $9.12161\times10^{-7}$ & 0.999750418342 \\
13 & Base & 0.999871485544 & $1.28514\times10^{-4}$ & $3.58426\times10^{-5}$ & 0.998457603175 \\
13 & PDRAG & 0.999927977309 & $7.20227\times10^{-5}$ & $1.66095\times10^{-6}$ & 0.998454161060 \\
\bottomrule
\end{tabular}
\end{table}

The state-error reduction factors are $5.10$, $2.18$, $5.99$, and $1.78$,
and the residual-leakage reduction factors are $7.13$, $3.35$, $82.5$, and
$21.6$ for $N=7,9,11,$ and $13$, respectively.
The time-averaged error is
$\overline\epsilon=T^{-1}\int_0^T[1-F_{\rm SSH}(t)]\,\dd t$.
Table~\ref{supp:tab:ssh:trajectory_metrics} gives these averages and
peak coupler populations from the full $0.05$-ns trajectories.

\begin{table}[!htbp]
\centering\footnotesize
\caption{Trajectory metrics at $T=350$ ns. Mean-error reduction compares
PDRAG with Base; peak coupler population is reported separately from
the residual leakage in Table~\ref{supp:tab:ssh:results}.}
\label{supp:tab:ssh:trajectory_metrics}
\begin{tabular}{rrrrrr}
\toprule
& \multicolumn{2}{c}{$\overline\epsilon$}
& \multicolumn{2}{c}{$\max_t L(t)$} & \\
$N$ & Base & PDRAG & Base & PDRAG & Mean-error reduction (\%) \\
\midrule
7 & 0.015035639 & 0.012736972 & 0.033205114 & 0.035857750 & 15.2881 \\
9 & 0.022277625 & 0.017861964 & 0.030669257 & 0.032775239 & 19.8211 \\
11 & 0.060827382 & 0.032287982 & 0.050056241 & 0.055940332 & 46.9187 \\
13 & 0.041023297 & 0.038152713 & 0.055192490 & 0.057798676 & 6.9974 \\
\bottomrule
\end{tabular}
\end{table}

With the selected parameters fixed, dressed-basis DOP853 propagation
uses relative tolerance $2\times10^{-10}$, absolute tolerance
$2\times10^{-12}$, and maximum step $0.06$ ns.
An independent implementation propagates the bare-basis equation
\begin{equation}
 \ii\partial_t\ket{\psi_b(t)}=
 \left[B(E-E_{Q_1})B^{\dagger}
 +\sum_m u_m(t)\ket{C_m}\bra{C_m}\right]\ket{\psi_b(t)},
 \label{supp:eq:ssh:independent}
\end{equation}
from $B_{\cdot,Q_1}$ with relative tolerance $3\times10^{-12}$,
absolute tolerance $3\times10^{-14}$, and maximum step $0.03$ ns.
Both results are transformed by Eq.~\eqref{supp:eq:ssh:tracked_frame}.
Table~\ref{supp:tab:ssh:validation} records the largest independent
propagation discrepancies, norm errors, and split-step diagnostics
across both pulse families for each chain length.
The source trajectories contain every state component at
$t=0,0.05,\ldots,350$ ns; their endpoints supply Fig.~6.
The complete controls, fixed Stark coefficients, dense trajectories,
and validation reports accompany the numerical source data.

\FloatBarrier

\section{Digital Floquet simulation and state transfer in SSH chains}
\label{supp:sec:alternating}

\subsection{Microscopic model and unequal-coupling schedule}
\label{supp:alternating:model}

The digital Floquet simulation in Fig.~7 uses the microscopic
Hamiltonian, device frequencies, and idle-dressed basis defined in
Eqs.~\eqref{supp:eq:ssh:device} and~\eqref{supp:eq:ssh:frequencies}.
Coherent evolution retains all $2N-1$ qubit-like and coupler-like modes in
the excitation-conserving single-excitation sector.
The two bond families of Eq.~\eqref{supp:eq:ssh:target} are activated
alternately, beginning with the odd bonds.
For parity $p=1,2$, let $s$ denote time within a layer and define
\begin{align}
 w_p(s)&=W(s/\tau_p)W[(T_p-s)/\tau_p],\notag\\
 J_p(s)&=J_p^{\max}w_p(s),\notag\\
 (T_1,T_2)&=(30,15)~\mathrm{ns},\qquad
 (\tau_1,\tau_2)=(5,2.5)~\mathrm{ns},\notag\\
 (J_1^{\max},J_2^{\max})/(2\pi)&=(10,20)~\mathrm{MHz}.
 \label{supp:eq:alternating:schedule}
\end{align}
Here $W$ is the switch in Eq.~\eqref{supp:eq:ssh:window}.
Thus $J_2(s)=2J_1(2s)$, and both layers have exchange area
$J_p^{\max}(T_p-\tau_p)=\pi/2$.
For $N=7,9,11,13$, the $N-1$ layers form $(N-1)/2$ Floquet periods
of $45$ ns each, giving $T_N=22.5(N-1)$ ns.
The ideal layer product maps $\ket{Q_1}_{\rm d}$ to
$(-\ii)^{N-1}\ket{Q_N}_{\rm d}$.
The microscopic state, including its coupler-like amplitudes, evolves
continuously across every layer boundary.

\subsection{Port commands and phase convention}
\label{supp:alternating:commands}

The nominal carrier $f_m^{(0)}$ and dressed coupler-frequency operator
$N_m$ follow Eq.~\eqref{supp:eq:ssh:seed}.
For an active port $m$ of parity $p$, define
\begin{align}
 M_{np}&=\max_{k=0,\ldots,4000}
 \left|w_p^{(n)}(kT_p/4000)\right|,\qquad n=1,\ldots,4,\notag\\
 \mathcal E_m(s)&=w_p(s)+\sum_{n=1}^{4}
 c_{mn}(-\ii)^n\frac{w_p^{(n)}(s)}{M_{np}},\notag\\
 \theta_m(s;t_\ell)&=2\pi[f_m^{(0)}+\delta f_m]s
 +\phi_m+2\pi f_m^{(0)}t_\ell,\notag\\
 \frac{u_m(t_\ell+s)}{2\pi}&=A_m\operatorname{Re}
 [\mathcal E_m(s)e^{-\ii\theta_m(s;t_\ell)}]
 +d_m w_p^2(s).
 \label{supp:eq:alternating:command}
\end{align}
The derivatives are evaluated analytically, with $M_{n2}=2^nM_{n1}$,
and the command is zero outside its active layer.
Time is expressed in ns and ordinary frequencies in GHz.
In the saved parameter vector
$(A_m,q_m,\phi_m,d_m,c_{m1},\ldots,c_{m4})$,
$\delta f_m=0.01q_m$ GHz, or $10q_m$ MHz.

At a layer beginning at $t_\ell$, the nominal carrier advances by
$2\pi f_m^{(0)}t_\ell$, whereas the calibrated frequency-offset phase
is referenced to the local time $s$.
For repetition $r=0,\ldots,(N-3)/2$, the actual start times are
$t_{r,1}=45r$ ns and $t_{r,2}=45r+30$ ns.
The common comparison frame is the idle interaction picture,
$\ket{\psi_I(t)}=e^{\ii Et}B^\dagger\ket{\psi_b(t)}$.
In this frame the microscopic Hamiltonian is
\begin{equation}
 \frac{H_I(t)}{\hbar}
 =\sum_m u_m(t)e^{\ii Et}N_m e^{-\ii Et}.
 \label{supp:eq:alternating:frame}
\end{equation}

\subsection{Independent parity-layer calibration and resource bounds}
\label{supp:alternating:calibration}

Each parity layer is calibrated on all $N$ logical input states.
Let $P_Q$ embed those states in the full dressed space, let
$X_p=U_{I,p}(T_p)P_Q$ be the propagated $(2N-1)\times N$ isometry
for a layer referenced to $t_\ell=0$, and let $V_p$ be its ideal
logical exchange matrix.
The objective is the phase-aligned Frobenius cost
\begin{align}
 \gamma_p&=\arg\Tr[V_p^\dagger P_Q^\dagger X_p],\notag\\
 C_p&=\frac{1}{N}\left\|e^{-\ii\gamma_p}X_p-P_QV_p\right\|_F^2.
 \label{supp:eq:alternating:objective}
\end{align}
The full-space residual includes both logical operation error and
coupler-like population for every logical input.
The odd and even controls are fitted independently and then held fixed
throughout the Floquet sequence, with the phase convention of
Eq.~\eqref{supp:eq:alternating:command}.
Each active port has four Base parameters $(A_m,\delta f_m,\phi_m,d_m)$;
PDRAG adds four real derivative coefficients $c_{mn}$.
Base fixes $c_{mn}=0$.

Both families use $0.001\leq A_m\leq1.9$ GHz,
$|\delta f_m|\leq160$ MHz, $|\phi_m|\leq2\pi$,
$|d_m|\leq0.6$ GHz, and, for PDRAG, $|c_{mn}|\leq6$.
The complete real command obeys
\begin{equation}
 \max_{m,t}\frac{|u_m(t)|}{2\pi}\leq1.9~\mathrm{GHz},
 \qquad
 \max_{m,t}\frac{|\dot u_m(t)|}{2\pi}
 \leq5~\mathrm{GHz/ns}.
 \label{supp:eq:alternating:resources}
\end{equation}
Optimization uses inner bounds of $1.899$ GHz and $4.995$ GHz/ns.
The slew is evaluated from the analytic derivative of the complete
carrier-modulated command, including the DC contribution.
Every actual start phase is included in the constraints.
Candidate extrema are located on a $0.01$-ns grid and refined with
analytic time derivatives; parameter derivatives of the attained
extrema supply the constraint Jacobian.

Starting from preliminary full-operation calibrations, each parity and
pulse family is screened with $64$ scrambled Sobol candidates
($32$ local and $32$ broader perturbations), followed by four SLSQP
starts with a cumulative cap of $240$ objective evaluations per start.
The Sobol seed is $20261030+10N+(p-1)+\mathbf1_{\rm PDRAG}$.
The starts comprise the preceding feasible solution, the preceding
optimizer endpoint (or a $1.5$-MHz offset perturbation when these coincide),
and the two lowest-cost screened candidates. The preceding feasible
solution is retained if its layer cost is lower.
SLSQP uses analytic state sensitivities and a stopping tolerance of
$10^{-10}$ on the scaled objective $100C_p$.
All $64$ local refinements across the four chain lengths, two parities,
and two pulse families terminate successfully.
Layer optimization uses DOP853 with relative/absolute tolerances
$3\times10^{-7}/3\times10^{-9}$ and maximum step $0.15$ ns.
The assembled sequence is reevaluated at
$10^{-10}/10^{-12}$ and maximum step $0.04$ ns.

\subsection{Endpoint observables and independent validation}
\label{supp:alternating:validation}

For initial state $\ket{Q_1}_{\rm d}$, the reported endpoint quantities are
\begin{align}
 1-F_{\rm tr}(T_N)
 &=1-\left|{}_{\rm d}\!\bra{Q_N}\psi_I(T_N)\rangle\right|^2,
 \notag\\
 L(T_N)&=\sum_{m=1}^{N-1}
 \left|{}_{\rm d}\!\bra{C_m}\psi_I(T_N)\rangle\right|^2.
 \label{supp:eq:alternating:metrics}
\end{align}
An independent implementation reconstructs the microscopic matrix,
dressed-mode assignment, port commands, and ideal exchange product
from the saved controls. It propagates all $N$ logical inputs directly
in the global idle interaction picture, using DOP853 with
relative/absolute tolerances $2\times10^{-12}/2\times10^{-14}$ and
maximum step $0.025$ ns.
Table~\ref{supp:tab:alternating:endpoints} lists the independently
evaluated endpoint metrics and their Base-to-PDRAG improvement factors.

\begin{table}[!htbp]
\centering\small
\caption{Independently validated endpoint results for state transfer
in the digital Floquet SSH simulation. Infidelity $I=1-F_{\rm tr}$ and leakage $L$ are
given in percent. Improvement factors are the corresponding Base
value divided by the PDRAG value, evaluated before rounding.}
\label{supp:tab:alternating:endpoints}
\begin{tabular}{rrrrrrrr}
\toprule
& & \multicolumn{3}{c}{Final infidelity $I$ (\%)}
& \multicolumn{3}{c}{Residual leakage $L$ (\%)} \\
\cmidrule(lr){3-5}\cmidrule(lr){6-8}
$N$ & $T_N$ (ns) & Base & PDRAG & Reduction
& Base & PDRAG & Reduction \\
\midrule
7  & 135 & 5.70410 & 1.49826 & 3.807 & 4.15259 & 0.185026 & 22.443 \\
9  & 180 & 7.36796 & 0.778164 & 9.468 & 5.85678 & 0.199470 & 29.362 \\
11 & 225 & 11.7287 & 1.30572 & 8.983 & 9.15148 & 0.118865 & 76.991 \\
13 & 270 & 13.6261 & 2.01540 & 6.761 & 9.91765 & 0.250857 & 39.535 \\
\bottomrule
\end{tabular}
\end{table}

Across all eight chains, absolute discrepancies between the two
propagators remain below $6.21\times10^{-11}$ in final infidelity and
$7.77\times10^{-12}$ in residual leakage.
The maximum matrix-element difference over the saved full-isometry
trajectories is below $2.77\times10^{-9}$, and the independent propagation
has maximum probability-norm error below $1.08\times10^{-12}$.
Resource validation evaluates analytic commands and slew on a
$0.0025$-ns grid, refines the largest sampled peaks locally, and checks
every repeated layer and its endpoints.
The verified peak excursions span $0.4614$--$0.4676$ GHz for Base and
$0.5941$--$0.6296$ GHz for PDRAG; their maximum slew values span
$2.5927$--$2.6060$ and reach $4.9950$ GHz/ns, respectively.
Both commands and slew vanish at layer endpoints.
Machine-readable controls, endpoint data, and independent validation
records accompany the source data.

\clearpage


\begin{thebibliography}{99}

\bibitem{Barends2014}
R. Barends \textit{et al.},
Superconducting quantum circuits at the surface code threshold for fault tolerance,
\href{https://doi.org/10.1038/nature13171}{Nature \textbf{508}, 500--503 (2014)}.

\bibitem{Bertet2006}
P. Bertet, C. J. P. M. Harmans, and J. E. Mooij,
Parametric coupling for superconducting qubits,
\href{https://doi.org/10.1103/PhysRevB.73.064512}{Phys. Rev. B \textbf{73}, 064512 (2006)}.

\bibitem{Didier2018}
N. Didier, E. A. Sete, M. P. da Silva, and C. Rigetti,
Analytical modeling of parametrically modulated transmon qubits,
\href{https://doi.org/10.1103/PhysRevA.97.022330}{Phys. Rev. A \textbf{97}, 022330 (2018)}.

\bibitem{Caldwell2018}
S. A. Caldwell \textit{et al.},
Parametrically activated entangling gates using transmon qubits,
\href{https://doi.org/10.1103/PhysRevApplied.10.034050}{Phys. Rev. Applied \textbf{10}, 034050 (2018)}.

\bibitem{Reagor2018}
M. Reagor \textit{et al.},
Demonstration of universal parametric entangling gates on a multi-qubit lattice,
\href{https://doi.org/10.1126/sciadv.aao3603}{Sci. Adv. \textbf{4}, eaao3603 (2018)}.

\bibitem{Li2018}
X. Li \textit{et al.},
Perfect quantum state transfer in a superconducting qubit chain with parametrically tunable couplings,
\href{https://doi.org/10.1103/PhysRevApplied.10.054009}{Phys. Rev. Applied \textbf{10}, 054009 (2018)}.

\bibitem{Salathe2015}
Y. Salath{\'e} \textit{et al.},
Digital quantum simulation of spin models with circuit quantum electrodynamics,
\href{https://doi.org/10.1103/PhysRevX.5.021027}{Phys. Rev. X \textbf{5}, 021027 (2015)}.

\bibitem{Barends2015}
R. Barends \textit{et al.},
Digital quantum simulation of fermionic models with a superconducting circuit,
\href{https://doi.org/10.1038/ncomms8654}{Nat. Commun. \textbf{6}, 7654 (2015)}.

\bibitem{Rosen2024}
I. T. Rosen \textit{et al.},
A synthetic magnetic vector potential in a 2D superconducting qubit array,
\href{https://doi.org/10.1038/s41567-024-02661-3}{Nat. Phys. \textbf{20}, 1881--1887 (2024)}.

\bibitem{Cai2019}
W. Cai \textit{et al.},
Observation of topological magnon insulator states in a superconducting circuit,
\href{https://doi.org/10.1103/PhysRevLett.123.080501}{Phys. Rev. Lett. \textbf{123}, 080501 (2019)}.

\bibitem{Qian2025Topology}
H. Qian \textit{et al.},
Programmable higher-order nonequilibrium topological phases on a superconducting quantum processor,
\href{https://doi.org/10.1126/science.adp6802}{Science \textbf{390}, 930--934 (2025)}.

\bibitem{Liu2026Prethermal}
Z.-H. Liu \textit{et al.},
Prethermalization by random multipolar driving on a 78-qubit processor,
\href{https://doi.org/10.1038/s41586-025-09977-x}{Nature \textbf{650}, 79--85 (2026)}.

\bibitem{Foxen2020}
B. Foxen \textit{et al.},
Demonstrating a continuous set of two-qubit gates for near-term quantum algorithms,
\href{https://doi.org/10.1103/PhysRevLett.125.120504}{Phys. Rev. Lett. \textbf{125}, 120504 (2020)}.

\bibitem{Koch2007}
J. Koch, T. M. Yu, J. Gambetta, A. A. Houck, D. I. Schuster,
J. Majer, A. Blais, M. H. Devoret, S. M. Girvin, and R. J. Schoelkopf,
Charge-insensitive qubit design derived from the Cooper pair box,
\href{https://doi.org/10.1103/PhysRevA.76.042319}{Phys. Rev. A \textbf{76}, 042319 (2007)}.

\bibitem{Wood2018}
C. J. Wood and J. M. Gambetta,
Quantification and characterization of leakage errors,
\href{https://doi.org/10.1103/PhysRevA.97.032306}{Phys. Rev. A \textbf{97}, 032306 (2018)}.

\bibitem{Chen2016}
Z. Chen \textit{et al.},
Measuring and suppressing quantum state leakage in a superconducting qubit,
\href{https://doi.org/10.1103/PhysRevLett.116.020501}{Phys. Rev. Lett. \textbf{116}, 020501 (2016)}.

\bibitem{McEwen2021Leakage}
M. McEwen \textit{et al.},
Removing leakage-induced correlated errors in superconducting quantum error correction,
\href{https://doi.org/10.1038/s41467-021-21982-y}{Nat. Commun. \textbf{12}, 1761 (2021)}.

\bibitem{Miao2023Leakage}
K. C. Miao \textit{et al.},
Overcoming leakage in quantum error correction,
\href{https://doi.org/10.1038/s41567-023-02226-w}{Nat. Phys. \textbf{19}, 1780--1786 (2023)}.

\bibitem{Karamlou2024Entanglement}
A. H. Karamlou \textit{et al.},
Probing entanglement in a 2D hard-core Bose--Hubbard lattice,
\href{https://doi.org/10.1038/s41586-024-07325-z}{Nature \textbf{629}, 561--566 (2024)}.

\bibitem{McKay2016}
D. C. McKay, S. Filipp, A. Mezzacapo, E. Magesan, J. M. Chow, and J. M. Gambetta,
Universal gate for fixed-frequency qubits via a tunable bus,
\href{https://doi.org/10.1103/PhysRevApplied.6.064007}{Phys. Rev. Applied \textbf{6}, 064007 (2016)}.

\bibitem{Yan2018}
F. Yan, P. Krantz, Y. Sung, M. Kjaergaard, D. L. Campbell, T. P. Orlando,
S. Gustavsson, and W. D. Oliver,
Tunable coupling scheme for implementing high-fidelity two-qubit gates,
\href{https://doi.org/10.1103/PhysRevApplied.10.054062}{Phys. Rev. Applied \textbf{10}, 054062 (2018)}.

\bibitem{Roth2017}
M. Roth, M. Ganzhorn, N. Moll, S. Filipp, G. Salis, and S. Schmidt,
Analysis of a parametrically driven exchange-type gate and a two-photon excitation gate between superconducting qubits,
\href{https://doi.org/10.1103/PhysRevA.96.062323}{Phys. Rev. A \textbf{96}, 062323 (2017)};
\href{https://doi.org/10.1103/PhysRevA.97.049903}{\textbf{97}, 049903(E) (2018)}.

\bibitem{Mundada2019}
P. Mundada, G. Zhang, T. Hazard, and A. Houck,
Suppression of qubit crosstalk in a tunable coupling superconducting circuit,
\href{https://doi.org/10.1103/PhysRevApplied.12.054023}{Phys. Rev. Applied \textbf{12}, 054023 (2019)}.

\bibitem{Stehlik2021}
J. Stehlik, D. M. Zajac, D. L. Underwood, T. Phung, J. Blair, S. Carnevale,
D. Klaus, G. A. Keefe, A. Carniol, M. Kumph, M. Steffen, and O. E. Dial,
Tunable coupling architecture for fixed-frequency transmon superconducting qubits,
\href{https://doi.org/10.1103/PhysRevLett.127.080505}{Phys. Rev. Lett. \textbf{127}, 080505 (2021)}.

\bibitem{Ganzhorn2019}
M. Ganzhorn, D. J. Egger, P. Barkoutsos, P. Ollitrault, G. Salis, N. Moll,
M. Roth, A. Fuhrer, P. Mueller, S. Woerner, I. Tavernelli, and S. Filipp,
Gate-efficient simulation of molecular eigenstates on a quantum computer,
\href{https://doi.org/10.1103/PhysRevApplied.11.044092}{Phys. Rev. Applied \textbf{11}, 044092 (2019)}.

\bibitem{Han2020}
X. Y. Han, T. Q. Cai, X. G. Li, Y. K. Wu, Y. W. Ma, Y. L. Ma,
J. H. Wang, H. Y. Zhang, Y. P. Song, and L. M. Duan,
Error analysis in suppression of unwanted qubit interactions for a parametric gate in a tunable superconducting circuit,
\href{https://doi.org/10.1103/PhysRevA.102.022619}{Phys. Rev. A \textbf{102}, 022619 (2020)}.

\bibitem{Sung2021}
Y. Sung \textit{et al.},
Realization of high-fidelity CZ and $ZZ$-free iSWAP gates with a tunable coupler,
\href{https://doi.org/10.1103/PhysRevX.11.021058}{Phys. Rev. X \textbf{11}, 021058 (2021)}.

\bibitem{Martinis2014}
J. M. Martinis and M. R. Geller,
Fast adiabatic qubit gates using only $\sigma_z$ control,
\href{https://doi.org/10.1103/PhysRevA.90.022307}{Phys. Rev. A \textbf{90}, 022307 (2014)}.

\bibitem{Rol2019}
M. A. Rol \textit{et al.},
Fast, high-fidelity conditional-phase gate exploiting leakage interference in weakly anharmonic superconducting qubits,
\href{https://doi.org/10.1103/PhysRevLett.123.120502}{Phys. Rev. Lett. \textbf{123}, 120502 (2019)}.

\bibitem{Motzoi2009}
F. Motzoi, J. M. Gambetta, P. Rebentrost, and F. K. Wilhelm,
Simple pulses for elimination of leakage in weakly nonlinear qubits,
\href{https://doi.org/10.1103/PhysRevLett.103.110501}{Phys. Rev. Lett. \textbf{103}, 110501 (2009)}.

\bibitem{Gambetta2011}
J. M. Gambetta, F. Motzoi, S. T. Merkel, and F. K. Wilhelm,
Analytic control methods for high-fidelity unitary operations in a weakly nonlinear oscillator,
\href{https://doi.org/10.1103/PhysRevA.83.012308}{Phys. Rev. A \textbf{83}, 012308 (2011)}.

\bibitem{Motzoi2013}
F. Motzoi and F. K. Wilhelm,
Improving frequency selection of driven pulses using derivative-based transition suppression,
\href{https://doi.org/10.1103/PhysRevA.88.062318}{Phys. Rev. A \textbf{88}, 062318 (2013)}.

\bibitem{Hyyppa2024}
E. Hyypp{\"a} \textit{et al.},
Reducing leakage of single-qubit gates for superconducting quantum processors using analytical control pulse envelopes,
\href{https://doi.org/10.1103/PRXQuantum.5.030353}{PRX Quantum \textbf{5}, 030353 (2024)}.

\bibitem{Jesus2026Blueprint}
J. D. Da Costa Jesus, B. Li, Y. Gao, R. Barends,
F. A. C{\'a}rdenas-L{\'o}pez, and F. Motzoi,
Analytical blueprint for 99.999\% fidelity X-gates on present superconducting hardware under strong driving,
\href{https://doi.org/10.1103/sdxb-v39h}{PRX Quantum \textbf{7}, 033040 (2026)}.

\bibitem{Jin2025}
X. Y. Jin, Z. Parrott, K. Cicak, S. Kotler, F. Lecocq, J. Teufel, J. Aumentado, E. Kapit, and R. W. Simmonds,
Superconducting architecture demonstrating fast, tunable high-fidelity CZ gates with parametric control of ZZ coupling,
\href{https://doi.org/10.1103/kmls-lgp5}{Phys. Rev. Applied \textbf{24}, 064026 (2025)}.

\bibitem{Georgiadis2026}
D. Georgiadis, B. Li, A. Galicia, R. Barends, F. A. C{\'a}r\-denas-L{\'o}pez, and F. Motzoi,
Simple analytical flux-tuned iSWAP pulses for leakage suppression,
\href{https://doi.org/10.48550/arXiv.2606.13052}{arXiv:2606.13052 (2026)}.

\bibitem{Heunisch2026}
L. Heunisch, M. J. Hartmann, and A. A. Clerk,
Analytic leakage suppression with a single control field: fast two-qubit gates with tunable couplers,
\href{https://arxiv.org/abs/2609.20766}{arXiv:2609.20766 (2026)}.

\bibitem{Weinberg2017Floquet}
P. Weinberg, M. Bukov, L. D'Alessio, A. Polkovnikov, S. Vajna, and M. Kolodrubetz,
Adiabatic perturbation theory and geometry of periodically-driven systems,
\href{https://doi.org/10.1016/j.physrep.2017.05.003}{Phys. Rep. \textbf{688}, 1--35 (2017)}.

\bibitem{Wang2018Topological}
T. Wang, Z. Zhang, L. Xiang, Z. Gong, J. Wu, and Y. Yin,
Simulating a topological transition in a superconducting phase qubit by fast adiabatic trajectories,
\href{https://doi.org/10.1007/s11433-017-9156-1}{Sci. China Phys. Mech. Astron. \textbf{61}, 047411 (2018)}.

\bibitem{Dangelis2020Transfer}
F. M. D'Angelis, F. A. Pinheiro, D. Gu{\'e}ry-Odelin,
S. Longhi, and F. Impens,
Fast and robust quantum state transfer in a topological
Su--Schrieffer--Heeger chain with next-to-nearest-neighbor interactions,
\href{https://doi.org/10.1103/PhysRevResearch.2.033475}{Phys. Rev. Research \textbf{2}, 033475 (2020)}.

\bibitem{Mei2018Transfer}
F. Mei, G. Chen, L. Tian, S.-L. Zhu, and S. Jia,
Robust quantum state transfer via topological edge states in superconducting qubit chains,
\href{https://doi.org/10.1103/PhysRevA.98.012331}{Phys. Rev. A \textbf{98}, 012331 (2018)}.

\bibitem{Asboth2014DrivenSSH}
J. K. Asb{\'o}th, B. Tarasinski, and P. Delplace,
Chiral symmetry and bulk-boundary correspondence in periodically driven one-dimensional systems,
\href{https://doi.org/10.1103/PhysRevB.90.125143}{Phys. Rev. B \textbf{90}, 125143 (2014)}.

\bibitem{SupplementalMaterial}
See Supplemental Material for device models, calibration procedures,
Floquet and charge-model derivations, convergence and resource checks,
the raised-cosine implementation, sinusoidal-envelope calibration,
and continuous and digital Floquet simulation of SSH chains.
It includes
Refs.~\cite{Wood2018,McKay2016,Roth2017,Hyyppa2024,Weinberg2017Floquet,Nielsen2002,McKay2017}.

\bibitem{Schutjens2013}
R. Schutjens, F. Abu Dagga, D. J. Egger, and F. K. Wilhelm,
Single-qubit gates in frequency-crowded transmon systems,
\href{https://doi.org/10.1103/PhysRevA.88.052330}{Phys. Rev. A \textbf{88}, 052330 (2013)}.

\bibitem{Theis2016}
L. S. Theis, F. Motzoi, and F. K. Wilhelm,
Simultaneous gates in frequency-crowded multilevel systems using fast, robust, analytic control shapes,
\href{https://doi.org/10.1103/PhysRevA.93.012324}{Phys. Rev. A \textbf{93}, 012324 (2016)}.

\bibitem{Li2025Qudit}
B. Li, F. A. C{\'a}rdenas-L{\'o}pez, A. Lupascu, and F. Motzoi,
Universal pulses for superconducting qudit ladder gates,
\href{https://doi.org/10.1103/9dxw-4c7y}{PRX Quantum \textbf{6}, 030357 (2025)}.

\bibitem{Wang2025Balanced}
R. Wang \textit{et al.},
Suppressing spurious transitions using spectrally balanced pulse,
\href{https://doi.org/10.1103/h4xf-vq2l}{Phys. Rev. Lett. \textbf{135}, 160804 (2025)}.

\bibitem{Nielsen2002}
M. A. Nielsen,
A simple formula for the average gate fidelity of a quantum dynamical operation,
\href{https://doi.org/10.1016/S0375-9601(02)01272-0}{Phys. Lett. A \textbf{303}, 249--252 (2002)}.

\bibitem{McKay2017}
D. C. McKay, C. J. Wood, S. Sheldon, J. M. Chow, and J. M. Gambetta,
Efficient $Z$ gates for quantum computing,
\href{https://doi.org/10.1103/PhysRevA.96.022330}{Phys. Rev. A \textbf{96}, 022330 (2017)}.

\bibitem{Zhang2022DigitalFloquet}
X. Zhang \textit{et al.},
Digital quantum simulation of Floquet symmetry-protected topological phases,
\href{https://doi.org/10.1038/s41586-022-04854-3}{Nature \textbf{607}, 468--473 (2022)}.

\bibitem{Kelly2014}
J. Kelly \textit{et al.},
Optimal quantum control using randomized benchmarking,
\href{https://doi.org/10.1103/PhysRevLett.112.240504}{Phys. Rev. Lett. \textbf{112}, 240504 (2014)}.

\bibitem{Werninghaus2021}
M. Werninghaus, D. J. Egger, F. Roy, S. Machnes, F. K. Wilhelm, and S. Filipp,
Leakage reduction in fast superconducting qubit gates via optimal control,
\href{https://doi.org/10.1038/s41534-020-00346-2}{npj Quantum Inf. \textbf{7}, 14 (2021)}.

\end{thebibliography}
\end{document}